\documentclass[manuscript, screen, nonacm]{acmart}

\AtBeginDocument{%
  \providecommand\BibTeX{{%
    \normalfont B\kern-0.5em{\scshape i\kern-0.25em b}\kern-0.8em\TeX}}}

\setcopyright{none}
\begin{document}

\title{WEIRD Audits? Research Trends, Linguistic and Geographical Disparities in the Algorithm Audits of Online Platforms - A Systematic Literature Review}

\author{Aleksandra Urman}
\email{urman@ifi.uzh.ch}
\affiliation{%
  \institution{University of Zurich}
  \city{Zurich}
  \country{Switzerland}
}
\author{Mykola Makhortykh}
\email{mykola.makhortykh@unibe.ch}
\affiliation{%
  \institution{University of Bern}
  \city{Bern}
  \country{Switzerland}
}

\author{Aniko Hannak}
\email{hannak@ifi.uzh.ch}
\affiliation{%
  \institution{University of Zurich}
  \city{Zurich}
  \country{Switzerland}
}

\renewcommand{\shortauthors}{Urman et al.}

\begin{abstract}
  The increasing reliance on complex algorithmic systems by online platforms has sparked a growing need for algorithm auditing, a methodology evaluating these systems' functionality and impact. In this paper, we systematically review 176 peer-reviewed online platform-focused algorithm auditing studies and identify trends in their methodological approaches, the geographic distribution of authors, and the selection of platforms, languages, geographies, and group-based attributes in the focus of the reviewed research. We find a significant skew of research focus towards few online platforms, Western contexts, particularly the US, and English language data. Additionally, our analysis indicates a tendency to focus on a narrow set of group-based attributes, often operationalized in simplified ways, which might obscure more nuanced aspects of algorithmic bias and discrimination. We provide a clearer understanding of the current state of the online platform-focused algorithm auditing and identify gaps to be addressed for a more inclusive and representative research landscape.
\end{abstract}





\maketitle

\section{Introduction}
The ubiquity of intransparent algorithms adopted by (online) platforms led to the emergence of a research methodology known as algorithm(ic)\footnote{Both terms are used in the field to refer to this methodology, we will use "algorithm auditing" in this paper for consistency.} auditing. It allows researchers to evaluate the functionality and/or impact of algorithmic systems and diagnose problems in algorithmic decision-making such as discrimination of social groups or misrepresentation of societal phenomena. Although the field is relatively young, it is developing quickly, and in an interdisciplinary manner: While auditing methodology emerged within the Computer Science community, it has been adopted by other disciplines such as Social Sciences \cite{haim_burst_2018,puschmann_beyond_2019}. The fast-paced and interdisciplinary nature of algorithm auditing makes it difficult to keep track of the field and identify the main trends in auditing research. Addressing this, \cite{bandy_problematic_2021} conducted the first and to date the only systematic review of the literature on algorithm audits carried out within academic research, identifying important imbalances in terms of the domains and platforms that are the focus of audit studies. However, other important trends in auditing literature - such as which national and linguistic contexts are the focus of auditing research - have not been examined before. 

It is widely recognized that conclusions drawn about the design and performance of technologies in specific contexts do not necessarily generalize to other contexts \cite{linxen_how_2021, metaxa_auditing_2021, septiandri_weird_2023}. Concerning algorithm auditing, for instance, if we find that a certain social media algorithm is more likely to prioritize right-leaning vs. left-leaning content in user feeds in the US, this does not automatically mean that the same effect will be observed in a Western European multiparty democracy or an authoritarian context. This is because, in the case of this example of a social media-based recommendation algorithm, even if the algorithms deployed by a platform work in the exact same ways in different countries or languages, the pool of content available for an algorithm to select from differs across contexts. Thus, the same algorithm will return (qualitatively) different sets of results depending on the linguistic and / or cultural context (e.g., national or geographic) in which it is deployed \cite{urman_right_2023}. Thus, it is highly relevant to evaluate which of these contexts are (under)researched in algorithm auditing. Furthermore, since imbalances in the contexts scrutinized by a research field can be associated with imbalances in the geographical distribution of researchers \cite{septiandri_weird_2023,linxen_how_2021}, it is important to establish what the corresponding distribution is in auditing research. 

In response to these gaps, in this paper, we conduct a systematic literature review of \textit{public-facing} algorithmic systems-focused academic algorithm audits and, building on \cite{bandy_problematic_2021}, update the review to catch up with the rapidly developing field, reviewing 176 research papers. Our analysis is structured around three research questions:

\textbf{RQ1:} \textit{What kinds of problematic machine behavior have been diagnosed by previous algorithm audits?}

\textbf{RQ2:}\textit{ What has been the focus of the previous algorithm audits in terms of (RQ2a) geography, (RQ2b) language, (RQ2c) 
group-based attributes?}

\textbf{RQ3:} \textit{What is the geographic distribution of the institutions where algorithm audit studies have been conducted?} 

This systematic literature review \textbf{focuses on audits of public-facing (online) platforms and their algorithms}, such as those used by social media, search engines, and e-commerce sites. The choice to limit our focus to the audits of such platforms is directly related to our explicit interest in \textbf{linguistic and geographical disparities}. Algorithmic systems on online platforms are different from more specialized systems, such as facial recognition systems used by law enforcement, that are not available to the general public and are often used locally. Public facing algorithmic systems are used globally, and are universally, regardless of the researchers' location, accessible for (closed-source) algorithm audits. Thus, they are not only a highly relevant target for audits, but any skews in the geographic and/or linguistic focus of auditing literature focusing on such systems can not be explained by the imbalances in the researchers' access to them across geographies. 

This paper makes several \textbf{contributions}. \textbf{First,} we reaffirm that the imbalances identified by \cite{bandy_problematic_2021} concerning the problems and domains examined in algorithm audits are still present and, in some cases, have become even more pronounced. \textbf{Second,} we provide empirical evidence showing that the field is disproportionately focused on the contexts of liberal democracies in Western Europe and North America, with an emphasis on English-language data. \textbf{Third,} we reveal that audits focusing on group-based attributes (e.g., gender or race) tend to concentrate on a limited set of these attributes and often operationalize them in overly simplistic ways, such as treating gender and race as binary categories. \textbf{Finally, }we establish that the authors of these auditing studies are predominantly affiliated with academic institutions in the US or a small number of European countries, a skew that only partially aligns with the observed geographical focus of the research.

\section{Related Work}

The field of algorithm auditing is rapidly evolving, but to date, the only systematic literature review is the work by \cite{bandy_problematic_2021}. This review identified key problematic algorithmic behaviors, evaluated the platforms most frequently studied, and highlighted critical gaps. For instance, it found a concentration of audits on platforms like Google, which, while influential, has led to the neglect of others, such as YouTube, that significantly shape public discourse and social behavior.

While \cite{bandy_problematic_2021} speculated about the lack of linguistic and geographic diversity in algorithm auditing, these assumptions were not empirically tested, leaving important questions about the global applicability of research findings. If audits are predominantly conducted in English-speaking, Western contexts, there is a risk that their findings may not translate to non-Western or non-English-speaking regions, which may face distinct algorithmic challenges due to cultural, linguistic, or regulatory differences.

The importance of context in understanding the generalizability of empirical findings has been well-established in other disciplines. For example, in the behavioral sciences, a landmark study by \cite{henrich_most_2010} revealed that behavioral sciences disproportionately focus on WEIRD—Western, Educated, Industrialized, Rich, and Democratic—societies, leading to a narrow understanding of human behavior that does not adequately account for the diversity of global experiences. Similarly, audits conducted in WEIRD contexts may not fully capture the diverse ways in which algorithms impact people across different cultural and societal landscapes \cite{urman_right_2023,metaxa_auditing_2021}.

Similar concerns have been raised in the field of computing broadly, particularly in areas related to human-computer interaction (HCI) and AI fairness and ethics. Researchers have found that findings from society-related computing work, such as those in HCI or algorithmic fairness, often do not generalize across different contexts. For example, \cite{ma_enthusiasts_2022} and \cite{bentvelzen_designing_2023} have shown that design and interaction patterns that work well in Western contexts may not be applicable in non-Western societies due to cultural differences.

The disproportionate focus on Western countries, particularly the United States, has also been linked to important deficits in computing fairness research. For instance, categories that dominate the US public discourse, such as race and gender, are scrutinized with regard to discrimination and fairness far more often than other attributes like age. Additionally, biases that are particularly relevant in non-Western cultures are often overlooked, as noted by \cite{sambasivan_re-imagining_2021}. This narrow focus can lead to a skewed understanding of algorithmic fairness, where the specific needs and challenges of non-Western societies are underrepresented.

These observations have prompted a growing number of researchers to scrutinize the national contexts that tend to dominate computing research. For example, \cite{linxen_how_2021} found that 73\% of studies presented at CHI, one of the leading conferences on human-computer interaction, are based on samples from WEIRD contexts. Furthermore, the authors of these studies are also disproportionately affiliated with institutions in industrialized, democratic, and rich countries. Similarly, a study by \cite{septiandri_weird_2023} on papers presented at FAccT, a key conference in computing ethics and fairness, and a major outlet for auditing work, revealed that between 2018 and 2022, 63\% of papers focused exclusively on US samples, 84\% on Western samples, and 65\% of the authors were affiliated with institutions in Western countries. 

Similarly, \cite{xivuri_systematic_2021} reported that AI fairness research predominantly originates from Europe and North America, raising concerns about the global applicability of its findings. While algorithm auditing shares these issues, no systematic evaluation of its geographic and linguistic dimensions has been conducted. Recent research that has examined the landscape of algorithm and AI auditing also did not address these critical questions. For example, studies by \cite{costanza-chock_who_2022} and \cite{birhane_ai_2024}, which explored who engages in algorithmic or AI auditing work and provided important insights on the topic, nonetheless did not systematically investigate the international distribution of auditors or the geographic and linguistic focus of audits. This further underscores the need for a more comprehensive analysis of the field, one that takes into account the global distribution of algorithm auditing research and its implications for the generalizability and relevance of the findings.

In summary, while \cite{bandy_problematic_2021} made significant contributions to mapping the field, key gaps remain regarding the geographic and linguistic diversity of algorithm auditing. Addressing these gaps is essential to ensure the research reflects the diverse contexts where algorithms operate. Expanding the scope of studies to include underrepresented regions and languages will lead to a more nuanced understanding of algorithmic impacts and better equip the field to tackle global challenges in ethics and fairness.

\section{Methodology}
As our work, similarly to \cite{bandy_problematic_2021}, entails a systematic literature review of algorithm audits, we have largely adopted the methodology employed by \cite{bandy_problematic_2021}. At the same time, we made several modificiations to \cite{bandy_problematic_2021} methodology in line with a) the different focus of our work; b) critical reflection on some of the terminologies and operationalizations used by \cite{bandy_problematic_2021}  that we did not fully agree with. Below, we describe in detail which steps of \cite{bandy_problematic_2021}'s pipeline were adopted by us directly, which were modified, and which were introduced by us, as well as provide the rationale for our methodological choices.

We rely on the Preferred Reporting Items for Systematic Reviews and Meta-Analyses (PRISMA) guidelines \cite{moher_preferred_2009} - an established methodology for systematic literature reviews. The review thus involves the following steps: \textit{identification} of relevant studies, \textit{screening} article metadata for potential relevance, \textit{assessing eligibility} through full-text review, and \textit{inclusion} of articles in the comprehensive analytic stage.

Within identification and screening steps, we rely on the definition of algorithm audit by \cite{bandy_problematic_2021} as "an empirical study investigating a public algorithmic system for potential problematic behavior" except the "public" part since, due to our focus on the geographic and linguistic skews in algorithm auditing, we decided to focus on systems directly accessible to the public. Specifically, unlike \cite{bandy_problematic_2021}, we regard systems as public only when people can interact with them directly, and to avoid confusion, we opted to use the term "public-facing." Hence, we do not include in our analysis systems used in public settings but inaccessible for ordinary citizens (e.g., those used by law enforcement for facial recognition).

\subsection{Identification, screening and analytical procedure}
To identify studies to review, we used Scopus database keyword search, using the search query from \cite{bandy_problematic_2021} as a base and further expanding it. We looked for studies referencing influential algorithm auditing papers listed by \cite{bandy_problematic_2021} and also included \cite{bandy_problematic_2021} paper itself. We added terms to define a) empirical studies ("experimental design", "agent-based" or "examination"); and b) studies relevant for algorithm auditing ("algorithm* personalization", "algorithm* diversity", "algorithm* recommendation"; we replaced the term "algorithmic bias" by Bandy with a broader one - "algorithm* bias*")\footnote{The full search query was the following: ( TITLE-ABS-KEY ( "algorithmic discrimination" ) OR TITLE-ABS-KEY ( "algorithmic fairness" ) OR TITLE-ABS-KEY ( "algorithmic accountability" ) OR ALL ( "algorithm audit*" ) OR ALL ( "algorithmic audit*" ) OR REF ( "auditing algorithms: research methods" ) OR REF ( "thinking critically about researching algorithms" ) OR REF ( "the relevance of algorithms" ) OR REF ( "problematic machine behavior" ) OR TITLE-ABS-KEY ( "algorithm* bias*" ) OR TITLE-ABS-KEY ( "algorithm* personalization" ) OR TITLE-ABS-KEY ( "algorithm* diversity" ) OR TITLE-ABS-KEY ( "algorithm* recommendation" ) AND TITLE-ABS-KEY ( "study" OR "audit" OR "analysis" OR "experiment" OR "experimental design" OR "agent-based" OR "examination" ) ).}. 

We acknowledge the limitations of using Scopus and an English-only query in our study. Scopus, while less geographically biased than other databases like Web of Science \cite{tennant_web_2020}, still exhibits certain biases. Similarly, focusing solely on English-language research introduces a limitation. However, it is important to note that using English as the primary language for our queries is also appropriate given the context of our research focus since key outlets in Human-Computer Interaction (HCI), computer science, and related fields that focus on fairness and algorithmic evaluations are predominantly published in English. These include leading conferences and journals in Computer Science like FAccT and CSCW, or, in other disciplines, Social Science Computer Review, Digital Journalism and Information, Communication \& Society, where algorithm auditing work is widely published. As a result, English serves as the lingua franca of academic discourse in these fields, ensuring that the studies we reviewed are part of the central body of research that informs global discussions on algorithmic fairness and ethics. Moreover, our search on Google Scholar and Scopus for algorithm audit studies in five widely spoken European languages (German, French, Spanish, Italian, Russian) and Mandarin Chinese yielded very few relevant results, none of which were \textit{empirical} algorithm auditing studies. This outcome, coupled with our broader knowledge of international algorithm audit communities, gives us confidence that including queries in other languages would not have significantly altered our results since academic algorithm auditing work is published primarily, if not exclusively, in English language outlets. 

For screening, we had a total of 3,590 papers included - papers retrieved using our query, published until the end of 2024. The title and abstract screening was performed by the first author in a similar procedure to \cite{bandy_problematic_2021}. This has led to the exclusion of 3,310 papers and the selection of 280 papers for the full-text eligibility check. A total of 104 papers were excluded at the full-text screening stage. Further details on this are provided in the Appendix \ref{appendix:methodology}. \textbf{A total of 176 papers were included in the final analysis}. The full list of papers along with the coding of key categories is in the Appendix \ref{appendix:papers}.  

For identifying coding categories, we followed and expanded \cite{bandy_problematic_2021}. The first author developed the initial codebook based on their knowledge of relevant scholarship, and the categories from \cite{bandy_problematic_2021}. Then, the codebook was refined through discussions with the second author, who also possesses extensive knowledge of the field. Next, we describe each category and the reasoning behind it in detail.

\subsubsection{Platform}
This category lists the name of the platform/website audited in the reviewed study 
(e.g., YouTube or Twitter\footnote{The overwhelming majority of audits in our collection 
analyzed the platform before it was renamed to X so we keep the original name here.}). In rare cases when the platform/website was not named or when there were too many platforms/websites (e.g., dozens of websites of a certain type), we listed the type of platform/website audited (e.g., News website; Job search websites).

\subsubsection{Problem}
This category is based on \cite{bandy_problematic_2021} where four common types of problems audited for were identified: Distortion, Discrimination, Misjudgement and Exploitation. See \cite{bandy_problematic_2021} for details. 

\subsubsection{More specific problem}
We added this category to provide a more fine-grained categorization of the audited problems. The initial list of options was drafted by the first author based on their knowledge of the field; this coding scheme was tested on 30 randomly selected studies  
and then further expanded and refined. The final list of problems 
includes the following: Personalization; Filter Buggle; News Distribution; Harmful Content; Group Misrepresentation; Price Discrimination; Discrimination (other); Information Quality; User Categorization. Multiple options could apply to one study. Detailed definitions of each category are listed in the Appendix \ref{appendix:methodology}. 

\subsubsection{Audit method}
Here we made several modifications to the categorization used by \cite{bandy_problematic_2021}. First, instead of the term "carrier puppet" we used "repurposing" due to the original term being confusing as the respective studies involve researchers (and not "puppets") repurposing platform functionalities to audit algorithmic systems. Second, we distinguished not between "sock puppets" and "direct scrape" but between "personalized scrape" (usually associated with "sock puppets" in \cite{sandvig_auditing_nodate,bandy_problematic_2021}) and "non-personalized scrape" (aligned with "direct scrape"). Instead of distinguishing between direct web scraping and the use of sock puppets (e.g., via browser automation tools), we were more interested in whether the authors collected system data under non-personalized conditions or modeled the behavior of users with specific characteristics. The complexity of contemporary platform architectures prompts the growing use of browser automation tools even for non-personalized scraping. Hence, the use of sock puppets often does not tell us much about the research design per se. In contrast, our amended categorization allows capturing the distinction between studies that do or do not model specific user behavior that is increasingly important under the conditions of behavior-based content personalization. Further, we included a category of platform-led studies denoting audits conducted by a company/online platform itself. Examples of such studies are \cite{geyik_fairness-aware_2019,huszar_algorithmic_2022,sun_value_2024}.

\subsubsection{Domain} 
This category followed \cite{bandy_problematic_2021} and listed a general domain in which an algorithm was deployed. Possible options included Ad delivery, E-commerce, Search, Recommendation, Spam detection, Generative AI, Monetization, Translation, User categorization.

\subsubsection{Geographical focus}
We included this category to examine the geographic focus of audit studies (RQ2a). This category listed countries where an audit was focused. Where a country selection was ambiguous, we used the term "Mixed". In one case, we also used the term "Comprehensive" as the authors included all 191 countries where the audited social media platform was operating \cite{lambrecht_algorithmic_2019}. In another case, we used "EU" as the authors focused on all EU countries \cite{cabanas_unveiling_2018}. Importantly, if an audit used IP addresses located in a certain country, and it may have affected the system outputs (e.g., in the context of search results that are known to be personalized based on location), we also listed that country as a geographical focus of the audit.

\subsubsection{Language of content}
In line with our aim to evaluate the focus of audits in terms of languages (RQ2b), we listed the language(s) relevant for the audit, such as the language of search queries used or of the news included. There was one study for which language was irrelevant, and this category was coded as NA - a study on the personalization of borders on Google Maps based on a user's location \cite{soeller_mapwatch_2016}. Further, there were 18 studies where the language was coded as "Mixed" because they included content in several unspecified languages. In all other cases, specific languages were listed in this category.

\subsubsection{Countries of author affiliations}
In line with our RQ3, we identified the countries of all institutional affiliations of the authors at the time when a given paper was published.

\subsubsection{Group-based attributes where relevant}
For studies dealing with discrimination and/or misrepresentation of people, we further specified which group attributes the study focused on and how those were operationalized. Specifically, such attributes included Gender (binary vs other - as listed by the authors); Race/ethnicity (as listed by the authors); Sexuality (as listed by the authors); Age (as listed by the authors); Nationality (as listed by the authors); Religion (as listed by the authors); Socio-economic background (as listed by the authors); Appearance (beyond race and gender - e.g., clothing, - as specified by the authors). These attributes were not pre-specified by us but rather were "snowballed" during the coding procedure - this way we made sure we included \textit{all} group-based attributes studied in at least 1 of the reviewed studies.

\subsection{Coding procedure and synthesis}
Two first authors coded all the studies selected for full-text analysis. Inter-coder reliability was 80\% agreement on language, 90\% agreement on country-context, general specific problem categories, 96\% agreement on audit method, and perfect agreement on all other categories. Then, based on the coding outcomes, the lead author synthesized and summarized the results.

\section{Results}
\subsection{RQ1: Overview of the field}

\subsubsection{Problems audited}

\begin{figure}[h]
  \centering
  \includegraphics[width=0.6\linewidth]{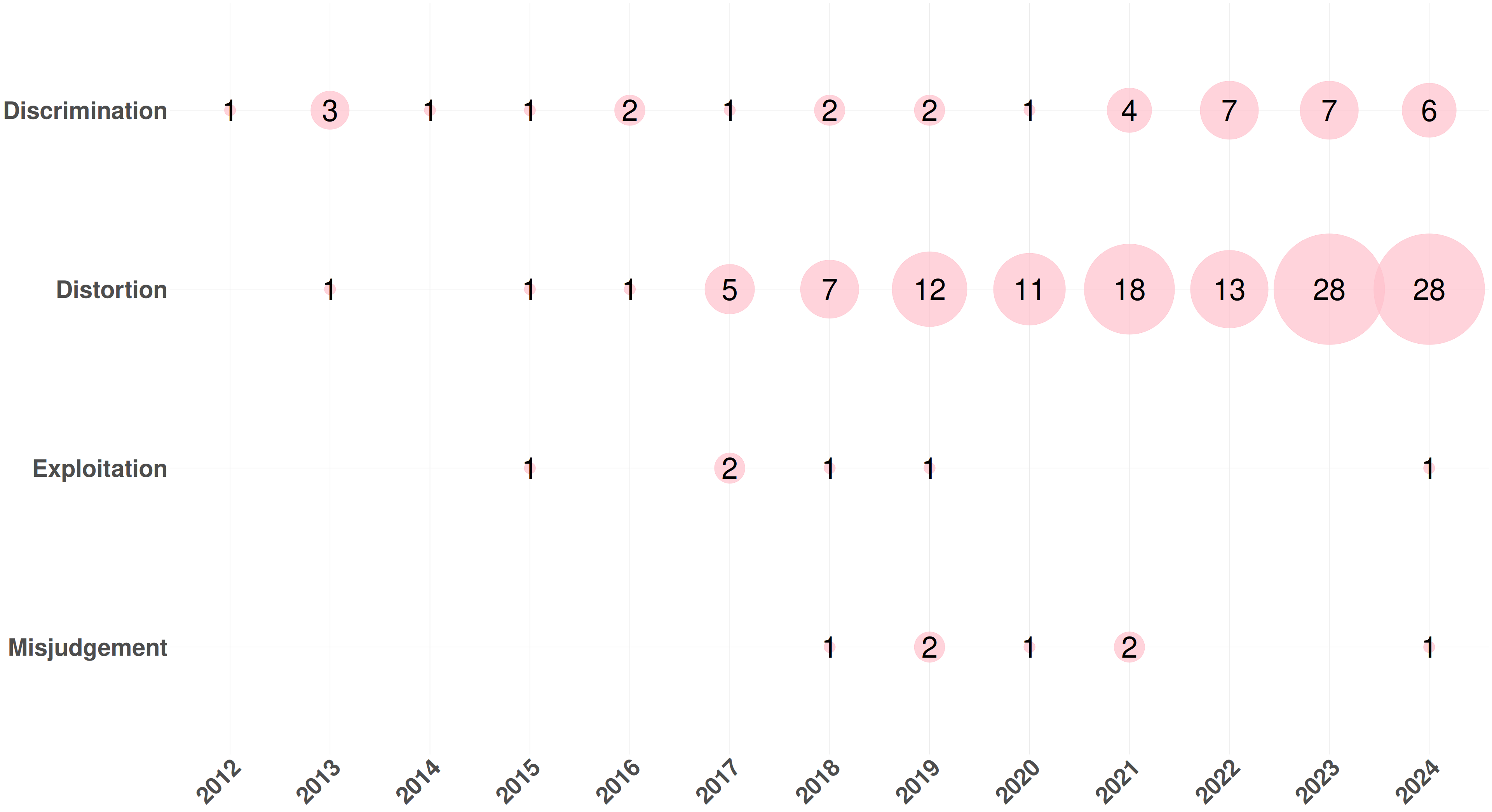}
  \caption{Number of audits examining each problem type by year}
  \label{fig:problems}
  \Description{this is an important figure}
\end{figure}
In Figure \ref{fig:problems}\footnote{Due to the absence of Adobe Acrobat Pro License we were not able to add the alt text to the figures. However, we provide figure descriptions in the Appendix \ref{appendix:figuredesc} to ensure accessibility, and apologize for the inconvenience associated with this solution.} we present an overview of the audit studies per year focused on the four overarching problems - discrimination, distortion, exploitation, and misjudgment. Up to 2020, our observations correspond to those of \cite{bandy_problematic_2021}. We manually verified that the marginal differences between our findings and \cite{bandy_problematic_2021} are 
attributed either to our focus on studies that audited public-facing algorithms or the inclusion of studies not included in \cite{bandy_problematic_2021} as we used a slightly broader search query. Similarly to \cite{bandy_problematic_2021}, we find that the number of studies has been increasing over the years, rising from less than 5 per year before 2017 to over 20 yearly since 2021, and over 30 in 2023 and 2024. This growth is mainly associated with the increased number of audits focused on Distortion - it already was the most common focus of audits before 2021 \cite{bandy_problematic_2021}, and the attention to this overarching issue has only increased, with Distortion-focused audits accounting for over two-thirds of all reviewed public-facing algorithm audits every year since 2018.

\begin{figure}[h]
  \centering
  \includegraphics[width=\linewidth]{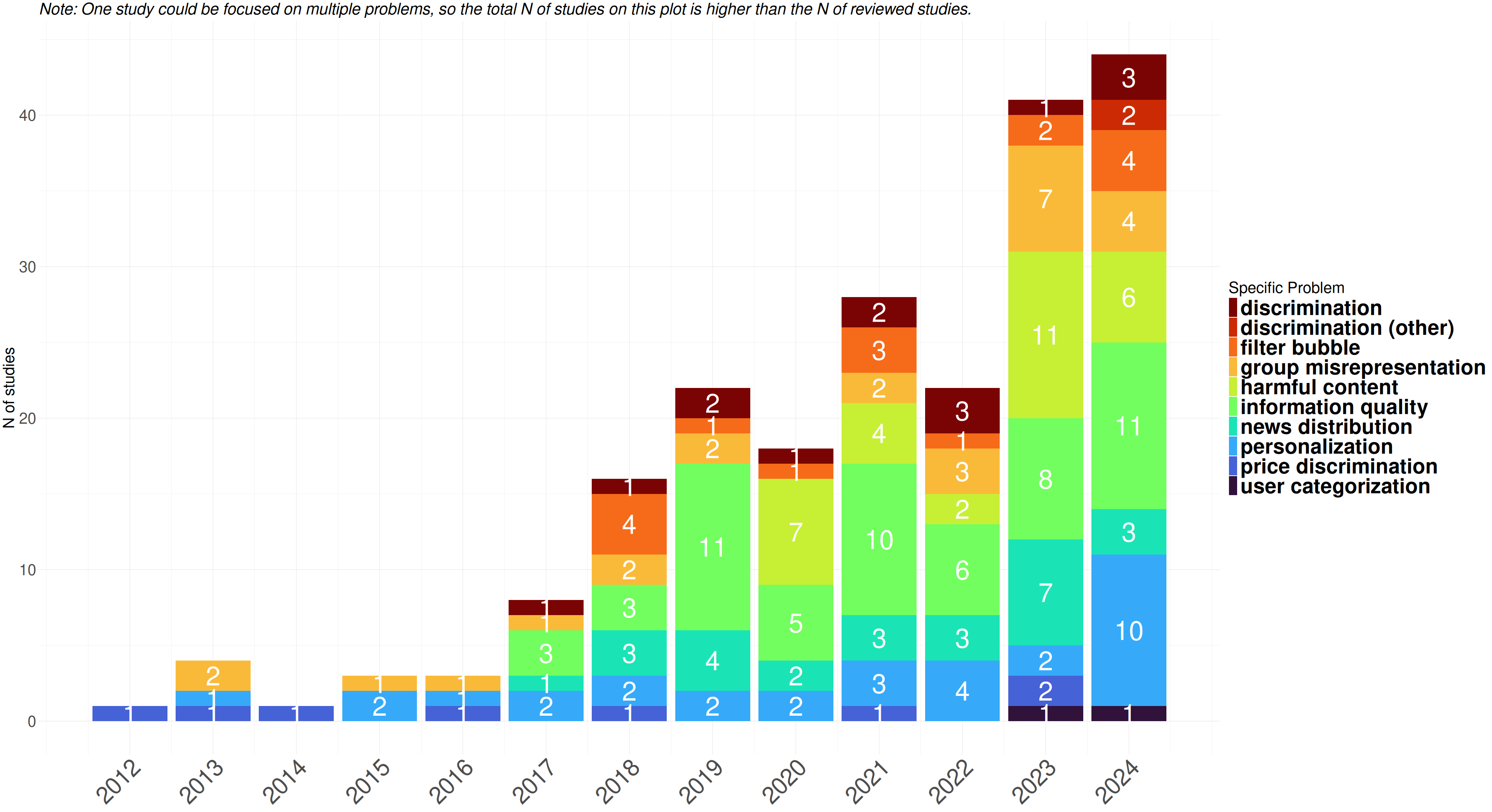}
  \caption{Number of audits examining each \textit{specific }problem type by year}
  \label{fig:specificproblems}
\end{figure}

In Figure \ref{fig:specificproblems} we present the number of auditing studies that focused on different types of more specific problems each year. The first few auditing studies - those published between 2012 and 2016 as the field was just emerging - focused on personalization, price discrimination and group misrepresentation. However, over time, as the number of audits was increasing, so was the diversity of specific problems they examined. One of the most common problems addressed since 2017 has become information quality. Additionally, since 2020 audits frequently focus on harmful content, with a record number of such studies - 11 - being published in 2023. Additionally, group misrepresentation, news distribution, filter bubbles, and personalization have been comparatively often the focus of audits between 2017 and 2024, though less commonly than information quality or harmful content. In contrast to that, (price) discrimination and user categorization have been receiving comparatively little attention in recent years.

\subsubsection{Methods used}
In terms of methodologies used by the reviewed audit studies, the most commonly utilized were two scraping-based methods: non-personalized scraping (utilized in 87 papers) and personalized scraping (48 studies). The third most common methodology was crowdsourcing (32 studies), followed by platform repurposing (10 studies, all focused on Ad delivery domain). Three studies involved platform-led experiments by Alibaba \cite{sun_value_2024}, Twitter \cite{huszar_algorithmic_2022}, and LinkedIn \cite{geyik_fairness-aware_2019}. One study used code audit as a method \cite{weber_coding_2018}. The fact that in our sample there were fewer code audits than in \cite{bandy_problematic_2021} is related to our focus on public-facing platforms. The only code audit study in our review \cite{weber_coding_2018} examined \textit{open source} mobile news apps, and we suggest the scarcity of code audits of public-facing algorithms is directly related to the predominantly closed-source nature of such algorithms, often deployed by large tech companies.

\subsubsection{Domains audited: Overview}

In the previous sections, we mentioned that \textit{search} was the domain that audits in the most "popular" categories of distortion and discrimination most commonly focused on. Hence, it is not surprising it is the most examined domain in the reviewed studies in general: 78 out of 176 reviewed papers audited search domain. Importantly, while most often this meant auditing web search engines, some studies that analyzed search functionalities on other platforms - e.g., on YouTube \cite{lutz_examining_2021}, - were also categorized as focusing on search. The second most commonly analyzed domain was recommendation with 55 studies, followed by ad delivery (20), generative AI (13), e-commerce (10), user categorization (2) and 1 each for spam, translation, and monetization.

In Figure \ref{fig:domains}, we also show how the number of audits focusing on different domains developed over time. In the early years of algorithm auditing, the few audits published between 2012 and 2016 focused on e-commerce, ad delivery, and search, with no major skew in the number of audits between these categories. In 2017, first recommendation-focused audits were published, and since 2017 search and recommendation have started to become the most commonly audited domains in the field. At the same time, since 2023 audits focused on generative AI have started to appear, and are becoming increasingly common \footnote{It is likely that generative AI tools are, in fact, audited even more often, but the terminology used in the evaluations of these tools differs from that common for the algorithm auditing domain, and hence, not all such evaluations were retrieved by us using an algorithm auditing-focused search query.}.

\subsubsection{Platforms audited: Overview}

\begin{figure}[h]
  \centering
  \begin{minipage}{0.48\linewidth}
    \centering
    \includegraphics[width=\linewidth]{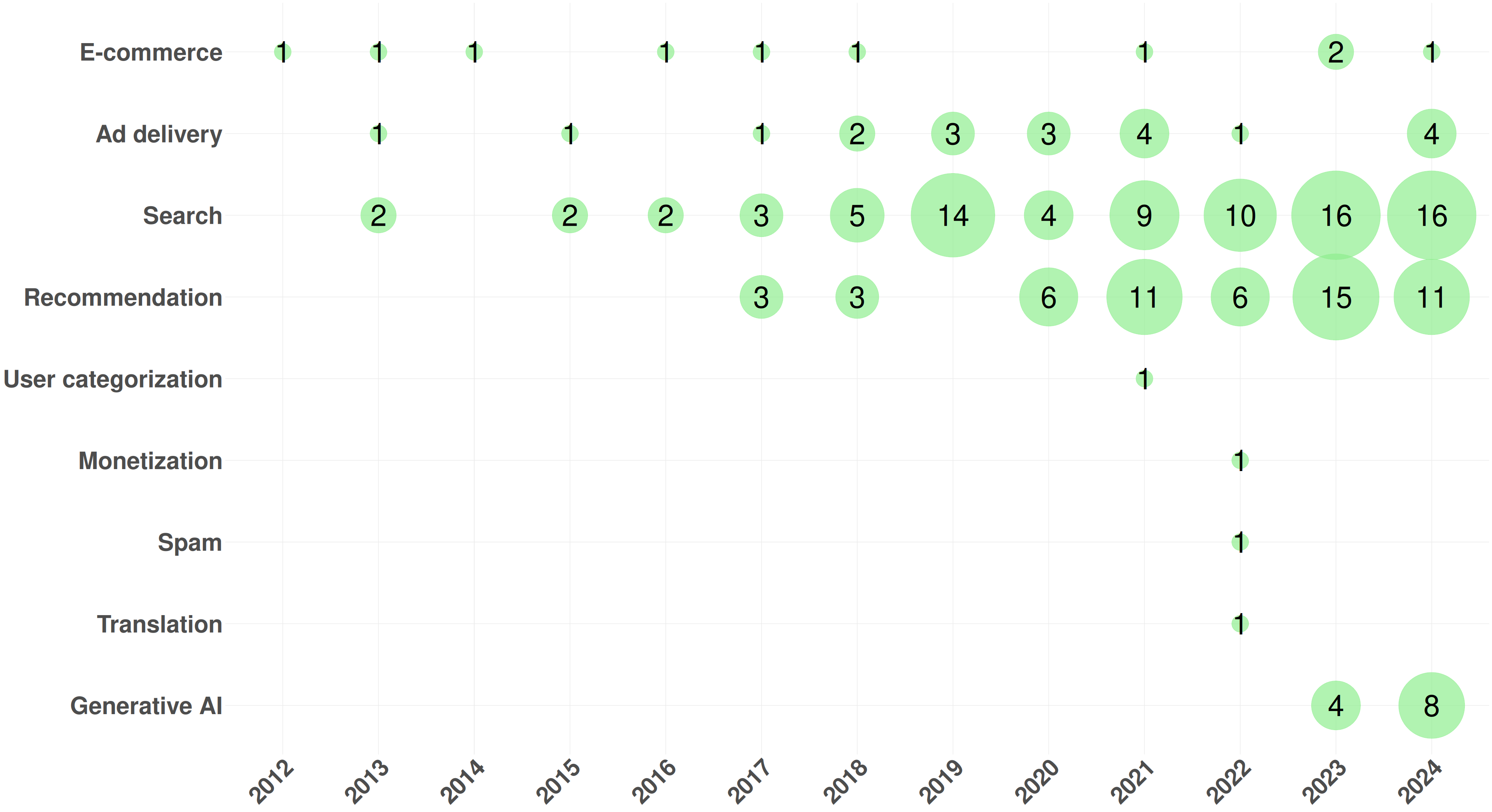}
    \caption{Number of audits examining each domain by year}
    \label{fig:domains}
  \end{minipage}
  \hfill
  \begin{minipage}{0.48\linewidth}
    \centering
    \includegraphics[width=\linewidth]{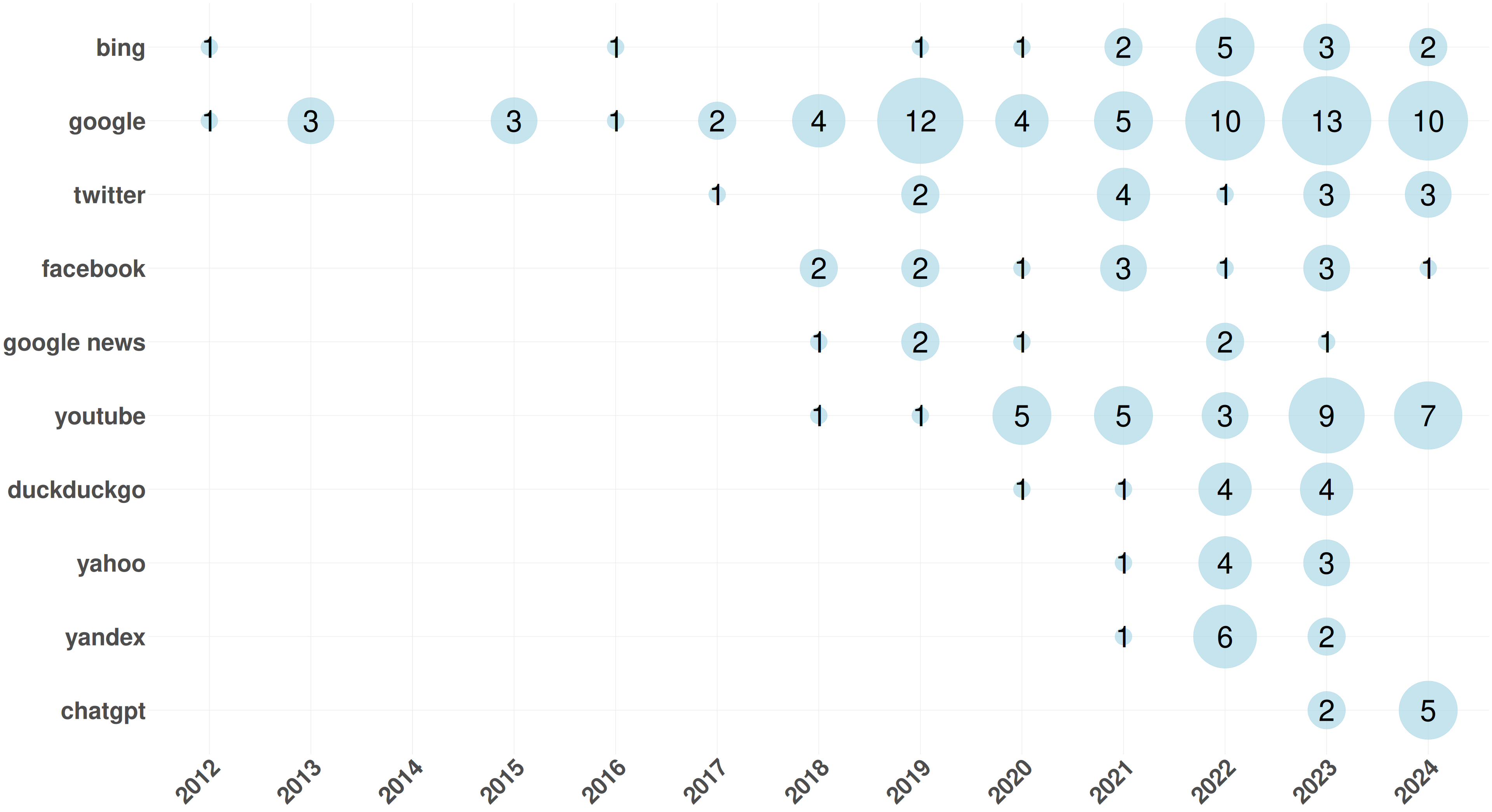}
    \caption{Number of audits examining each of the top 10 most frequently audited platforms by year}
    \label{fig:platforms}
  \end{minipage}
\end{figure}

In terms of specific platforms that were audited, the focus of the field is concentrated around a small selection of large platforms, most often search engines: 67 out of 176 included audits of Google Search (incl. advertising in search), 31 audited YouTube, 16 Bing Search, 14 Twitter (X), 13 Facebook (and 1 Meta more broadly), 10 DuckDuckGo, 9 Yandex Search, 8 Yahoo Search, 7 Google News, 7 ChatGPT, 6 Amazon (E-commerce platform), 5 TikTok, 4 Baidu, 3 Spotify. This was followed by a long tail of platforms audited by only 1 or 2 studies. This long tail included such widely popular platforms and e-commerce websites with algorithmic content distribution as Instagram \cite{kollyri_-coding_2021, lambrecht_algorithmic_2019}, LinkedIn \cite{imana_auditing_2021,geyik_fairness-aware_2019}, Alibaba \cite{sun_value_2024}, Douyin \cite{shi_new_2024}, Booking.com \cite{hupperich_empirical_2018,eslami_be_2017}, Bing Chat (Copilot) and Perplexity \cite{li_generative_2024}, Google Scholar and Semantic Scholar \cite{kacperski_examining_2024} or Google Maps \cite{smets_does_2019,soeller_mapwatch_2016}. These platforms thus are, compared to their popularity, underresearched in the field.

In Figure \ref{fig:platforms} we present an overview of the number of audits per year focused on each of the 10 most frequently audited platforms. Between 2012 and 2016 these included only audits of Bing and Google. Starting 2018, Google started to become the most frequently - compared to other platforms - audited platform. YouTube audits started to become more common in 2020. Overall, the set of platforms audited has been becoming more diverse over the years, especially since 2021.

\subsubsection{Summary}
Our analysis demonstrates that algorithm auditing has experienced major growth in recent years in terms of the number of published papers. We also identify several imbalances in the focus areas of studies in the field. Specifically, we observe that the studies tend to be most often focused on distortion as a type of problem, and search is by far the most audited domain, with almost half of all auditing studies focusing on search. We also find that audits tend to focus on major platforms like Google and (since 2020) YouTube, while other highly popular platforms like Instagram remain underexplored. Additionally, we find that the only platforms based outside of Western liberal democratic countries that were audited are Baidu, Douyin, Alibaba, and Yandex, indicating that platforms created and/or headquartered in regions outside the West remain understudied.

\subsection{RQ2: country contexts, languages, and group-based characteristics analyzed in algorithm audits}

\subsubsection{Country contexts}

\begin{figure}[h]
  \centering
  \includegraphics[width=0.7\linewidth]{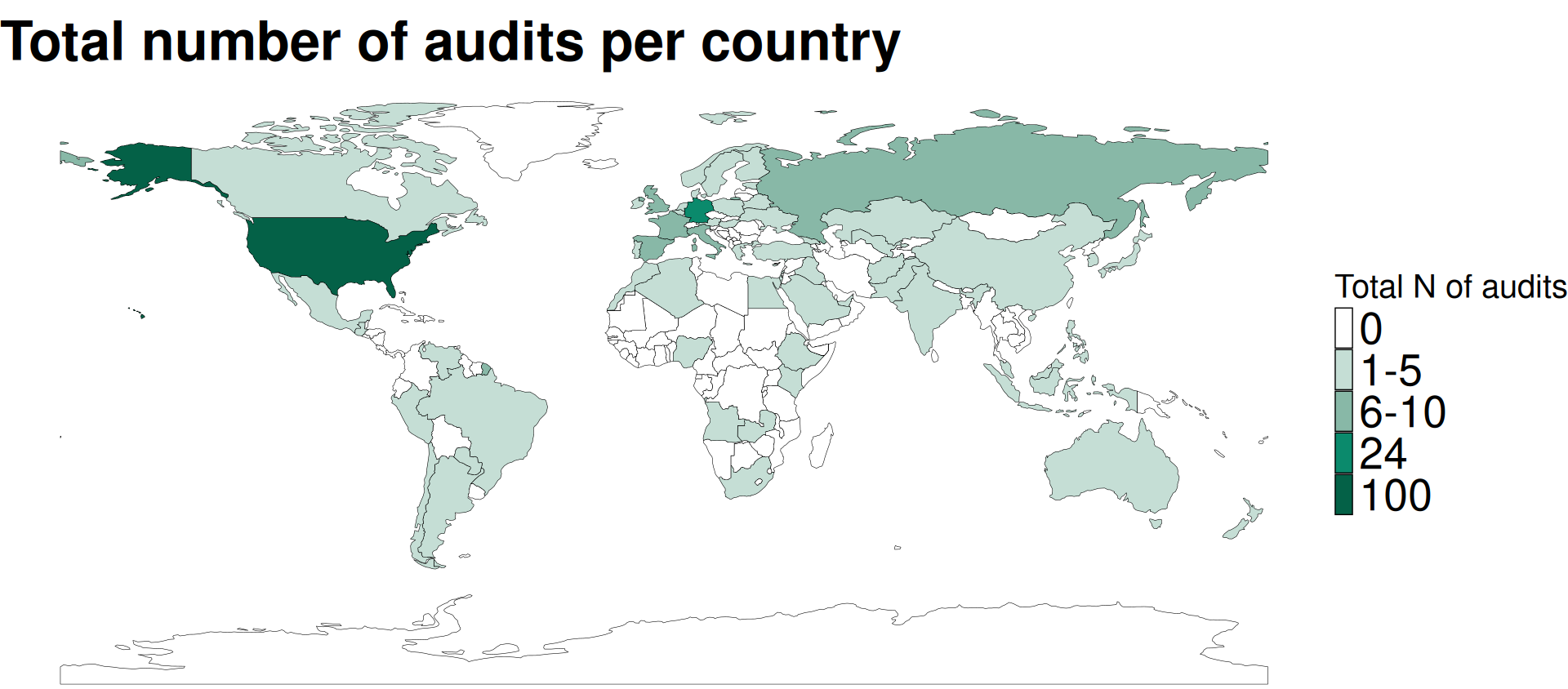}
  \caption{Countries color-coded by the N of auditing studies that included data from a given country context.}
  \label{fig:countries}
\end{figure}

Figure \ref{fig:countries} shows countries color-coded by the N of auditing studies that included data from each country-context. It does not include 44 studies (25\% of the collection) with the Mixed country contexts. The counts include studies that focused on each context explicitly. Two more studies that covered a wide array of countries - comprehensive \cite{lambrecht_algorithmic_2019} EU-focused \cite{cabanas_unveiling_2018} - are \textit{not} included in the map. Top countries by the number of auditing studies are, with the exception of Russia, Western liberal democracies, with the US accounting for over half of all auditing studies (100 papers). The other countries from the top 5 were analyzed $>$4 times less often than the US. These are Germany (24 papers), the UK (10), France (8) and Russia (7).

Other countries explicitly analyzed in the auditing studies (excluding \cite{lambrecht_algorithmic_2019, cabanas_unveiling_2018}) were Italy and Spain (6 papers each), and Brazil (5). The following countries were included 4 times: Ukraine, Canada, Denmark, India. 3 times: Belarus, China, Japan. 2 times: Argentina, Australia, Austria, Belgium, Egypt, Greece, Iraq, Ireland, Mexico, Morocco, Netherlands, Nigeria, Pakistan, Portugal, Saudi Arabia, Sweden. Countries included once: Afghanistan, Algeria, Angola, Chile, Comprehensive, Cyprus, Estonia, Ethiopia, Finland, Georgia, Guatemala, Hungary, Indonesia, Israel, Kazakhstan, Kenya, Malaysia, New Zealand, Norway, Paraguay, Peru, Philippines, Poland, South Africa, South Korea, Turkey, UAE, Uzbekistan, Venezuela, Zambia. All other countries, if included at all, were only included in the two larger-scale analyses \cite{lambrecht_algorithmic_2019, cabanas_unveiling_2018}.

This summary shows that algorithm audits to date have been disproportionately focused on the US and Western European liberal democracies. Among non-democratic countries, the one most commonly analyzed was Russia, yet only 7 studies (3.98\% of the reviewed papers) focused on it. Furthermore, out of these, only \cite{makhortykh_story_2022} focused on Russia explicitly while other studies included Russia as part of larger-scale (4+ countries) comparative analyses. The same applies to the second most commonly studied authoritarian context - Belarus - that was examined in 4 papers but only \cite{kravets_different_2023} focused on it specifically, not within a larger-scale comparative research. The only other authoritarian country analyzed in more than 2 audits was China \cite{shi_new_2024,sun_value_2024,soeller_mapwatch_2016}. 

Within the studies focused on democratic contexts, we also observe that countries outside North America and Western Europe are understudied. For instance, despite India, Brazil, and Indonesia being among the top 10 largest democracies by population, only 3 papers included India in their analysis, all of them being 4+ country comparative studies \cite{boratto_crowdsourcing_2022,soeller_mapwatch_2016,spiro_identifying_2016};\cite{onyepunuka_multilingual_nodate} focused on Indonesia; Brazilian context was included in 5 papers \cite{spiro_identifying_2016,boratto_crowdsourcing_2022,silva_facebook_2020,venkatadri_auditing_2019,santini_recommending_2023}. Thus, within democratic contexts, the skew in the focus of the studies does not correspond to the skews in the sizes of digital markets of different countries. Instead, we observe a strong geographic skew, with North American and Western European - WEIRD - contexts being analyzed disproportionately more often than others.

\subsubsection{Languages}
\begin{figure}[h]
  \centering
  \includegraphics[width=0.7\linewidth]{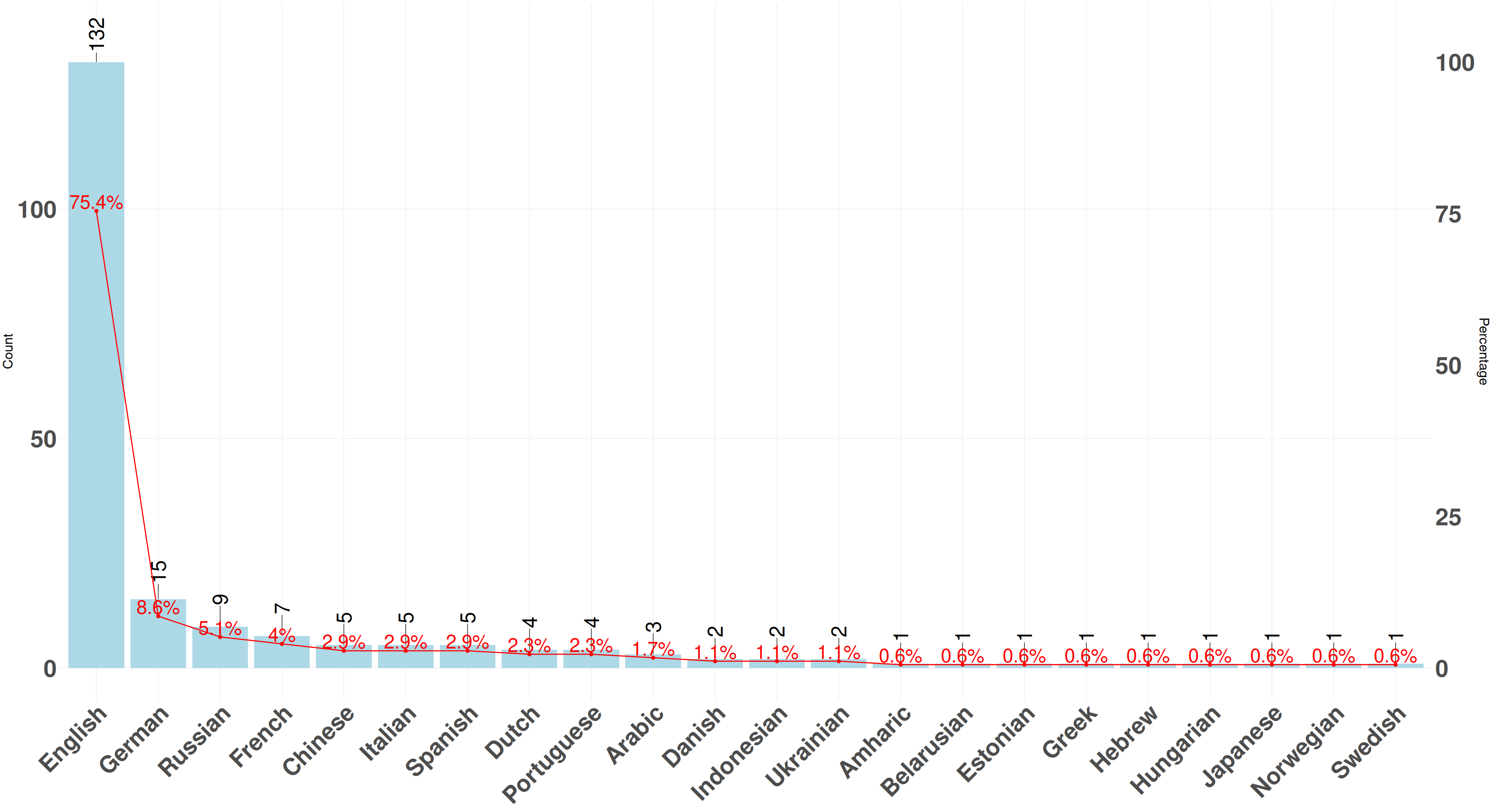}
  \caption{Number of audits focused on each language and their share among the reviewed studies. One study where language was irrelevant is excluded from the counts.}
  \label{fig:langs}
\end{figure}
As we demonstrate in Figure \ref{fig:langs}, the linguistic focus of audits is highly skewed, even more so than the selection of country contexts. 132 studies (75.4\% of the collection\footnote{Note that percentages for languages are counted as a share of 175 studies, not 176, as language was irrelevant for \cite{soeller_mapwatch_2016}}) focused explicitly on English (this does not include 17 papers where the language was coded as Mixed). The second most common language - German - was analyzed by 15 studies, $>$8 times less frequently than content in English, followed by Russian (9 papers), French (7), Chinese, Spanish and Italian (5 each), Portuguese and Dutch (4 each). Other languages were
featured in 3 or fewer studies. These included Arabic (3 papers); Ukrainian, Indonesian, Danish (2 papers); Amharic, Belarusian, Estonian, Greek, Hebrew, Japanese, Norwegian, Swedish (1 paper each).

The distribution of languages in the collection
is different from the distribution of
the number of total speakers of the languages \cite{ethnologue_what_2023} or
of online content in different languages \cite{w3techs_usage_2023}\footnote{Both \cite{ethnologue_what_2023, w3techs_usage_2023} were consulted in early January 2025; as \cite{w3techs_usage_2023} is updated daily, the current data on the website may differ from the one used by us.}. For example, while English is the most spoken language worldwide with ca. 1.5 billion speakers (ca. 18\% of the world population), it is followed by Mandarin Chinese (ca. 1.1 billion), Hindi (609 million), Spanish (559 million) and French (310 million). However, only 5 studies in our collection analyzed data in Mandarin Chinese and none focused on Hindi. The share of studies that focused on Spanish is also much lower than the estimated share of Spanish speakers among the world population. At the same time, the shares of papers focused on French, German, and Russian are higher than the estimated share of French-, German- and Russian speakers worlwide. 

We find a similar skew if we use the share of online content in a given language as a baseline. English is the most popular language on the internet with ca.49\% of all content created in this language \cite{w3techs_usage_2023}. According to \cite{w3techs_usage_2023}, the other most common languages on the internet as of early January 2025 are Spanish (6\% of online content), German (5.6\%), Japanese (5.1\%), French (4.4\%), and Russian (3.9\%). Thus, among these "common languages of the internet", the shares of studies focused on English, German, and Russian are higher than the shares of online content in corresponding languages, while Spanish and Japanese are understudied.

To sum up, algorithm auditing research is disproportionately focused on system performance in English and, to a much lower degree, German and Russian. It is skewed compared to the distribution of the total N of speakers of these languages or of the shares of online content in these languages, especially in the cases of English and, to a lower extent, German.

\subsubsection{Group-based attributes}
There were 46 studies that focused on at least one of the group attributes. Out of those, 38 focused on gender - often in combination with other attributes such as age or race/ethnicity. The most common specific problems audited for among gender-focused studies were group misrepresentation (21 studies) and discrimination (9 studies). Misrepresentation studies often looked at stereotyping, including gender-based stereotypes in image search \cite{kay_unequal_2015} or biases in generative Ai \cite{ovalle_im_2023}, whereas discrimination research examined discriminatory treatment in ad delivery \cite{imana_auditing_2021, lambrecht_algorithmic_2019} and online freelance marketplaces \cite{hannak_bias_2017} \cite{karan_your_2023, fabris_algorithmic_2021}.

Out of 38 studies that included \textbf{gender} category, 32 treated gender as binary. The exceptions were \cite{leidinger_which_2023} which also included transgender identities; \cite{ovalle_im_2023} which included a diverse set of identities such as genderqueer or genderfluid; \cite{ungless_stereotypes_2023} also included a diverse set of non-cisgender identities; finally, \cite{fosch-villaronga_little_2021,gleason_perceptions_2024} used non-binary as an umbrella term. Notably, 12 of 38 reviewed studies focused on gender in isolation without considering other group-based categories \cite{sun_smiling_2023, ungless_stereotypes_2023, ovalle_im_2023, haak_auditing_2022, farkas_how_2022, imana_auditing_2021, van_es_gendered_2021, geyik_fairness-aware_2019, chen_investigating_2018, eriksson_tracking_2017, kay_unequal_2015, datta_automated_nodate}. 

The second and third most common attributes were \textbf{age} and \textbf{race} (17 studies each). Most studies examining \textbf{age} treated it based on age groups (e.g., 18-25 y.o.; 26-35 y.o., etc.), albeit there was a variation in how the groups were defined. Only \cite{spiro_identifying_2016} operationalized age as a discrete variable. Furthermore, age was never the only group attribute in focus (unlike gender), and all studies focused on it also included at least gender as another attribute. Out of the 17 studies that included \textbf{race} as a group attribute, 5 studies treated this category as binary - white vs. non-white \cite{urman_auditing_2022,metaxa_image_2021} - or white vs Black \cite{sweeney_discrimination_2013,imana_auditing_2024,lambrecht_apparent_2024}. \cite{sapiezynski_use_2024} additionally included Hispanic. \cite{hannak_bias_2017, spiro_identifying_2016,lin_trapped_2023,apiola_first_2024} focused on white, Asian and Black, while \cite{noble_google_2013} further added Latino, and \cite{neumann_data_2024} Hispanic to these three categories. 
\cite{papakyriakopoulos_beyond_2023} instead added Indian and Other, while \cite{becerril-arreola_method_2023} included Hispanic and Native American. \cite{iqbal_left_2022} included white, African-American, Hispanic, Asian, and South Asian; \cite{asplund_auditing_2020} African-American, Asian, Hispanic, Caucasian; \cite{leidinger_which_2023} had the most comprehensive list with African, Asian, Black, American, Hispanic, Latinos, Native Americans, white categories. Notably, 11 out of 17 studies that examined race/ethnicity focused exclusively on the US context; further 4 studies had Mixed country-context \cite{papakyriakopoulos_beyond_2023, leidinger_which_2023, apiola_first_2024, lin_trapped_2023}, \cite{urman_auditing_2022} focused on Germany, and \cite{spiro_identifying_2016} focused on multicountry context.

Other attributes were examined less often. 4 papers, all published in 2022-2023 and focused on (mis)representation of different groups in search outputs, included \textbf{nationalities} as a specific category \cite{urman_foreign_2022,leidinger_which_2023,ieracitano_im_2023,mbalaka_epistemically_2023}. \cite{ieracitano_im_2023} focused only on Romanians, Albanians, Indians, Polish, Algerians, Moroccans, and Turks, whereas \cite{leidinger_which_2023,mbalaka_epistemically_2023} had more extensive (but still not comprehensive) lists of nationalities, while \cite{urman_foreign_2022} used a comprehensive list of national groups. 5 studies, also published since 2022, included \textbf{socio-economic situation} \cite{leidinger_which_2023,buda_foundations_2022,becerril-arreola_method_2023,shi_new_2024, neumann_data_2024}. 3 papers focused on \textbf{sexuality} as one of the attributes - \cite{leidinger_which_2023, fosch-villaronga_little_2021,lin_trapped_2023}. \cite{leidinger_which_2023} included asexual, bisexual, gay, homosexual, queer categories; \cite{fosch-villaronga_little_2021} included straight, gay, lesbian, asexual, bisexual, questioning, other; \cite{lin_trapped_2023} included heterosexual and homosexual. Finally, 3 studies \cite{leidinger_which_2023,lambrecht_apparent_2024,samuel-azran_analyzing_2024} focused on religion and included diverse religious groups. In addition, \cite{apiola_first_2024} included in their analysis differences in people's clothing, visual outlook and view of the family, while \cite{gleason_perceptions_2024} examined different skin tone categories based on the Monk's skin tone scale \cite{monk_monk_2023}.

\subsection{RQ3: Geographical distribution of affiliations of audit studies' authors' affiliations}

Among the countries with the highest number of authored auditing papers (i.e., those with which at least one author of the reviewed study was affiliated), the US is a clear leader with 54.55\% of all auditing studies (96 papers) being authored by local researchers. Other countries in the top 5 are European, with Germany accounting for 18.19\% (32 studies) of the papers, Switzerland - 8.52\% (15 studies), the Netherlands - 7.96\% (14 studies), Spain - 5.49\% (9 studies). Other countries were distributed the following way: Spain (9 papers), the UK (7); Brazil, Australia (5 each); Italy, Sweden (4 each); France, China, India, Canada (3 each); Belgium, Cyprus, Denmark, Finland, Israel, Norway (2 each); and Chile, Hong Kong, Hungary, Japan, New Zealand, Pakistan, Portugal, Saudi Arabia, Singapore, Slovakia, South Africa, Taiwan, Turkey, UAE (1 each).

There is thus a clear geographic imbalance in the affiliations of the authors of auditing studies. This imbalance is also only partially aligned with the imbalances in geographic and linguistic foci of the reviewed studies. On one hand, the most frequently examined national contexts - the US and Germany - are also the countries where audit authors are most often based. On another hand, while researchers from Switzerland and the Netherlands have authored relatively many auditing studies, Switzerland has not been the focus of any of them, and the Netherlands was always studied as part of larger-scale national comparisons except in \cite{van_es_gendered_2021,van_hoof_it_2024}. On the contrary, while France and Russia have been among the most frequently examined countries, with 7 papers examining each, none of the papers were by authors with Russian affiliations, and only 3 were from French institutions. Hence, the geographical skew in the author affiliations does not fully correspond to or explain the skew in the focus of auditing studies.

\section{Discussion and Recommendations}

Our systematic literature review uncovers significant imbalances in algorithm auditing that limit its scope, efficacy, and global applicability. These imbalances highlight the urgent need for a more inclusive, comprehensive, and context-sensitive approach to ensure the field evolves to address the full range of algorithmic challenges. Below, we summarize key findings, discuss their implications for research, policy, and practice, and propose strategic recommendations.

\subsection{Key Findings and Implications}

\textbf{Narrow Research Scope:} Algorithm auditing research has predominantly focused on issues like discrimination and distortion, neglecting other critical harms such as exploitation and misjudgment \cite{bandy_problematic_2021}. This limited focus risks creating an incomplete understanding of the societal impacts of algorithmic systems, sidelining harms that may disproportionately affect vulnerable populations. Expanding the scope of research to include a broader array of algorithmic harms is essential to fully capture and mitigate the diverse risks these systems pose.

\textbf{Platform Bias:} Research disproportionately targets major platforms like Google Search and YouTube while neglecting less-studied platforms, including Instagram, LinkedIn, or regionally significant platforms such as Baidu and Yandex. This focus creates blind spots in understanding the broader digital ecosystem and hinders the development of regulatory and technical solutions applicable across diverse contexts. Addressing this imbalance is critical to capture the full range of algorithmic behaviors and ensure equitable scrutiny across platforms, especially those with unique challenges or underexplored impacts.

\textbf{Geographic and Cultural Bias:} The overwhelming focus on Western liberal democracies, particularly the US, undermines the field’s ability to account for algorithmic impacts in other socio-political contexts. Algorithms optimized for democratic societies may behave differently or cause unintended harms in authoritarian regimes, where they could exacerbate censorship, surveillance, or repression. This geographic bias also reinforces Western-centric perspectives, limiting the global applicability of research findings and solutions. Expanding geographic diversity in research is essential to develop globally relevant insights and address algorithmic harms in diverse settings.

\textbf{Linguistic Limitations:} The dominance of English-language content in algorithm auditing overlooks the impact of algorithms on other languages, particularly those spoken by large populations such as Mandarin Chinese, Arabic, or Hindi. This bias risks missing language-specific algorithmic behaviors, such as failures in content moderation or increased bias in non-English contexts. Expanding the linguistic scope of algorithm audits is vital to ensure fair and accurate assessments across linguistic and cultural contexts.

\textbf{Simplistic Treatment of Group-Based Attributes:} The field often reduces attributes like race and gender to binary categories, ignoring the complexity of intersectional identities. This reductionist approach leads to incomplete evaluations of algorithmic bias and neglects the experiences of non-binary, transgender, and multi-marginalized individuals. Adopting more nuanced and intersectional methodologies is necessary to better reflect real-world diversity and uncover harms that may otherwise go unnoticed.

\textbf{Limited Authorship Diversity:} The concentration of research authorship in the US and a few European countries perpetuates Western-centric viewpoints and neglects regionally specific issues. This imbalance risks overlooking cultural factors that influence algorithmic outcomes. Increasing the participation of researchers from underrepresented regions is crucial to enrich the field with diverse perspectives and ensure it addresses global algorithmic challenges.

\subsection{Implications for Policy and Practice}

\textbf{Policy Development:} Policymakers rely heavily on academic research to inform regulations. The narrow focus of current research risks producing policies well-suited to Western democracies but inadequate or even harmful in non-Western or authoritarian contexts. For example, regulations targeting algorithmic bias in the US may fail to address legal and cultural differences in other regions. Broadening the focus of algorithm auditing research is critical for developing globally relevant policies.

\textbf{Corporate Accountability and Transparency:} Algorithm audits have shown potential to drive corporate change, as demonstrated by companies adapting systems in response to audit findings \cite{raji_actionable_2019}. However, the focus on major Western platforms limits the ability to hold companies in less-scrutinized markets accountable. Expanding the scope of audits can increase transparency and accountability across the global tech industry, empowering civil society and consumers even in regions with weaker regulatory frameworks.

\textbf{Ethical Considerations:} The field risks perpetuating algorithmic colonialism by imposing Western norms on non-Western contexts. Algorithms designed with Western notions of fairness may fail to align with the cultural and ethical values of other societies, leading to unintended harms. Ensuring that algorithm auditing reflects diverse cultural perspectives is essential to avoid these ethical pitfalls and promote equitable algorithmic systems.

\subsection{Strategic Recommendations}

\textbf{Diversify the Research Agenda:} Broaden the focus of algorithm auditing to include underexplored harms like exploitation and misjudgment, expand studies to cover diverse platforms, and prioritize investigations in non-Western and underrepresented contexts. This includes analyzing regionally significant platforms and addressing language-specific impacts to capture the full scope of algorithmic effects.

\textbf{Foster Global Collaboration:} Encourage partnerships between researchers from different regions and disciplines to address geographic and cultural biases. Collaborative initiatives involving underrepresented regions can uncover new insights and address blind spots in current research.

\textbf{Enhance Methodological Rigor:} Adopt intersectional and inclusive approaches that reflect the complexity of identities. Move beyond binary categorizations and use context-specific measures for race, gender, and other attributes. Researchers should also ensure that generalizations are supported by robust evidence and consider the cultural and linguistic diversity of algorithmic effects.

\textbf{Promote Structural Change:} Address systemic biases in the organization and leadership of algorithm auditing conferences and publications. Increase representation of researchers from non-Western regions in editorial boards, program committees, and leadership roles at conferences such as FAccT, CSCW, and CHI. Drawing lessons from fields like behavioral science, where similar biases exist \cite{henrich_most_2010, arnett_neglected_2008}, can provide guidance for fostering greater inclusivity in algorithm auditing.

\bibliographystyle{ACM-Reference-Format}
\bibliography{audits}


\begin{thebibliography}{196}


\ifx \showCODEN    \undefined \def \showCODEN     #1{\unskip}     \fi
\ifx \showDOI      \undefined \def \showDOI       #1{#1}\fi
\ifx \showISBNx    \undefined \def \showISBNx     #1{\unskip}     \fi
\ifx \showISBNxiii \undefined \def \showISBNxiii  #1{\unskip}     \fi
\ifx \showISSN     \undefined \def \showISSN      #1{\unskip}     \fi
\ifx \showLCCN     \undefined \def \showLCCN      #1{\unskip}     \fi
\ifx \shownote     \undefined \def \shownote      #1{#1}          \fi
\ifx \showarticletitle \undefined \def \showarticletitle #1{#1}   \fi
\ifx \showURL      \undefined \def \showURL       {\relax}        \fi
\providecommand\bibfield[2]{#2}
\providecommand\bibinfo[2]{#2}
\providecommand\natexlab[1]{#1}
\providecommand\showeprint[2][]{arXiv:#2}

\bibitem[Abul-Fottouh et~al\mbox{.}(2020)]%
        {abul-fottouh_examining_2020}
\bibfield{author}{\bibinfo{person}{Deena Abul-Fottouh}, \bibinfo{person}{Melodie~Yunju Song}, {and} \bibinfo{person}{Anatoliy Gruzd}.} \bibinfo{year}{2020}\natexlab{}.
\newblock \showarticletitle{Examining algorithmic biases in {YouTube}’s recommendations of vaccine videos}.
\newblock \bibinfo{journal}{\emph{International Journal of Medical Informatics}}  \bibinfo{volume}{140} (\bibinfo{date}{Aug.} \bibinfo{year}{2020}), \bibinfo{pages}{104175}.
\newblock
\showISSN{13865056}
\urldef\tempurl%
\url{https://doi.org/10.1016/j.ijmedinf.2020.104175}
\showDOI{\tempurl}


\bibitem[Albadi et~al\mbox{.}(2022)]%
        {albadi_deradicalizing_2022}
\bibfield{author}{\bibinfo{person}{Nuha Albadi}, \bibinfo{person}{Maram Kurdi}, {and} \bibinfo{person}{Shivakant Mishra}.} \bibinfo{year}{2022}\natexlab{}.
\newblock \showarticletitle{Deradicalizing {YouTube}: {Characterization}, {Detection}, and {Personalization} of {Religiously} {Intolerant} {Arabic} {Videos}}.
\newblock \bibinfo{journal}{\emph{Proceedings of the ACM on Human-Computer Interaction}} \bibinfo{volume}{6}, \bibinfo{number}{CSCW2} (\bibinfo{date}{Nov.} \bibinfo{year}{2022}), \bibinfo{pages}{1--25}.
\newblock
\showISSN{2573-0142}
\urldef\tempurl%
\url{https://doi.org/10.1145/3555618}
\showDOI{\tempurl}


\bibitem[Ali et~al\mbox{.}(2021)]%
        {ali_ad_2021}
\bibfield{author}{\bibinfo{person}{Muhammad Ali}, \bibinfo{person}{Piotr Sapiezynski}, \bibinfo{person}{Aleksandra Korolova}, \bibinfo{person}{Alan Mislove}, {and} \bibinfo{person}{Aaron Rieke}.} \bibinfo{year}{2021}\natexlab{}.
\newblock \showarticletitle{Ad {Delivery} {Algorithms}: {The} {Hidden} {Arbiters} of {Political} {Messaging}}. In \bibinfo{booktitle}{\emph{Proceedings of the 14th {ACM} {International} {Conference} on {Web} {Search} and {Data} {Mining}}}. \bibinfo{publisher}{ACM}, \bibinfo{address}{Virtual Event Israel}, \bibinfo{pages}{13--21}.
\newblock
\showISBNx{978-1-4503-8297-7}
\urldef\tempurl%
\url{https://doi.org/10.1145/3437963.3441801}
\showDOI{\tempurl}


\bibitem[Amirova et~al\mbox{.}(2024)]%
        {amirova_framework-based_2024}
\bibfield{author}{\bibinfo{person}{Aliya Amirova}, \bibinfo{person}{Theodora Fteropoulli}, \bibinfo{person}{Nafiso Ahmed}, \bibinfo{person}{Martin~R. Cowie}, {and} \bibinfo{person}{Joel~Z. Leibo}.} \bibinfo{year}{2024}\natexlab{}.
\newblock \showarticletitle{Framework-based qualitative analysis of free responses of {Large} {Language} {Models}: {Algorithmic} fidelity}.
\newblock \bibinfo{journal}{\emph{PLOS ONE}} \bibinfo{volume}{19}, \bibinfo{number}{3} (\bibinfo{date}{March} \bibinfo{year}{2024}), \bibinfo{pages}{e0300024}.
\newblock
\showISSN{1932-6203}
\urldef\tempurl%
\url{https://doi.org/10.1371/journal.pone.0300024}
\showDOI{\tempurl}


\bibitem[Apiola et~al\mbox{.}(2024)]%
        {apiola_first_2024}
\bibfield{author}{\bibinfo{person}{Mikko Apiola}, \bibinfo{person}{Henriikka Vartiainen}, {and} \bibinfo{person}{Matti Tedre}.} \bibinfo{year}{2024}\natexlab{}.
\newblock \showarticletitle{First {Year} {CS} {Students} {Exploring} {And} {Identifying} {Biases} and {Social} {Injustices} in {Text}-to-{Image} {Generative} {AI}}. In \bibinfo{booktitle}{\emph{Proceedings of the 2024 on {Innovation} and {Technology} in {Computer} {Science} {Education} {V}. 1}}. \bibinfo{publisher}{ACM}, \bibinfo{address}{Milan Italy}, \bibinfo{pages}{485--491}.
\newblock
\showISBNx{9798400706004}
\urldef\tempurl%
\url{https://doi.org/10.1145/3649217.3653596}
\showDOI{\tempurl}


\bibitem[Araújo et~al\mbox{.}(2016)]%
        {spiro_identifying_2016}
\bibfield{author}{\bibinfo{person}{Camila~Souza Araújo}, \bibinfo{person}{Wagner Meira}, {and} \bibinfo{person}{Virgilio Almeida}.} \bibinfo{year}{2016}\natexlab{}.
\newblock \showarticletitle{Identifying {Stereotypes} in the {Online} {Perception} of {Physical} {Attractiveness}}.
\newblock In \bibinfo{booktitle}{\emph{Social {Informatics}}}, \bibfield{editor}{\bibinfo{person}{Emma Spiro} {and} \bibinfo{person}{Yong-Yeol Ahn}} (Eds.). Vol.~\bibinfo{volume}{10046}. \bibinfo{publisher}{Springer International Publishing}, \bibinfo{address}{Cham}, \bibinfo{pages}{419--437}.
\newblock
\showISBNx{978-3-319-47879-1 978-3-319-47880-7}
\urldef\tempurl%
\url{https://doi.org/10.1007/978-3-319-47880-7_26}
\showDOI{\tempurl}
\newblock
\shownote{Series Title: Lecture Notes in Computer Science}.


\bibitem[Arnett(2008)]%
        {arnett_neglected_2008}
\bibfield{author}{\bibinfo{person}{Jeffrey~J. Arnett}.} \bibinfo{year}{2008}\natexlab{}.
\newblock \showarticletitle{The neglected 95\%: why {American} psychology needs to become less {American}}.
\newblock \bibinfo{journal}{\emph{The American Psychologist}} \bibinfo{volume}{63}, \bibinfo{number}{7} (\bibinfo{date}{Oct.} \bibinfo{year}{2008}), \bibinfo{pages}{602--614}.
\newblock
\showISSN{0003-066X}
\urldef\tempurl%
\url{https://doi.org/10.1037/0003-066X.63.7.602}
\showDOI{\tempurl}


\bibitem[Asplund et~al\mbox{.}(2020)]%
        {asplund_auditing_2020}
\bibfield{author}{\bibinfo{person}{Joshua Asplund}, \bibinfo{person}{Motahhare Eslami}, \bibinfo{person}{Hari Sundaram}, \bibinfo{person}{Christian Sandvig}, {and} \bibinfo{person}{Karrie Karahalios}.} \bibinfo{year}{2020}\natexlab{}.
\newblock \showarticletitle{Auditing {Race} and {Gender} {Discrimination} in {Online} {Housing} {Markets}}.
\newblock \bibinfo{journal}{\emph{Proceedings of the International AAAI Conference on Web and Social Media}}  \bibinfo{volume}{14} (\bibinfo{date}{May} \bibinfo{year}{2020}), \bibinfo{pages}{24--35}.
\newblock
\showISSN{2334-0770}
\urldef\tempurl%
\url{https://doi.org/10.1609/icwsm.v14i1.7276}
\showDOI{\tempurl}


\bibitem[Bandy(2021)]%
        {bandy_problematic_2021}
\bibfield{author}{\bibinfo{person}{Jack Bandy}.} \bibinfo{year}{2021}\natexlab{}.
\newblock \showarticletitle{Problematic {Machine} {Behavior}: {A} {Systematic} {Literature} {Review} of {Algorithm} {Audits}}.
\newblock \bibinfo{journal}{\emph{Proceedings of the ACM on Human-Computer Interaction}} \bibinfo{volume}{5}, \bibinfo{number}{CSCW1} (\bibinfo{date}{April} \bibinfo{year}{2021}), \bibinfo{pages}{74:1--74:34}.
\newblock
\urldef\tempurl%
\url{https://doi.org/10.1145/3449148}
\showDOI{\tempurl}


\bibitem[Bandy and Diakopoulos(2020)]%
        {bandy_auditing_2020}
\bibfield{author}{\bibinfo{person}{Jack Bandy} {and} \bibinfo{person}{Nicholas Diakopoulos}.} \bibinfo{year}{2020}\natexlab{}.
\newblock \showarticletitle{Auditing {News} {Curation} {Systems}: {A} {Case} {Study} {Examining} {Algorithmic} and {Editorial} {Logic} in {Apple} {News}}.
\newblock \bibinfo{journal}{\emph{Proceedings of the International AAAI Conference on Web and Social Media}}  \bibinfo{volume}{14} (\bibinfo{date}{May} \bibinfo{year}{2020}), \bibinfo{pages}{36--47}.
\newblock
\showISSN{2334-0770}
\urldef\tempurl%
\url{https://doi.org/10.1609/icwsm.v14i1.7277}
\showDOI{\tempurl}


\bibitem[Bandy and Diakopoulos(2021a)]%
        {bandy_curating_2021}
\bibfield{author}{\bibinfo{person}{Jack Bandy} {and} \bibinfo{person}{Nicholas Diakopoulos}.} \bibinfo{year}{2021}\natexlab{a}.
\newblock \showarticletitle{Curating {Quality}? {How} {Twitter}’s {Timeline} {Algorithm} {Treats} {Different} {Types} of {News}}.
\newblock \bibinfo{journal}{\emph{Social Media + Society}} \bibinfo{volume}{7}, \bibinfo{number}{3} (\bibinfo{date}{July} \bibinfo{year}{2021}), \bibinfo{pages}{205630512110416}.
\newblock
\showISSN{2056-3051, 2056-3051}
\urldef\tempurl%
\url{https://doi.org/10.1177/20563051211041648}
\showDOI{\tempurl}


\bibitem[Bandy and Diakopoulos(2021b)]%
        {bandy_more_2021}
\bibfield{author}{\bibinfo{person}{Jack Bandy} {and} \bibinfo{person}{Nicholas Diakopoulos}.} \bibinfo{year}{2021}\natexlab{b}.
\newblock \showarticletitle{More {Accounts}, {Fewer} {Links}: {How} {Algorithmic} {Curation} {Impacts} {Media} {Exposure} in {Twitter} {Timelines}}.
\newblock \bibinfo{journal}{\emph{Proceedings of the ACM on Human-Computer Interaction}} \bibinfo{volume}{5}, \bibinfo{number}{CSCW1} (\bibinfo{date}{April} \bibinfo{year}{2021}), \bibinfo{pages}{1--28}.
\newblock
\showISSN{2573-0142}
\urldef\tempurl%
\url{https://doi.org/10.1145/3449152}
\showDOI{\tempurl}


\bibitem[Bandy and Diakopoulos(2023)]%
        {bandy_facebooks_2023}
\bibfield{author}{\bibinfo{person}{Jack Bandy} {and} \bibinfo{person}{Nicholas Diakopoulos}.} \bibinfo{year}{2023}\natexlab{}.
\newblock \showarticletitle{Facebook’s {News} {Feed} {Algorithm} and the 2020 {US} {Election}}.
\newblock \bibinfo{journal}{\emph{Social Media + Society}} \bibinfo{volume}{9}, \bibinfo{number}{3} (\bibinfo{date}{July} \bibinfo{year}{2023}), \bibinfo{pages}{20563051231196898}.
\newblock
\showISSN{2056-3051, 2056-3051}
\urldef\tempurl%
\url{https://doi.org/10.1177/20563051231196898}
\showDOI{\tempurl}


\bibitem[Bandy and Hecht(2021)]%
        {bandy_errors_2021}
\bibfield{author}{\bibinfo{person}{Jack Bandy} {and} \bibinfo{person}{Brent Hecht}.} \bibinfo{year}{2021}\natexlab{}.
\newblock \showarticletitle{Errors in {Geotargeted} {Display} {Advertising}: {Good} {News} for {Local} {Journalism}?}
\newblock \bibinfo{journal}{\emph{Proceedings of the ACM on Human-Computer Interaction}} \bibinfo{volume}{5}, \bibinfo{number}{CSCW1} (\bibinfo{date}{April} \bibinfo{year}{2021}), \bibinfo{pages}{1--19}.
\newblock
\showISSN{2573-0142}
\urldef\tempurl%
\url{https://doi.org/10.1145/3449166}
\showDOI{\tempurl}


\bibitem[Bartley et~al\mbox{.}(2021)]%
        {bartley_auditing_2021}
\bibfield{author}{\bibinfo{person}{Nathan Bartley}, \bibinfo{person}{Andres Abeliuk}, \bibinfo{person}{Emilio Ferrara}, {and} \bibinfo{person}{Kristina Lerman}.} \bibinfo{year}{2021}\natexlab{}.
\newblock \showarticletitle{Auditing {Algorithmic} {Bias} on {Twitter}}. In \bibinfo{booktitle}{\emph{13th {ACM} {Web} {Science} {Conference} 2021}}. \bibinfo{publisher}{ACM}, \bibinfo{address}{Virtual Event United Kingdom}, \bibinfo{pages}{65--73}.
\newblock
\showISBNx{978-1-4503-8330-1}
\urldef\tempurl%
\url{https://doi.org/10.1145/3447535.3462491}
\showDOI{\tempurl}


\bibitem[Bashir et~al\mbox{.}(2019)]%
        {bashir_quantity_2019}
\bibfield{author}{\bibinfo{person}{Muhammad~Ahmad Bashir}, \bibinfo{person}{Umar Farooq}, \bibinfo{person}{Maryam Shahid}, \bibinfo{person}{Muhammad~Fareed Zaffar}, {and} \bibinfo{person}{Christo Wilson}.} \bibinfo{year}{2019}\natexlab{}.
\newblock \showarticletitle{Quantity vs. {Quality}: {Evaluating} {User} {Interest} {Profiles} {Using} {Ad} {Preference} {Managers}.}. In \bibinfo{booktitle}{\emph{{NDSS}}}.
\newblock
\urldef\tempurl%
\url{https://www.cbw.sh/static/pdf/bashir-2019-ndss.pdf}
\showURL{%
\tempurl}


\bibitem[Becerril-Arreola(2023)]%
        {becerril-arreola_method_2023}
\bibfield{author}{\bibinfo{person}{Rafael Becerril-Arreola}.} \bibinfo{year}{2023}\natexlab{}.
\newblock \showarticletitle{A {Method} to {Assess} and {Explain} {Disparate} {Impact} in {Online} {Retailing}}. In \bibinfo{booktitle}{\emph{Proceedings of the {ACM} {Web} {Conference} 2023}}. \bibinfo{publisher}{ACM}, \bibinfo{address}{Austin TX USA}, \bibinfo{pages}{3670--3679}.
\newblock
\showISBNx{978-1-4503-9416-1}
\urldef\tempurl%
\url{https://doi.org/10.1145/3543507.3583270}
\showDOI{\tempurl}


\bibitem[Bechmann and Nielbo(2018)]%
        {bechmann_are_2018}
\bibfield{author}{\bibinfo{person}{Anja Bechmann} {and} \bibinfo{person}{Kristoffer~L. Nielbo}.} \bibinfo{year}{2018}\natexlab{}.
\newblock \showarticletitle{Are {We} {Exposed} to the {Same} “{News}” in the {News} {Feed}?: {An} empirical analysis of filter bubbles as information similarity for {Danish} {Facebook} users}.
\newblock \bibinfo{journal}{\emph{Digital Journalism}} \bibinfo{volume}{6}, \bibinfo{number}{8} (\bibinfo{date}{Sept.} \bibinfo{year}{2018}), \bibinfo{pages}{990--1002}.
\newblock
\showISSN{2167-0811, 2167-082X}
\urldef\tempurl%
\url{https://doi.org/10.1080/21670811.2018.1510741}
\showDOI{\tempurl}


\bibitem[Bentvelzen et~al\mbox{.}(2023)]%
        {bentvelzen_designing_2023}
\bibfield{author}{\bibinfo{person}{Marit Bentvelzen}, \bibinfo{person}{Jasmin Niess}, {and} \bibinfo{person}{Paweł~W. Woźniak}.} \bibinfo{year}{2023}\natexlab{}.
\newblock \showarticletitle{Designing {Reflective} {Derived} {Metrics} for {Fitness} {Trackers}}.
\newblock \bibinfo{journal}{\emph{Proceedings of the ACM on Interactive, Mobile, Wearable and Ubiquitous Technologies}} \bibinfo{volume}{6}, \bibinfo{number}{4} (\bibinfo{date}{Jan.} \bibinfo{year}{2023}), \bibinfo{pages}{158:1--158:19}.
\newblock
\urldef\tempurl%
\url{https://doi.org/10.1145/3569475}
\showDOI{\tempurl}


\bibitem[Birhane et~al\mbox{.}(2024)]%
        {birhane_ai_2024}
\bibfield{author}{\bibinfo{person}{Abeba Birhane}, \bibinfo{person}{Ryan Steed}, \bibinfo{person}{Victor Ojewale}, \bibinfo{person}{Briana Vecchione}, {and} \bibinfo{person}{Inioluwa~Deborah Raji}.} \bibinfo{year}{2024}\natexlab{}.
\newblock \showarticletitle{{AI} auditing: {The} {Broken} {Bus} on the {Road} to {AI} {Accountability}}. In \bibinfo{booktitle}{\emph{2024 {IEEE} {Conference} on {Secure} and {Trustworthy} {Machine} {Learning} ({SaTML})}}. \bibinfo{publisher}{IEEE}, \bibinfo{address}{Toronto, ON, Canada}, \bibinfo{pages}{612--643}.
\newblock
\showISBNx{9798350349504}
\urldef\tempurl%
\url{https://doi.org/10.1109/SaTML59370.2024.00037}
\showDOI{\tempurl}


\bibitem[Boeker and Urman(2022)]%
        {boeker_empirical_2022}
\bibfield{author}{\bibinfo{person}{Maximilian Boeker} {and} \bibinfo{person}{Aleksandra Urman}.} \bibinfo{year}{2022}\natexlab{}.
\newblock \showarticletitle{An {Empirical} {Investigation} of {Personalization} {Factors} on {TikTok}}. In \bibinfo{booktitle}{\emph{Proceedings of the {ACM} {Web} {Conference} 2022}}. \bibinfo{publisher}{ACM}, \bibinfo{address}{Virtual Event, Lyon France}, \bibinfo{pages}{2298--2309}.
\newblock
\showISBNx{978-1-4503-9096-5}
\urldef\tempurl%
\url{https://doi.org/10.1145/3485447.3512102}
\showDOI{\tempurl}


\bibitem[Bonart et~al\mbox{.}(2019)]%
        {bonart_investigation_2019}
\bibfield{author}{\bibinfo{person}{Malte Bonart}, \bibinfo{person}{Anastasiia Samokhina}, \bibinfo{person}{Gernot Heisenberg}, {and} \bibinfo{person}{Philipp Schaer}.} \bibinfo{year}{2019}\natexlab{}.
\newblock \showarticletitle{An investigation of biases in web search engine query suggestions}.
\newblock \bibinfo{journal}{\emph{Online Information Review}} \bibinfo{volume}{44}, \bibinfo{number}{2} (\bibinfo{date}{Dec.} \bibinfo{year}{2019}), \bibinfo{pages}{365--381}.
\newblock
\showISSN{1468-4527}
\urldef\tempurl%
\url{https://doi.org/10.1108/OIR-11-2018-0341}
\showDOI{\tempurl}


\bibitem[Bouchaud(2024)]%
        {cherifi_algorithmic_2024}
\bibfield{author}{\bibinfo{person}{Paul Bouchaud}.} \bibinfo{year}{2024}\natexlab{}.
\newblock \showarticletitle{Algorithmic {Amplification} of {Politics} and {Engagement} {Maximization} on {Social} {Media}}.
\newblock In \bibinfo{booktitle}{\emph{Complex {Networks} \& {Their} {Applications} {XII}}}, \bibfield{editor}{\bibinfo{person}{Hocine Cherifi}, \bibinfo{person}{Luis~M. Rocha}, \bibinfo{person}{Chantal Cherifi}, {and} \bibinfo{person}{Murat Donduran}} (Eds.). Vol.~\bibinfo{volume}{1144}. \bibinfo{publisher}{Springer Nature Switzerland}, \bibinfo{address}{Cham}, \bibinfo{pages}{131--142}.
\newblock
\showISBNx{978-3-031-53502-4 978-3-031-53503-1}
\urldef\tempurl%
\url{https://doi.org/10.1007/978-3-031-53503-1_11}
\showDOI{\tempurl}
\newblock
\shownote{Series Title: Studies in Computational Intelligence}.


\bibitem[Buda et~al\mbox{.}(2022)]%
        {buda_foundations_2022}
\bibfield{author}{\bibinfo{person}{Teodora~Sandra Buda}, \bibinfo{person}{João Guerreiro}, \bibinfo{person}{Jesus Omana~Iglesias}, \bibinfo{person}{Carlos Castillo}, \bibinfo{person}{Oliver Smith}, {and} \bibinfo{person}{Aleksandar Matic}.} \bibinfo{year}{2022}\natexlab{}.
\newblock \showarticletitle{Foundations for fairness in digital health apps}.
\newblock \bibinfo{journal}{\emph{Frontiers in Digital Health}}  \bibinfo{volume}{4} (\bibinfo{date}{Aug.} \bibinfo{year}{2022}), \bibinfo{pages}{943514}.
\newblock
\showISSN{2673-253X}
\urldef\tempurl%
\url{https://doi.org/10.3389/fdgth.2022.943514}
\showDOI{\tempurl}


\bibitem[Cabañas et~al\mbox{.}(2018)]%
        {cabanas_unveiling_2018}
\bibfield{author}{\bibinfo{person}{José~González Cabañas}, \bibinfo{person}{Ángel Cuevas}, {and} \bibinfo{person}{Rubén Cuevas}.} \bibinfo{year}{2018}\natexlab{}.
\newblock \showarticletitle{Unveiling and {Quantifying} {Facebook} {Exploitation} of {Sensitive} {Personal} {Data} for {Advertising} {Purposes}}. \bibinfo{pages}{479--495}.
\newblock
\showISBNx{978-1-939133-04-5}
\urldef\tempurl%
\url{https://www.usenix.org/conference/usenixsecurity18/presentation/cabanas}
\showURL{%
\tempurl}


\bibitem[Cakmak et~al\mbox{.}(2024)]%
        {cakmak_bias_2024}
\bibfield{author}{\bibinfo{person}{Mert~Can Cakmak}, \bibinfo{person}{Nitin Agarwal}, {and} \bibinfo{person}{Remi Oni}.} \bibinfo{year}{2024}\natexlab{}.
\newblock \showarticletitle{The bias beneath: analyzing drift in {YouTube}’s algorithmic recommendations}.
\newblock \bibinfo{journal}{\emph{Social Network Analysis and Mining}} \bibinfo{volume}{14}, \bibinfo{number}{1} (\bibinfo{date}{Aug.} \bibinfo{year}{2024}), \bibinfo{pages}{171}.
\newblock
\showISSN{1869-5469}
\urldef\tempurl%
\url{https://doi.org/10.1007/s13278-024-01343-5}
\showDOI{\tempurl}


\bibitem[Cano-Orón(2019)]%
        {cano-oron_dr_2019}
\bibfield{author}{\bibinfo{person}{Lorena Cano-Orón}.} \bibinfo{year}{2019}\natexlab{}.
\newblock \showarticletitle{Dr. {Google}, what can you tell me about homeopathy? {Comparative} study of the top10 websites in the {United} {States}, {United} {Kingdom}, {France}, {Mexico} and {Spain}}.
\newblock \bibinfo{journal}{\emph{El Profesional de la Información}} \bibinfo{volume}{28}, \bibinfo{number}{2} (\bibinfo{date}{March} \bibinfo{year}{2019}).
\newblock
\showISSN{1699-2407, 1386-6710}
\urldef\tempurl%
\url{https://doi.org/10.3145/epi.2019.mar.13}
\showDOI{\tempurl}


\bibitem[Chakraborty and Ganguly(2018)]%
        {chakraborty_analyzing_2018}
\bibfield{author}{\bibinfo{person}{Abhijnan Chakraborty} {and} \bibinfo{person}{Niloy Ganguly}.} \bibinfo{year}{2018}\natexlab{}.
\newblock \showarticletitle{Analyzing the {News} {Coverage} of {Personalized} {Newspapers}}. In \bibinfo{booktitle}{\emph{2018 {IEEE}/{ACM} {International} {Conference} on {Advances} in {Social} {Networks} {Analysis} and {Mining} ({ASONAM})}}. \bibinfo{publisher}{IEEE}, \bibinfo{address}{Barcelona}, \bibinfo{pages}{540--543}.
\newblock
\showISBNx{978-1-5386-6051-5}
\urldef\tempurl%
\url{https://doi.org/10.1109/ASONAM.2018.8508812}
\showDOI{\tempurl}


\bibitem[Chen et~al\mbox{.}(2018)]%
        {chen_investigating_2018}
\bibfield{author}{\bibinfo{person}{Le Chen}, \bibinfo{person}{Ruijun Ma}, \bibinfo{person}{Anikó Hannák}, {and} \bibinfo{person}{Christo Wilson}.} \bibinfo{year}{2018}\natexlab{}.
\newblock \showarticletitle{Investigating the {Impact} of {Gender} on {Rank} in {Resume} {Search} {Engines}}. In \bibinfo{booktitle}{\emph{Proceedings of the 2018 {CHI} {Conference} on {Human} {Factors} in {Computing} {Systems}}}. \bibinfo{publisher}{ACM}, \bibinfo{address}{Montreal QC Canada}, \bibinfo{pages}{1--14}.
\newblock
\showISBNx{978-1-4503-5620-6}
\urldef\tempurl%
\url{https://doi.org/10.1145/3173574.3174225}
\showDOI{\tempurl}


\bibitem[Chen et~al\mbox{.}(2016)]%
        {chen_empirical_2016}
\bibfield{author}{\bibinfo{person}{Le Chen}, \bibinfo{person}{Alan Mislove}, {and} \bibinfo{person}{Christo Wilson}.} \bibinfo{year}{2016}\natexlab{}.
\newblock \showarticletitle{An {Empirical} {Analysis} of {Algorithmic} {Pricing} on {Amazon} {Marketplace}}. In \bibinfo{booktitle}{\emph{Proceedings of the 25th {International} {Conference} on {World} {Wide} {Web}}}. \bibinfo{publisher}{International World Wide Web Conferences Steering Committee}, \bibinfo{address}{Montréal Québec Canada}, \bibinfo{pages}{1339--1349}.
\newblock
\showISBNx{978-1-4503-4143-1}
\urldef\tempurl%
\url{https://doi.org/10.1145/2872427.2883089}
\showDOI{\tempurl}


\bibitem[Costanza-Chock et~al\mbox{.}(2022)]%
        {costanza-chock_who_2022}
\bibfield{author}{\bibinfo{person}{Sasha Costanza-Chock}, \bibinfo{person}{Inioluwa~Deborah Raji}, {and} \bibinfo{person}{Joy Buolamwini}.} \bibinfo{year}{2022}\natexlab{}.
\newblock \showarticletitle{Who {Audits} the {Auditors}? {Recommendations} from a field scan of the algorithmic auditing ecosystem}. In \bibinfo{booktitle}{\emph{Proceedings of the 2022 {ACM} {Conference} on {Fairness}, {Accountability}, and {Transparency}}} \emph{(\bibinfo{series}{{FAccT} '22})}. \bibinfo{publisher}{Association for Computing Machinery}, \bibinfo{address}{New York, NY, USA}, \bibinfo{pages}{1571--1583}.
\newblock
\showISBNx{978-1-4503-9352-2}
\urldef\tempurl%
\url{https://doi.org/10.1145/3531146.3533213}
\showDOI{\tempurl}


\bibitem[Courtois et~al\mbox{.}(2018)]%
        {courtois_challenging_2018}
\bibfield{author}{\bibinfo{person}{Cédric Courtois}, \bibinfo{person}{Laura Slechten}, {and} \bibinfo{person}{Lennert Coenen}.} \bibinfo{year}{2018}\natexlab{}.
\newblock \showarticletitle{Challenging {Google} {Search} filter bubbles in social and political information: {Disconforming} evidence from a digital methods case study}.
\newblock \bibinfo{journal}{\emph{Telematics and Informatics}} \bibinfo{volume}{35}, \bibinfo{number}{7} (\bibinfo{date}{Oct.} \bibinfo{year}{2018}), \bibinfo{pages}{2006--2015}.
\newblock
\showISSN{07365853}
\urldef\tempurl%
\url{https://doi.org/10.1016/j.tele.2018.07.004}
\showDOI{\tempurl}


\bibitem[Dabran-Zivan et~al\mbox{.}(2023)]%
        {dabran-zivan_is_2023}
\bibfield{author}{\bibinfo{person}{Shakked Dabran-Zivan}, \bibinfo{person}{Ayelet Baram-Tsabari}, \bibinfo{person}{Roni Shapira}, \bibinfo{person}{Miri Yitshaki}, \bibinfo{person}{Daria Dvorzhitskaia}, {and} \bibinfo{person}{Nir Grinberg}.} \bibinfo{year}{2023}\natexlab{}.
\newblock \showarticletitle{“{Is} {COVID}-19 a hoax?”: auditing the quality of {COVID}-19 conspiracy-related information and misinformation in {Google} search results in four languages}.
\newblock \bibinfo{journal}{\emph{Internet Research}} \bibinfo{volume}{33}, \bibinfo{number}{5} (\bibinfo{date}{Nov.} \bibinfo{year}{2023}), \bibinfo{pages}{1774--1801}.
\newblock
\showISSN{1066-2243}
\urldef\tempurl%
\url{https://doi.org/10.1108/INTR-07-2022-0560}
\showDOI{\tempurl}


\bibitem[Dambanemuya and Diakopoulos(2021)]%
        {dambanemuya_auditing_2021}
\bibfield{author}{\bibinfo{person}{Henry~Kudzanai Dambanemuya} {and} \bibinfo{person}{Nicholas Diakopoulos}.} \bibinfo{year}{2021}\natexlab{}.
\newblock \showarticletitle{Auditing the {Information} {Quality} of {News}-{Related} {Queries} on the {Alexa} {Voice} {Assistant}}.
\newblock \bibinfo{journal}{\emph{Proceedings of the ACM on Human-Computer Interaction}} \bibinfo{volume}{5}, \bibinfo{number}{CSCW1} (\bibinfo{date}{April} \bibinfo{year}{2021}), \bibinfo{pages}{1--21}.
\newblock
\showISSN{2573-0142}
\urldef\tempurl%
\url{https://doi.org/10.1145/3449157}
\showDOI{\tempurl}


\bibitem[Dash et~al\mbox{.}(2021)]%
        {dash_when_2021}
\bibfield{author}{\bibinfo{person}{Abhisek Dash}, \bibinfo{person}{Abhijnan Chakraborty}, \bibinfo{person}{Saptarshi Ghosh}, \bibinfo{person}{Animesh Mukherjee}, {and} \bibinfo{person}{Krishna~P. Gummadi}.} \bibinfo{year}{2021}\natexlab{}.
\newblock \showarticletitle{When the {Umpire} is also a {Player}: {Bias} in {Private} {Label} {Product} {Recommendations} on {E}-commerce {Marketplaces}}. In \bibinfo{booktitle}{\emph{Proceedings of the 2021 {ACM} {Conference} on {Fairness}, {Accountability}, and {Transparency}}}. \bibinfo{publisher}{ACM}, \bibinfo{address}{Virtual Event Canada}, \bibinfo{pages}{873--884}.
\newblock
\showISBNx{978-1-4503-8309-7}
\urldef\tempurl%
\url{https://doi.org/10.1145/3442188.3445944}
\showDOI{\tempurl}


\bibitem[Dash et~al\mbox{.}(2024)]%
        {dash_investigating_2024}
\bibfield{author}{\bibinfo{person}{Abhisek Dash}, \bibinfo{person}{Abhijnan Chakraborty}, \bibinfo{person}{Saptarshi Ghosh}, \bibinfo{person}{Animesh Mukherjee}, {and} \bibinfo{person}{Krishna~P. Gummadi}.} \bibinfo{year}{2024}\natexlab{}.
\newblock \showarticletitle{Investigating {Nudges} toward {Related} {Sellers} on {E}-commerce {Marketplaces}: {A} {Case} {Study} on {Amazon}}.
\newblock \bibinfo{journal}{\emph{Proceedings of the ACM on Human-Computer Interaction}} \bibinfo{volume}{8}, \bibinfo{number}{CSCW2} (\bibinfo{date}{Nov.} \bibinfo{year}{2024}), \bibinfo{pages}{1--31}.
\newblock
\showISSN{2573-0142}
\urldef\tempurl%
\url{https://doi.org/10.1145/3686994}
\showDOI{\tempurl}


\bibitem[Datta et~al\mbox{.}(2015)]%
        {datta_automated_nodate}
\bibfield{author}{\bibinfo{person}{Amit Datta}, \bibinfo{person}{Michael~Carl Tschantz}, {and} \bibinfo{person}{Anupam Datta}.} \bibinfo{year}{2015}\natexlab{}.
\newblock \showarticletitle{Automated {Experiments} on {Ad} {Privacy} {Settings}: {A} {Tale} of {Opacity}, {Choice}, and {Discrimination}}.
\newblock  (\bibinfo{year}{2015}).
\newblock


\bibitem[Dunna et~al\mbox{.}(2022)]%
        {dunna_paying_2022}
\bibfield{author}{\bibinfo{person}{Arun Dunna}, \bibinfo{person}{Katherine~A. Keith}, \bibinfo{person}{Ethan Zuckerman}, \bibinfo{person}{Narseo Vallina-Rodriguez}, \bibinfo{person}{Brendan O'Connor}, {and} \bibinfo{person}{Rishab Nithyanand}.} \bibinfo{year}{2022}\natexlab{}.
\newblock \showarticletitle{Paying {Attention} to the {Algorithm} {Behind} the {Curtain}: {Bringing} {Transparency} to {YouTube}'s {Demonetization} {Algorithms}}.
\newblock \bibinfo{journal}{\emph{Proceedings of the ACM on Human-Computer Interaction}} \bibinfo{volume}{6}, \bibinfo{number}{CSCW2} (\bibinfo{date}{Nov.} \bibinfo{year}{2022}), \bibinfo{pages}{1--31}.
\newblock
\showISSN{2573-0142}
\urldef\tempurl%
\url{https://doi.org/10.1145/3555209}
\showDOI{\tempurl}


\bibitem[Duskin et~al\mbox{.}(2024)]%
        {duskin_echo_2024}
\bibfield{author}{\bibinfo{person}{Kayla Duskin}, \bibinfo{person}{Joseph~S Schafer}, \bibinfo{person}{Jevin~D West}, {and} \bibinfo{person}{Emma~S Spiro}.} \bibinfo{year}{2024}\natexlab{}.
\newblock \showarticletitle{Echo {Chambers} in the {Age} of {Algorithms}: {An} {Audit} of {Twitter}’s {Friend} {Recommender} {System}}. In \bibinfo{booktitle}{\emph{{ACM} {Web} {Science} {Conference}}}. \bibinfo{publisher}{ACM}, \bibinfo{address}{Stuttgart Germany}, \bibinfo{pages}{11--21}.
\newblock
\showISBNx{9798400703348}
\urldef\tempurl%
\url{https://doi.org/10.1145/3614419.3643996}
\showDOI{\tempurl}


\bibitem[Eriksson and Johansson(2017)]%
        {eriksson_tracking_2017}
\bibfield{author}{\bibinfo{person}{Maria~C. Eriksson} {and} \bibinfo{person}{Anna Johansson}.} \bibinfo{year}{2017}\natexlab{}.
\newblock \showarticletitle{Tracking {Gendered} {Streams}}.
\newblock \bibinfo{journal}{\emph{Culture Unbound}} \bibinfo{volume}{9}, \bibinfo{number}{2} (\bibinfo{date}{Oct.} \bibinfo{year}{2017}), \bibinfo{pages}{163--183}.
\newblock
\showISSN{2000-1525}
\urldef\tempurl%
\url{https://doi.org/10.3384/cu.2000.1525.1792163}
\showDOI{\tempurl}


\bibitem[Eslami et~al\mbox{.}(2017)]%
        {eslami_be_2017}
\bibfield{author}{\bibinfo{person}{Motahhare Eslami}, \bibinfo{person}{Kristen Vaccaro}, \bibinfo{person}{Karrie Karahalios}, {and} \bibinfo{person}{Kevin Hamilton}.} \bibinfo{year}{2017}\natexlab{}.
\newblock \showarticletitle{“{Be} {Careful}; {Things} {Can} {Be} {Worse} than {They} {Appear}”: {Understanding} {Biased} {Algorithms} and {Users}’ {Behavior} {Around} {Them} in {Rating} {Platforms}}.
\newblock \bibinfo{journal}{\emph{Proceedings of the International AAAI Conference on Web and Social Media}} \bibinfo{volume}{11}, \bibinfo{number}{1} (\bibinfo{date}{May} \bibinfo{year}{2017}), \bibinfo{pages}{62--71}.
\newblock
\showISSN{2334-0770}
\urldef\tempurl%
\url{https://doi.org/10.1609/icwsm.v11i1.14898}
\showDOI{\tempurl}
\newblock
\shownote{Number: 1}.


\bibitem[Ethnologue(2023)]%
        {ethnologue_what_2023}
\bibfield{author}{\bibinfo{person}{Ethnologue}.} \bibinfo{year}{2023}\natexlab{}.
\newblock \bibinfo{title}{What are the top 200 most spoken languages?}
\newblock
\newblock
\urldef\tempurl%
\url{https://www.ethnologue.com/insights/ethnologue200/}
\showURL{%
\tempurl}


\bibitem[Evans et~al\mbox{.}(2023)]%
        {evans_google_2023}
\bibfield{author}{\bibinfo{person}{Ryan Evans}, \bibinfo{person}{Daniel Jackson}, {and} \bibinfo{person}{Jaron Murphy}.} \bibinfo{year}{2023}\natexlab{}.
\newblock \showarticletitle{Google {News} and {Machine} {Gatekeepers}: {Algorithmic} {Personalisation} and {News} {Diversity} in {Online} {News} {Search}}.
\newblock \bibinfo{journal}{\emph{Digital Journalism}} \bibinfo{volume}{11}, \bibinfo{number}{9} (\bibinfo{date}{Oct.} \bibinfo{year}{2023}), \bibinfo{pages}{1682--1700}.
\newblock
\showISSN{2167-0811, 2167-082X}
\urldef\tempurl%
\url{https://doi.org/10.1080/21670811.2022.2055596}
\showDOI{\tempurl}


\bibitem[Fabris et~al\mbox{.}(2021)]%
        {fabris_algorithmic_2021}
\bibfield{author}{\bibinfo{person}{Alessandro Fabris}, \bibinfo{person}{Alan Mishler}, \bibinfo{person}{Stefano Gottardi}, \bibinfo{person}{Mattia Carletti}, \bibinfo{person}{Matteo Daicampi}, \bibinfo{person}{Gian~Antonio Susto}, {and} \bibinfo{person}{Gianmaria Silvello}.} \bibinfo{year}{2021}\natexlab{}.
\newblock \showarticletitle{Algorithmic {Audit} of {Italian} {Car} {Insurance}: {Evidence} of {Unfairness} in {Access} and {Pricing}}. In \bibinfo{booktitle}{\emph{Proceedings of the 2021 {AAAI}/{ACM} {Conference} on {AI}, {Ethics}, and {Society}}}. \bibinfo{publisher}{ACM}, \bibinfo{address}{Virtual Event USA}, \bibinfo{pages}{458--468}.
\newblock
\showISBNx{978-1-4503-8473-5}
\urldef\tempurl%
\url{https://doi.org/10.1145/3461702.3462569}
\showDOI{\tempurl}


\bibitem[Faddoul et~al\mbox{.}(2020)]%
        {faddoul_longitudinal_2020}
\bibfield{author}{\bibinfo{person}{Marc Faddoul}, \bibinfo{person}{Guillaume Chaslot}, {and} \bibinfo{person}{H. Farid}.} \bibinfo{year}{2020}\natexlab{}.
\newblock \showarticletitle{A {Longitudinal} {Analysis} of {YouTube}'s {Promotion} of {Conspiracy} {Videos}}.
\newblock \bibinfo{journal}{\emph{ArXiv}} (\bibinfo{date}{March} \bibinfo{year}{2020}).
\newblock
\urldef\tempurl%
\url{https://www.semanticscholar.org/paper/A-Longitudinal-Analysis-of-YouTube%27s-Promotion-of-Faddoul-Chaslot/e9b8266ec5d94660dde5f89b4e01586dcddf2515}
\showURL{%
\tempurl}


\bibitem[Farkas and Németh(2022)]%
        {farkas_how_2022}
\bibfield{author}{\bibinfo{person}{Anna Farkas} {and} \bibinfo{person}{Renáta Németh}.} \bibinfo{year}{2022}\natexlab{}.
\newblock \showarticletitle{How to measure gender bias in machine translation: {Real}-world oriented machine translators, multiple reference points}.
\newblock \bibinfo{journal}{\emph{Social Sciences \& Humanities Open}} \bibinfo{volume}{5}, \bibinfo{number}{1} (\bibinfo{year}{2022}), \bibinfo{pages}{100239}.
\newblock
\showISSN{25902911}
\urldef\tempurl%
\url{https://doi.org/10.1016/j.ssaho.2021.100239}
\showDOI{\tempurl}


\bibitem[Fischer et~al\mbox{.}(2020)]%
        {fischer_auditing_2020}
\bibfield{author}{\bibinfo{person}{Sean Fischer}, \bibinfo{person}{Kokil Jaidka}, {and} \bibinfo{person}{Yphtach Lelkes}.} \bibinfo{year}{2020}\natexlab{}.
\newblock \showarticletitle{Auditing local news presence on {Google} {News}}.
\newblock \bibinfo{journal}{\emph{Nature Human Behaviour}} \bibinfo{volume}{4}, \bibinfo{number}{12} (\bibinfo{date}{Sept.} \bibinfo{year}{2020}), \bibinfo{pages}{1236--1244}.
\newblock
\showISSN{2397-3374}
\urldef\tempurl%
\url{https://doi.org/10.1038/s41562-020-00954-0}
\showDOI{\tempurl}


\bibitem[Fosch-Villaronga et~al\mbox{.}(2021)]%
        {fosch-villaronga_little_2021}
\bibfield{author}{\bibinfo{person}{E. Fosch-Villaronga}, \bibinfo{person}{A. Poulsen}, \bibinfo{person}{R.A. Søraa}, {and} \bibinfo{person}{B.H.M. Custers}.} \bibinfo{year}{2021}\natexlab{}.
\newblock \showarticletitle{A little bird told me your gender: {Gender} inferences in social media}.
\newblock \bibinfo{journal}{\emph{Information Processing \& Management}} \bibinfo{volume}{58}, \bibinfo{number}{3} (\bibinfo{date}{May} \bibinfo{year}{2021}), \bibinfo{pages}{102541}.
\newblock
\showISSN{03064573}
\urldef\tempurl%
\url{https://doi.org/10.1016/j.ipm.2021.102541}
\showDOI{\tempurl}


\bibitem[Fujimoto and Takemoto(2023)]%
        {fujimoto_revisiting_2023}
\bibfield{author}{\bibinfo{person}{Sasuke Fujimoto} {and} \bibinfo{person}{Kazuhiro Takemoto}.} \bibinfo{year}{2023}\natexlab{}.
\newblock \showarticletitle{Revisiting the political biases of {ChatGPT}}.
\newblock \bibinfo{journal}{\emph{Frontiers in Artificial Intelligence}}  \bibinfo{volume}{6} (\bibinfo{date}{Oct.} \bibinfo{year}{2023}), \bibinfo{pages}{1232003}.
\newblock
\showISSN{2624-8212}
\urldef\tempurl%
\url{https://doi.org/10.3389/frai.2023.1232003}
\showDOI{\tempurl}


\bibitem[George et~al\mbox{.}(2024)]%
        {george_use_2024}
\bibfield{author}{\bibinfo{person}{Rachel~Surrage George}, \bibinfo{person}{Hannah Goodey}, \bibinfo{person}{Maria~Antonietta Russo}, \bibinfo{person}{Rovena Tula}, {and} \bibinfo{person}{Pietro Ghezzi}.} \bibinfo{year}{2024}\natexlab{}.
\newblock \showarticletitle{Use of immunology in news and {YouTube} videos in the context of {COVID}-19: politicisation and information bubbles}.
\newblock \bibinfo{journal}{\emph{Frontiers in Public Health}}  \bibinfo{volume}{12} (\bibinfo{date}{Feb.} \bibinfo{year}{2024}), \bibinfo{pages}{1327704}.
\newblock
\showISSN{2296-2565}
\urldef\tempurl%
\url{https://doi.org/10.3389/fpubh.2024.1327704}
\showDOI{\tempurl}


\bibitem[Geyik et~al\mbox{.}(2019)]%
        {geyik_fairness-aware_2019}
\bibfield{author}{\bibinfo{person}{Sahin~Cem Geyik}, \bibinfo{person}{Stuart Ambler}, {and} \bibinfo{person}{Krishnaram Kenthapadi}.} \bibinfo{year}{2019}\natexlab{}.
\newblock \showarticletitle{Fairness-{Aware} {Ranking} in {Search} \& {Recommendation} {Systems} with {Application} to {LinkedIn} {Talent} {Search}}. In \bibinfo{booktitle}{\emph{Proceedings of the 25th {ACM} {SIGKDD} {International} {Conference} on {Knowledge} {Discovery} \& {Data} {Mining}}}. \bibinfo{publisher}{ACM}, \bibinfo{address}{Anchorage AK USA}, \bibinfo{pages}{2221--2231}.
\newblock
\showISBNx{978-1-4503-6201-6}
\urldef\tempurl%
\url{https://doi.org/10.1145/3292500.3330691}
\showDOI{\tempurl}


\bibitem[Gezici et~al\mbox{.}(2021)]%
        {gezici_evaluation_2021}
\bibfield{author}{\bibinfo{person}{Gizem Gezici}, \bibinfo{person}{Aldo Lipani}, \bibinfo{person}{Yucel Saygin}, {and} \bibinfo{person}{Emine Yilmaz}.} \bibinfo{year}{2021}\natexlab{}.
\newblock \showarticletitle{Evaluation metrics for measuring bias in search engine results}.
\newblock \bibinfo{journal}{\emph{Information Retrieval Journal}} \bibinfo{volume}{24}, \bibinfo{number}{2} (\bibinfo{date}{April} \bibinfo{year}{2021}), \bibinfo{pages}{85--113}.
\newblock
\showISSN{1386-4564, 1573-7659}
\urldef\tempurl%
\url{https://doi.org/10.1007/s10791-020-09386-w}
\showDOI{\tempurl}


\bibitem[Glaesener(2022)]%
        {glaesener_exploring_2022}
\bibfield{author}{\bibinfo{person}{Tim Glaesener}.} \bibinfo{year}{2022}\natexlab{}.
\newblock \showarticletitle{Exploring {Siri}’s {Content} {Diversity} {Using} a {Crowdsourced} {Audit}}.
\newblock \bibinfo{journal}{\emph{Journal of Digital Social Research}} \bibinfo{volume}{4}, \bibinfo{number}{1} (\bibinfo{date}{April} \bibinfo{year}{2022}), \bibinfo{pages}{128--151}.
\newblock
\showISSN{2003-1998}
\urldef\tempurl%
\url{https://doi.org/10.33621/jdsr.v4i1.115}
\showDOI{\tempurl}
\newblock
\shownote{Number: 1}.


\bibitem[Gleason et~al\mbox{.}(2024)]%
        {gleason_perceptions_2024}
\bibfield{author}{\bibinfo{person}{Jeffrey Gleason}, \bibinfo{person}{Avijit Ghosh}, \bibinfo{person}{Ronald~E. Robertson}, {and} \bibinfo{person}{Christo Wilson}.} \bibinfo{year}{2024}\natexlab{}.
\newblock \showarticletitle{Perceptions in {Pixels}: {Analyzing} {Perceived} {Gender} and {Skin} {Tone} in {Real}-world {Image} {Search} {Results}}. In \bibinfo{booktitle}{\emph{Proceedings of the {ACM} on {Web} {Conference} 2024}}. \bibinfo{publisher}{ACM}, \bibinfo{address}{Singapore Singapore}, \bibinfo{pages}{1249--1259}.
\newblock
\showISBNx{9798400701719}
\urldef\tempurl%
\url{https://doi.org/10.1145/3589334.3645666}
\showDOI{\tempurl}


\bibitem[Godinez and Mustafaraj(2024)]%
        {godinez_youtube_2024}
\bibfield{author}{\bibinfo{person}{Lillie Godinez} {and} \bibinfo{person}{Eni Mustafaraj}.} \bibinfo{year}{2024}\natexlab{}.
\newblock \showarticletitle{{YouTube} and {Conspiracy} {Theories}: {A} {Longitudinal} {Audit} of {Information} {Panels}}. In \bibinfo{booktitle}{\emph{Proceedings of the 35th {ACM} {Conference} on {Hypertext} and {Social} {Media}}}. \bibinfo{publisher}{ACM}, \bibinfo{address}{Poznan Poland}, \bibinfo{pages}{273--284}.
\newblock
\showISBNx{9798400705953}
\urldef\tempurl%
\url{https://doi.org/10.1145/3648188.3675128}
\showDOI{\tempurl}


\bibitem[Gong et~al\mbox{.}(2024)]%
        {gong_does_2024}
\bibfield{author}{\bibinfo{person}{Yaqi Gong}, \bibinfo{person}{Ashley Schroeder}, \bibinfo{person}{Bing Pan}, \bibinfo{person}{S.~Shyam Sundar}, {and} \bibinfo{person}{Andrew~J. Mowen}.} \bibinfo{year}{2024}\natexlab{}.
\newblock \showarticletitle{Does algorithmic filtering lead to filter bubbles in online tourist information searches?}
\newblock \bibinfo{journal}{\emph{Information Technology \& Tourism}} \bibinfo{volume}{26}, \bibinfo{number}{1} (\bibinfo{date}{March} \bibinfo{year}{2024}), \bibinfo{pages}{183--217}.
\newblock
\showISSN{1098-3058, 1943-4294}
\urldef\tempurl%
\url{https://doi.org/10.1007/s40558-023-00279-4}
\showDOI{\tempurl}


\bibitem[Haak and Schaer(2022)]%
        {haak_auditing_2022}
\bibfield{author}{\bibinfo{person}{Fabian Haak} {and} \bibinfo{person}{Philipp Schaer}.} \bibinfo{year}{2022}\natexlab{}.
\newblock \showarticletitle{Auditing {Search} {Query} {Suggestion} {Bias} {Through} {Recursive} {Algorithm} {Interrogation}}. In \bibinfo{booktitle}{\emph{14th {ACM} {Web} {Science} {Conference} 2022}}. \bibinfo{publisher}{ACM}, \bibinfo{address}{Barcelona Spain}, \bibinfo{pages}{219--227}.
\newblock
\showISBNx{978-1-4503-9191-7}
\urldef\tempurl%
\url{https://doi.org/10.1145/3501247.3531567}
\showDOI{\tempurl}


\bibitem[Hagar and Diakopoulos(2023)]%
        {hagar_algorithmic_2023}
\bibfield{author}{\bibinfo{person}{Nick Hagar} {and} \bibinfo{person}{Nicholas Diakopoulos}.} \bibinfo{year}{2023}\natexlab{}.
\newblock \showarticletitle{Algorithmic indifference: {The} dearth of news recommendations on {TikTok}}.
\newblock \bibinfo{journal}{\emph{New Media \& Society}} (\bibinfo{date}{Aug.} \bibinfo{year}{2023}), \bibinfo{pages}{14614448231192964}.
\newblock
\showISSN{1461-4448, 1461-7315}
\urldef\tempurl%
\url{https://doi.org/10.1177/14614448231192964}
\showDOI{\tempurl}


\bibitem[Haim et~al\mbox{.}(2018)]%
        {haim_burst_2018}
\bibfield{author}{\bibinfo{person}{Mario Haim}, \bibinfo{person}{Andreas Graefe}, {and} \bibinfo{person}{Hans-Bernd Brosius}.} \bibinfo{year}{2018}\natexlab{}.
\newblock \showarticletitle{Burst of the {Filter} {Bubble}?: {Effects} of personalization on the diversity of \textit{{Google} {News}}}.
\newblock \bibinfo{journal}{\emph{Digital Journalism}} \bibinfo{volume}{6}, \bibinfo{number}{3} (\bibinfo{date}{March} \bibinfo{year}{2018}), \bibinfo{pages}{330--343}.
\newblock
\showISSN{2167-0811, 2167-082X}
\urldef\tempurl%
\url{https://doi.org/10.1080/21670811.2017.1338145}
\showDOI{\tempurl}


\bibitem[Hannak et~al\mbox{.}(2013)]%
        {hannak_measuring_2013}
\bibfield{author}{\bibinfo{person}{Aniko Hannak}, \bibinfo{person}{Piotr Sapiezynski}, \bibinfo{person}{Arash Molavi~Kakhki}, \bibinfo{person}{Balachander Krishnamurthy}, \bibinfo{person}{David Lazer}, \bibinfo{person}{Alan Mislove}, {and} \bibinfo{person}{Christo Wilson}.} \bibinfo{year}{2013}\natexlab{}.
\newblock \showarticletitle{Measuring personalization of web search}. In \bibinfo{booktitle}{\emph{Proceedings of the 22nd international conference on {World} {Wide} {Web}}}. \bibinfo{publisher}{ACM}, \bibinfo{address}{Rio de Janeiro Brazil}, \bibinfo{pages}{527--538}.
\newblock
\showISBNx{978-1-4503-2035-1}
\urldef\tempurl%
\url{https://doi.org/10.1145/2488388.2488435}
\showDOI{\tempurl}


\bibitem[Hannak et~al\mbox{.}(2014)]%
        {hannak_measuring_2014}
\bibfield{author}{\bibinfo{person}{Aniko Hannak}, \bibinfo{person}{Gary Soeller}, \bibinfo{person}{David Lazer}, \bibinfo{person}{Alan Mislove}, {and} \bibinfo{person}{Christo Wilson}.} \bibinfo{year}{2014}\natexlab{}.
\newblock \showarticletitle{Measuring {Price} {Discrimination} and {Steering} on {E}-commerce {Web} {Sites}}. In \bibinfo{booktitle}{\emph{Proceedings of the 2014 {Conference} on {Internet} {Measurement} {Conference}}}. \bibinfo{publisher}{ACM}, \bibinfo{address}{Vancouver BC Canada}, \bibinfo{pages}{305--318}.
\newblock
\showISBNx{978-1-4503-3213-2}
\urldef\tempurl%
\url{https://doi.org/10.1145/2663716.2663744}
\showDOI{\tempurl}


\bibitem[Hannák et~al\mbox{.}(2017)]%
        {hannak_bias_2017}
\bibfield{author}{\bibinfo{person}{Anikó Hannák}, \bibinfo{person}{Claudia Wagner}, \bibinfo{person}{David Garcia}, \bibinfo{person}{Alan Mislove}, \bibinfo{person}{Markus Strohmaier}, {and} \bibinfo{person}{Christo Wilson}.} \bibinfo{year}{2017}\natexlab{}.
\newblock \showarticletitle{Bias in {Online} {Freelance} {Marketplaces}: {Evidence} from {TaskRabbit} and {Fiverr}}. In \bibinfo{booktitle}{\emph{Proceedings of the 2017 {ACM} {Conference} on {Computer} {Supported} {Cooperative} {Work} and {Social} {Computing}}}. \bibinfo{publisher}{ACM}, \bibinfo{address}{Portland Oregon USA}, \bibinfo{pages}{1914--1933}.
\newblock
\showISBNx{978-1-4503-4335-0}
\urldef\tempurl%
\url{https://doi.org/10.1145/2998181.2998327}
\showDOI{\tempurl}


\bibitem[Henrich et~al\mbox{.}(2010)]%
        {henrich_most_2010}
\bibfield{author}{\bibinfo{person}{Joseph Henrich}, \bibinfo{person}{Steven~J. Heine}, {and} \bibinfo{person}{Ara Norenzayan}.} \bibinfo{year}{2010}\natexlab{}.
\newblock \showarticletitle{Most people are not {WEIRD}}.
\newblock \bibinfo{journal}{\emph{Nature}} \bibinfo{volume}{466}, \bibinfo{number}{7302} (\bibinfo{date}{July} \bibinfo{year}{2010}), \bibinfo{pages}{29--29}.
\newblock
\showISSN{1476-4687}
\urldef\tempurl%
\url{https://doi.org/10.1038/466029a}
\showDOI{\tempurl}
\newblock
\shownote{Number: 7302 Publisher: Nature Publishing Group}.


\bibitem[Heuer et~al\mbox{.}(2021)]%
        {heuer_auditing_2021}
\bibfield{author}{\bibinfo{person}{Hendrik Heuer}, \bibinfo{person}{Hendrik Hoch}, \bibinfo{person}{Andreas Breiter}, {and} \bibinfo{person}{Yannis Theocharis}.} \bibinfo{year}{2021}\natexlab{}.
\newblock \showarticletitle{Auditing the {Biases} {Enacted} by {YouTube} for {Political} {Topics} in {Germany}}. In \bibinfo{booktitle}{\emph{Mensch und {Computer} 2021}}. \bibinfo{publisher}{ACM}, \bibinfo{address}{Ingolstadt Germany}, \bibinfo{pages}{456--468}.
\newblock
\showISBNx{978-1-4503-8645-6}
\urldef\tempurl%
\url{https://doi.org/10.1145/3473856.3473864}
\showDOI{\tempurl}


\bibitem[Hosseinmardi et~al\mbox{.}(2024)]%
        {hosseinmardi_causally_2024}
\bibfield{author}{\bibinfo{person}{Homa Hosseinmardi}, \bibinfo{person}{Amir Ghasemian}, \bibinfo{person}{Miguel Rivera-Lanas}, \bibinfo{person}{Manoel Horta~Ribeiro}, \bibinfo{person}{Robert West}, {and} \bibinfo{person}{Duncan~J. Watts}.} \bibinfo{year}{2024}\natexlab{}.
\newblock \showarticletitle{Causally estimating the effect of {YouTube}’s recommender system using counterfactual bots}.
\newblock \bibinfo{journal}{\emph{Proceedings of the National Academy of Sciences}} \bibinfo{volume}{121}, \bibinfo{number}{8} (\bibinfo{date}{Feb.} \bibinfo{year}{2024}), \bibinfo{pages}{e2313377121}.
\newblock
\showISSN{0027-8424, 1091-6490}
\urldef\tempurl%
\url{https://doi.org/10.1073/pnas.2313377121}
\showDOI{\tempurl}


\bibitem[Hu et~al\mbox{.}(2019)]%
        {hu_auditing_2019}
\bibfield{author}{\bibinfo{person}{Desheng Hu}, \bibinfo{person}{Shan Jiang}, \bibinfo{person}{Ronald E.~Robertson}, {and} \bibinfo{person}{Christo Wilson}.} \bibinfo{year}{2019}\natexlab{}.
\newblock \showarticletitle{Auditing the {Partisanship} of {Google} {Search} {Snippets}}. In \bibinfo{booktitle}{\emph{The {World} {Wide} {Web} {Conference}}}. \bibinfo{publisher}{ACM}, \bibinfo{address}{San Francisco CA USA}, \bibinfo{pages}{693--704}.
\newblock
\showISBNx{978-1-4503-6674-8}
\urldef\tempurl%
\url{https://doi.org/10.1145/3308558.3313654}
\showDOI{\tempurl}


\bibitem[Huang and Yang(2024)]%
        {huang_auditing_2024}
\bibfield{author}{\bibinfo{person}{Shengchun Huang} {and} \bibinfo{person}{Tian Yang}.} \bibinfo{year}{2024}\natexlab{}.
\newblock \showarticletitle{Auditing {Entertainment} {Traps} on {YouTube}: {How} {Do} {Recommendation} {Algorithms} {Pull} {Users} {Away} from {News}}.
\newblock \bibinfo{journal}{\emph{Political Communication}} (\bibinfo{date}{April} \bibinfo{year}{2024}), \bibinfo{pages}{1--19}.
\newblock
\showISSN{1058-4609, 1091-7675}
\urldef\tempurl%
\url{https://doi.org/10.1080/10584609.2024.2343769}
\showDOI{\tempurl}


\bibitem[Hupperich et~al\mbox{.}(2018)]%
        {hupperich_empirical_2018}
\bibfield{author}{\bibinfo{person}{Thomas Hupperich}, \bibinfo{person}{Dennis Tatang}, \bibinfo{person}{Nicolai Wilkop}, {and} \bibinfo{person}{Thorsten Holz}.} \bibinfo{year}{2018}\natexlab{}.
\newblock \showarticletitle{An {Empirical} {Study} on {Online} {Price} {Differentiation}}. In \bibinfo{booktitle}{\emph{Proceedings of the {Eighth} {ACM} {Conference} on {Data} and {Application} {Security} and {Privacy}}}. \bibinfo{publisher}{ACM}, \bibinfo{address}{Tempe AZ USA}, \bibinfo{pages}{76--83}.
\newblock
\showISBNx{978-1-4503-5632-9}
\urldef\tempurl%
\url{https://doi.org/10.1145/3176258.3176338}
\showDOI{\tempurl}


\bibitem[Hussein et~al\mbox{.}(2020)]%
        {hussein_measuring_2020}
\bibfield{author}{\bibinfo{person}{Eslam Hussein}, \bibinfo{person}{Prerna Juneja}, {and} \bibinfo{person}{Tanushree Mitra}.} \bibinfo{year}{2020}\natexlab{}.
\newblock \showarticletitle{Measuring {Misinformation} in {Video} {Search} {Platforms}: {An} {Audit} {Study} on {YouTube}}.
\newblock \bibinfo{journal}{\emph{Proceedings of the ACM on Human-Computer Interaction}} \bibinfo{volume}{4}, \bibinfo{number}{CSCW1} (\bibinfo{date}{May} \bibinfo{year}{2020}), \bibinfo{pages}{1--27}.
\newblock
\showISSN{2573-0142}
\urldef\tempurl%
\url{https://doi.org/10.1145/3392854}
\showDOI{\tempurl}


\bibitem[Huszár et~al\mbox{.}(2022)]%
        {huszar_algorithmic_2022}
\bibfield{author}{\bibinfo{person}{Ferenc Huszár}, \bibinfo{person}{Sofia~Ira Ktena}, \bibinfo{person}{Conor O’Brien}, \bibinfo{person}{Luca Belli}, \bibinfo{person}{Andrew Schlaikjer}, {and} \bibinfo{person}{Moritz Hardt}.} \bibinfo{year}{2022}\natexlab{}.
\newblock \showarticletitle{Algorithmic amplification of politics on {Twitter}}.
\newblock \bibinfo{journal}{\emph{Proceedings of the National Academy of Sciences}} \bibinfo{volume}{119}, \bibinfo{number}{1} (\bibinfo{date}{Jan.} \bibinfo{year}{2022}), \bibinfo{pages}{e2025334119}.
\newblock
\showISSN{0027-8424, 1091-6490}
\urldef\tempurl%
\url{https://doi.org/10.1073/pnas.2025334119}
\showDOI{\tempurl}


\bibitem[Ibrahim et~al\mbox{.}(2023)]%
        {ibrahim_youtubes_2023}
\bibfield{author}{\bibinfo{person}{Hazem Ibrahim}, \bibinfo{person}{Nouar AlDahoul}, \bibinfo{person}{Sangjin Lee}, \bibinfo{person}{Talal Rahwan}, {and} \bibinfo{person}{Yasir Zaki}.} \bibinfo{year}{2023}\natexlab{}.
\newblock \showarticletitle{{YouTube}’s recommendation algorithm is left-leaning in the {United} {States}}.
\newblock \bibinfo{journal}{\emph{PNAS Nexus}} \bibinfo{volume}{2}, \bibinfo{number}{8} (\bibinfo{date}{Aug.} \bibinfo{year}{2023}), \bibinfo{pages}{pgad264}.
\newblock
\showISSN{2752-6542}
\urldef\tempurl%
\url{https://doi.org/10.1093/pnasnexus/pgad264}
\showDOI{\tempurl}


\bibitem[Ieracitano et~al\mbox{.}(2023)]%
        {ieracitano_im_2023}
\bibfield{author}{\bibinfo{person}{Francesca Ieracitano}, \bibinfo{person}{Francesco Vigneri}, {and} \bibinfo{person}{Francesca Comunello}.} \bibinfo{year}{2023}\natexlab{}.
\newblock \showarticletitle{‘{I}’m not bad, {I}’m just … drawn that way’: media and algorithmic systems logics in the {Italian} {Google} {Images} construction of (cr)immigrants’ communities}.
\newblock \bibinfo{journal}{\emph{Information, Communication \& Society}} (\bibinfo{date}{April} \bibinfo{year}{2023}), \bibinfo{pages}{1--20}.
\newblock
\showISSN{1369-118X, 1468-4462}
\urldef\tempurl%
\url{https://doi.org/10.1080/1369118X.2023.2205928}
\showDOI{\tempurl}


\bibitem[Imana et~al\mbox{.}(2021)]%
        {imana_auditing_2021}
\bibfield{author}{\bibinfo{person}{Basileal Imana}, \bibinfo{person}{Aleksandra Korolova}, {and} \bibinfo{person}{John Heidemann}.} \bibinfo{year}{2021}\natexlab{}.
\newblock \showarticletitle{Auditing for {Discrimination} in {Algorithms} {Delivering} {Job} {Ads}}. In \bibinfo{booktitle}{\emph{Proceedings of the {Web} {Conference} 2021}}. \bibinfo{publisher}{ACM}, \bibinfo{address}{Ljubljana Slovenia}, \bibinfo{pages}{3767--3778}.
\newblock
\showISBNx{978-1-4503-8312-7}
\urldef\tempurl%
\url{https://doi.org/10.1145/3442381.3450077}
\showDOI{\tempurl}


\bibitem[Imana et~al\mbox{.}(2024)]%
        {imana_auditing_2024}
\bibfield{author}{\bibinfo{person}{Basileal Imana}, \bibinfo{person}{Aleksandra Korolova}, {and} \bibinfo{person}{John Heidemann}.} \bibinfo{year}{2024}\natexlab{}.
\newblock \showarticletitle{Auditing for {Racial} {Discrimination} in the {Delivery} of {Education} {Ads}}. In \bibinfo{booktitle}{\emph{The 2024 {ACM} {Conference} on {Fairness}, {Accountability}, and {Transparency}}}. \bibinfo{publisher}{ACM}, \bibinfo{address}{Rio de Janeiro Brazil}, \bibinfo{pages}{2348--2361}.
\newblock
\showISBNx{9798400704505}
\urldef\tempurl%
\url{https://doi.org/10.1145/3630106.3659041}
\showDOI{\tempurl}


\bibitem[Iqbal et~al\mbox{.}(2022)]%
        {iqbal_left_2022}
\bibfield{author}{\bibinfo{person}{Hassan Iqbal}, \bibinfo{person}{Usman~Mahmood Khan}, \bibinfo{person}{Hassan~Ali Khan}, {and} \bibinfo{person}{Muhammad Shahzad}.} \bibinfo{year}{2022}\natexlab{}.
\newblock \showarticletitle{Left or {Right}: {A} {Peek} into the {Political} {Biases} in {Email} {Spam} {Filtering} {Algorithms} {During} {US} {Election} 2020}. In \bibinfo{booktitle}{\emph{Proceedings of the {ACM} {Web} {Conference} 2022}}. \bibinfo{publisher}{ACM}, \bibinfo{address}{Virtual Event, Lyon France}, \bibinfo{pages}{2491--2500}.
\newblock
\showISBNx{978-1-4503-9096-5}
\urldef\tempurl%
\url{https://doi.org/10.1145/3485447.3512121}
\showDOI{\tempurl}


\bibitem[Jaidka et~al\mbox{.}(2023)]%
        {jaidka_silenced_2023}
\bibfield{author}{\bibinfo{person}{Kokil Jaidka}, \bibinfo{person}{Subhayan Mukerjee}, {and} \bibinfo{person}{Yphtach Lelkes}.} \bibinfo{year}{2023}\natexlab{}.
\newblock \showarticletitle{Silenced on social media: the gatekeeping functions of shadowbans in the {American} {Twitterverse}}.
\newblock \bibinfo{journal}{\emph{Journal of Communication}} \bibinfo{volume}{73}, \bibinfo{number}{2} (\bibinfo{date}{April} \bibinfo{year}{2023}), \bibinfo{pages}{163--178}.
\newblock
\showISSN{0021-9916, 1460-2466}
\urldef\tempurl%
\url{https://doi.org/10.1093/joc/jqac050}
\showDOI{\tempurl}


\bibitem[Jokubauskaitė et~al\mbox{.}(2023)]%
        {jokubauskaite_winner-take-all_2023}
\bibfield{author}{\bibinfo{person}{Emilija Jokubauskaitė}, \bibinfo{person}{Bernhard Rieder}, {and} \bibinfo{person}{Sarah Burkhardt}.} \bibinfo{year}{2023}\natexlab{}.
\newblock \showarticletitle{Winner-{Take}-{All}? {Visibility}, {Availability}, and {Heterogeneity} on {Webcam} {Sex} {Platforms}}.
\newblock \bibinfo{journal}{\emph{Social Media + Society}} \bibinfo{volume}{9}, \bibinfo{number}{4} (\bibinfo{date}{Oct.} \bibinfo{year}{2023}), \bibinfo{pages}{20563051231214807}.
\newblock
\showISSN{2056-3051, 2056-3051}
\urldef\tempurl%
\url{https://doi.org/10.1177/20563051231214807}
\showDOI{\tempurl}


\bibitem[Juneja et~al\mbox{.}(2023)]%
        {juneja_assessing_2023}
\bibfield{author}{\bibinfo{person}{Prerna Juneja}, \bibinfo{person}{Md~Momen Bhuiyan}, {and} \bibinfo{person}{Tanushree Mitra}.} \bibinfo{year}{2023}\natexlab{}.
\newblock \showarticletitle{Assessing enactment of content regulation policies: {A} post hoc crowd-sourced audit of election misinformation on {YouTube}}. In \bibinfo{booktitle}{\emph{Proceedings of the 2023 {CHI} {Conference} on {Human} {Factors} in {Computing} {Systems}}}. \bibinfo{publisher}{ACM}, \bibinfo{address}{Hamburg Germany}, \bibinfo{pages}{1--22}.
\newblock
\showISBNx{978-1-4503-9421-5}
\urldef\tempurl%
\url{https://doi.org/10.1145/3544548.3580846}
\showDOI{\tempurl}


\bibitem[Juneja and Mitra(2021)]%
        {juneja_auditing_2021}
\bibfield{author}{\bibinfo{person}{Prerna Juneja} {and} \bibinfo{person}{Tanushree Mitra}.} \bibinfo{year}{2021}\natexlab{}.
\newblock \showarticletitle{Auditing {E}-{Commerce} {Platforms} for {Algorithmically} {Curated} {Vaccine} {Misinformation}}. In \bibinfo{booktitle}{\emph{Proceedings of the 2021 {CHI} {Conference} on {Human} {Factors} in {Computing} {Systems}}}. \bibinfo{publisher}{ACM}, \bibinfo{address}{Yokohama Japan}, \bibinfo{pages}{1--27}.
\newblock
\showISBNx{978-1-4503-8096-6}
\urldef\tempurl%
\url{https://doi.org/10.1145/3411764.3445250}
\showDOI{\tempurl}


\bibitem[Kacperski et~al\mbox{.}(2024)]%
        {kacperski_examining_2024}
\bibfield{author}{\bibinfo{person}{Celina Kacperski}, \bibinfo{person}{Mona Bielig}, \bibinfo{person}{Mykola Makhortykh}, \bibinfo{person}{Maryna Sydorova}, {and} \bibinfo{person}{Roberto Ulloa}.} \bibinfo{year}{2024}\natexlab{}.
\newblock \showarticletitle{Examining bias perpetuation in academic search engines: {An} algorithm audit of {Google} and {Semantic} {Scholar}}.
\newblock \bibinfo{journal}{\emph{First Monday}} (\bibinfo{date}{Nov.} \bibinfo{year}{2024}).
\newblock
\showISSN{1396-0466}
\urldef\tempurl%
\url{https://doi.org/10.5210/fm.v29i11.13730}
\showDOI{\tempurl}


\bibitem[Kaiser and Rauchfleisch(2020)]%
        {kaiser_birds_2020}
\bibfield{author}{\bibinfo{person}{Jonas Kaiser} {and} \bibinfo{person}{Adrian Rauchfleisch}.} \bibinfo{year}{2020}\natexlab{}.
\newblock \showarticletitle{Birds of a {Feather} {Get} {Recommended} {Together}: {Algorithmic} {Homophily} in {YouTube}’s {Channel} {Recommendations} in the {United} {States} and {Germany}}.
\newblock \bibinfo{journal}{\emph{Social Media + Society}} \bibinfo{volume}{6}, \bibinfo{number}{4} (\bibinfo{date}{Oct.} \bibinfo{year}{2020}), \bibinfo{pages}{205630512096991}.
\newblock
\showISSN{2056-3051, 2056-3051}
\urldef\tempurl%
\url{https://doi.org/10.1177/2056305120969914}
\showDOI{\tempurl}


\bibitem[Kaplan and Sapiezynski(2024)]%
        {kaplan_comprehensively_2024}
\bibfield{author}{\bibinfo{person}{Levi Kaplan} {and} \bibinfo{person}{Piotr Sapiezynski}.} \bibinfo{year}{2024}\natexlab{}.
\newblock \showarticletitle{Comprehensively {Auditing} the {TikTok} {Mobile} {App}}. In \bibinfo{booktitle}{\emph{Companion {Proceedings} of the {ACM} on {Web} {Conference} 2024}}. \bibinfo{publisher}{ACM}, \bibinfo{address}{Singapore Singapore}, \bibinfo{pages}{1198--1201}.
\newblock
\showISBNx{9798400701726}
\urldef\tempurl%
\url{https://doi.org/10.1145/3589335.3651260}
\showDOI{\tempurl}


\bibitem[Karan et~al\mbox{.}(2023)]%
        {karan_your_2023}
\bibfield{author}{\bibinfo{person}{Aditya Karan}, \bibinfo{person}{Naina Balepur}, {and} \bibinfo{person}{Hari Sundaram}.} \bibinfo{year}{2023}\natexlab{}.
\newblock \showarticletitle{Your {Browsing} {History} {May} {Cost} {You}: {A} {Framework} for {Discovering} {Differential} {Pricing} in {Non}-{Transparent} {Markets}}. In \bibinfo{booktitle}{\emph{2023 {ACM} {Conference} on {Fairness}, {Accountability}, and {Transparency}}}. \bibinfo{publisher}{ACM}, \bibinfo{address}{Chicago IL USA}, \bibinfo{pages}{717--735}.
\newblock
\showISBNx{9798400701924}
\urldef\tempurl%
\url{https://doi.org/10.1145/3593013.3594038}
\showDOI{\tempurl}


\bibitem[Kay et~al\mbox{.}(2015)]%
        {kay_unequal_2015}
\bibfield{author}{\bibinfo{person}{Matthew Kay}, \bibinfo{person}{Cynthia Matuszek}, {and} \bibinfo{person}{Sean~A. Munson}.} \bibinfo{year}{2015}\natexlab{}.
\newblock \showarticletitle{Unequal {Representation} and {Gender} {Stereotypes} in {Image} {Search} {Results} for {Occupations}}. In \bibinfo{booktitle}{\emph{Proceedings of the 33rd {Annual} {ACM} {Conference} on {Human} {Factors} in {Computing} {Systems}}}. \bibinfo{publisher}{ACM}, \bibinfo{address}{Seoul Republic of Korea}, \bibinfo{pages}{3819--3828}.
\newblock
\showISBNx{978-1-4503-3145-6}
\urldef\tempurl%
\url{https://doi.org/10.1145/2702123.2702520}
\showDOI{\tempurl}


\bibitem[Kim et~al\mbox{.}(2024)]%
        {kim_exploring_2024}
\bibfield{author}{\bibinfo{person}{Junghwan Kim}, \bibinfo{person}{Jinhyung Lee}, \bibinfo{person}{Kee~Moon Jang}, {and} \bibinfo{person}{Ismini Lourentzou}.} \bibinfo{year}{2024}\natexlab{}.
\newblock \showarticletitle{Exploring the limitations in how {ChatGPT} introduces environmental justice issues in the {United} {States}: {A} case study of 3,108 counties}.
\newblock \bibinfo{journal}{\emph{Telematics and Informatics}}  \bibinfo{volume}{86} (\bibinfo{date}{Feb.} \bibinfo{year}{2024}), \bibinfo{pages}{102085}.
\newblock
\showISSN{07365853}
\urldef\tempurl%
\url{https://doi.org/10.1016/j.tele.2023.102085}
\showDOI{\tempurl}


\bibitem[Kliman-Silver et~al\mbox{.}(2015)]%
        {kliman-silver_location_2015}
\bibfield{author}{\bibinfo{person}{Chloe Kliman-Silver}, \bibinfo{person}{Aniko Hannak}, \bibinfo{person}{David Lazer}, \bibinfo{person}{Christo Wilson}, {and} \bibinfo{person}{Alan Mislove}.} \bibinfo{year}{2015}\natexlab{}.
\newblock \showarticletitle{Location, {Location}, {Location}: {The} {Impact} of {Geolocation} on {Web} {Search} {Personalization}}. In \bibinfo{booktitle}{\emph{Proceedings of the 2015 {Internet} {Measurement} {Conference}}}. \bibinfo{publisher}{ACM}, \bibinfo{address}{Tokyo Japan}, \bibinfo{pages}{121--127}.
\newblock
\showISBNx{978-1-4503-3848-6}
\urldef\tempurl%
\url{https://doi.org/10.1145/2815675.2815714}
\showDOI{\tempurl}


\bibitem[Kollyri(2021)]%
        {kollyri_-coding_2021}
\bibfield{author}{\bibinfo{person}{Lydia Kollyri}.} \bibinfo{year}{2021}\natexlab{}.
\newblock \showarticletitle{De-coding {Instagram} as a {Spectacle}}.
\newblock \bibinfo{journal}{\emph{Medi\&\#225;ln\&\#237; studia}} \bibinfo{volume}{15}, \bibinfo{number}{02} (\bibinfo{year}{2021}), \bibinfo{pages}{103--124}.
\newblock
\showISSN{1801-9978, 2464-4846}
\urldef\tempurl%
\url{https://www.ceeol.com/search/article-detail?id=1006361}
\showURL{%
\tempurl}
\newblock
\shownote{Publisher: Univerzita Karlova v Praze, Fakulta soci\&\#225;ln\&\#237;ch věd}.


\bibitem[Koronska and Rogers(2024)]%
        {koronska_fact_2024}
\bibfield{author}{\bibinfo{person}{Kamila Koronska} {and} \bibinfo{person}{Richard Rogers}.} \bibinfo{year}{2024}\natexlab{}.
\newblock \showarticletitle{Fact checks versus problematic content in search rankings: {SEO} effects and the question of {Google}’s content moderation}. In \bibinfo{booktitle}{\emph{{ACM} {Web} {Science} {Conference}}}. \bibinfo{publisher}{ACM}, \bibinfo{address}{Stuttgart Germany}, \bibinfo{pages}{170--180}.
\newblock
\showISBNx{9798400703348}
\urldef\tempurl%
\url{https://doi.org/10.1145/3614419.3644017}
\showDOI{\tempurl}


\bibitem[Kravets et~al\mbox{.}(2023)]%
        {kravets_different_2023}
\bibfield{author}{\bibinfo{person}{Daria Kravets}, \bibinfo{person}{Anna Ryzhova}, \bibinfo{person}{Florian Toepfl}, {and} \bibinfo{person}{Arista Beseler}.} \bibinfo{year}{2023}\natexlab{}.
\newblock \showarticletitle{Different platforms, different plots? {The} {Kremlin}-controlled search engine {Yandex} as a resource for {Russia}’s informational influence in {Belarus} during the {COVID}-19 pandemic}.
\newblock \bibinfo{journal}{\emph{Journalism}} \bibinfo{volume}{24}, \bibinfo{number}{12} (\bibinfo{date}{Dec.} \bibinfo{year}{2023}), \bibinfo{pages}{2762--2780}.
\newblock
\showISSN{1464-8849, 1741-3001}
\urldef\tempurl%
\url{https://doi.org/10.1177/14648849231157845}
\showDOI{\tempurl}


\bibitem[Kravets and Toepfl(2022)]%
        {kravets_gauging_2022}
\bibfield{author}{\bibinfo{person}{Daria Kravets} {and} \bibinfo{person}{F. Toepfl}.} \bibinfo{year}{2022}\natexlab{}.
\newblock \showarticletitle{Gauging reference and source bias over time: how {Russia}’s partially state-controlled search engine {Yandex} mediated an anti-regime protest event}.
\newblock \bibinfo{journal}{\emph{Information, Communication \& Society}} \bibinfo{volume}{25}, \bibinfo{number}{15} (\bibinfo{date}{Nov.} \bibinfo{year}{2022}), \bibinfo{pages}{2207--2223}.
\newblock
\showISSN{1369-118X, 1468-4462}
\urldef\tempurl%
\url{https://doi.org/10.1080/1369118X.2021.1933563}
\showDOI{\tempurl}


\bibitem[Kulshrestha et~al\mbox{.}(2017)]%
        {kulshrestha_quantifying_2017}
\bibfield{author}{\bibinfo{person}{Juhi Kulshrestha}, \bibinfo{person}{Motahhare Eslami}, \bibinfo{person}{Johnnatan Messias}, \bibinfo{person}{Muhammad~Bilal Zafar}, \bibinfo{person}{Saptarshi Ghosh}, \bibinfo{person}{Krishna~P. Gummadi}, {and} \bibinfo{person}{Karrie Karahalios}.} \bibinfo{year}{2017}\natexlab{}.
\newblock \showarticletitle{Quantifying {Search} {Bias}: {Investigating} {Sources} of {Bias} for {Political} {Searches} in {Social} {Media}}. In \bibinfo{booktitle}{\emph{Proceedings of the 2017 {ACM} {Conference} on {Computer} {Supported} {Cooperative} {Work} and {Social} {Computing}}}. \bibinfo{publisher}{ACM}, \bibinfo{address}{Portland Oregon USA}, \bibinfo{pages}{417--432}.
\newblock
\showISBNx{978-1-4503-4335-0}
\urldef\tempurl%
\url{https://doi.org/10.1145/2998181.2998321}
\showDOI{\tempurl}


\bibitem[Kulshrestha et~al\mbox{.}(2019)]%
        {kulshrestha_search_2019}
\bibfield{author}{\bibinfo{person}{Juhi Kulshrestha}, \bibinfo{person}{Motahhare Eslami}, \bibinfo{person}{Johnnatan Messias}, \bibinfo{person}{Muhammad~Bilal Zafar}, \bibinfo{person}{Saptarshi Ghosh}, \bibinfo{person}{Krishna~P. Gummadi}, {and} \bibinfo{person}{Karrie Karahalios}.} \bibinfo{year}{2019}\natexlab{}.
\newblock \showarticletitle{Search bias quantification: investigating political bias in social media and web search}.
\newblock \bibinfo{journal}{\emph{Information Retrieval Journal}} \bibinfo{volume}{22}, \bibinfo{number}{1-2} (\bibinfo{date}{April} \bibinfo{year}{2019}), \bibinfo{pages}{188--227}.
\newblock
\showISSN{1386-4564, 1573-7659}
\urldef\tempurl%
\url{https://doi.org/10.1007/s10791-018-9341-2}
\showDOI{\tempurl}


\bibitem[Kuznetsova and Makhortykh(2023)]%
        {kuznetsova_blame_2023}
\bibfield{author}{\bibinfo{person}{Elizaveta Kuznetsova} {and} \bibinfo{person}{Mykola Makhortykh}.} \bibinfo{year}{2023}\natexlab{}.
\newblock \showarticletitle{Blame {It} on the {Algorithm}? {Russian} {Government}-{Sponsored} {Media} and {Algorithmic} {Curation} of {Political} {Information} on {Facebook}}.
\newblock \bibinfo{journal}{\emph{International Journal of Communication}} \bibinfo{volume}{17}, \bibinfo{number}{0} (\bibinfo{date}{Jan.} \bibinfo{year}{2023}), \bibinfo{pages}{22}.
\newblock
\showISSN{1932-8036}
\urldef\tempurl%
\url{https://ijoc.org/index.php/ijoc/article/view/18687}
\showURL{%
\tempurl}
\newblock
\shownote{Number: 0}.


\bibitem[Lai and Luczak-Roesch(2019)]%
        {weber_you_2019}
\bibfield{author}{\bibinfo{person}{Cameron Lai} {and} \bibinfo{person}{Markus Luczak-Roesch}.} \bibinfo{year}{2019}\natexlab{}.
\newblock \showarticletitle{You {Can}’t {See} {What} {You} {Can}’t {See}: {Experimental} {Evidence} for {How} {Much} {Relevant} {Information} {May} {Be} {Missed} {Due} to {Google}’s {Web} {Search} {Personalisation}}.
\newblock In \bibinfo{booktitle}{\emph{Social {Informatics}}}, \bibfield{editor}{\bibinfo{person}{Ingmar Weber}, \bibinfo{person}{Kareem~M. Darwish}, \bibinfo{person}{Claudia Wagner}, \bibinfo{person}{Emilio Zagheni}, \bibinfo{person}{Laura Nelson}, \bibinfo{person}{Samin Aref}, {and} \bibinfo{person}{Fabian Flöck}} (Eds.). Vol.~\bibinfo{volume}{11864}. \bibinfo{publisher}{Springer International Publishing}, \bibinfo{address}{Cham}, \bibinfo{pages}{253--266}.
\newblock
\showISBNx{978-3-030-34970-7 978-3-030-34971-4}
\urldef\tempurl%
\url{https://doi.org/10.1007/978-3-030-34971-4_17}
\showDOI{\tempurl}
\newblock
\shownote{Series Title: Lecture Notes in Computer Science}.


\bibitem[Lambrecht and Tucker(2019)]%
        {lambrecht_algorithmic_2019}
\bibfield{author}{\bibinfo{person}{Anja Lambrecht} {and} \bibinfo{person}{Catherine Tucker}.} \bibinfo{year}{2019}\natexlab{}.
\newblock \showarticletitle{Algorithmic {Bias}? {An} {Empirical} {Study} of {Apparent} {Gender}-{Based} {Discrimination} in the {Display} of {STEM} {Career} {Ads}}.
\newblock \bibinfo{journal}{\emph{Management Science}} \bibinfo{volume}{65}, \bibinfo{number}{7} (\bibinfo{date}{July} \bibinfo{year}{2019}), \bibinfo{pages}{2966--2981}.
\newblock
\showISSN{0025-1909, 1526-5501}
\urldef\tempurl%
\url{https://doi.org/10.1287/mnsc.2018.3093}
\showDOI{\tempurl}


\bibitem[Lambrecht and Tucker(2024)]%
        {lambrecht_apparent_2024}
\bibfield{author}{\bibinfo{person}{Anja Lambrecht} {and} \bibinfo{person}{Catherine Tucker}.} \bibinfo{year}{2024}\natexlab{}.
\newblock \showarticletitle{Apparent algorithmic discrimination and real-time algorithmic learning in digital search advertising}.
\newblock \bibinfo{journal}{\emph{Quantitative Marketing and Economics}} (\bibinfo{date}{July} \bibinfo{year}{2024}).
\newblock
\showISSN{1570-7156, 1573-711X}
\urldef\tempurl%
\url{https://doi.org/10.1007/s11129-024-09286-z}
\showDOI{\tempurl}


\bibitem[Le et~al\mbox{.}(2022)]%
        {boratto_crowdsourcing_2022}
\bibfield{author}{\bibinfo{person}{Binh Le}, \bibinfo{person}{Damiano Spina}, \bibinfo{person}{Falk Scholer}, {and} \bibinfo{person}{Hui Chia}.} \bibinfo{year}{2022}\natexlab{}.
\newblock \showarticletitle{A {Crowdsourcing} {Methodology} to {Measure} {Algorithmic} {Bias} in {Black}-{Box} {Systems}: {A} {Case} {Study} with {COVID}-{Related} {Searches}}.
\newblock In \bibinfo{booktitle}{\emph{Advances in {Bias} and {Fairness} in {Information} {Retrieval}}}, \bibfield{editor}{\bibinfo{person}{Ludovico Boratto}, \bibinfo{person}{Stefano Faralli}, \bibinfo{person}{Mirko Marras}, {and} \bibinfo{person}{Giovanni Stilo}} (Eds.). Vol.~\bibinfo{volume}{1610}. \bibinfo{publisher}{Springer International Publishing}, \bibinfo{address}{Cham}, \bibinfo{pages}{43--55}.
\newblock
\showISBNx{978-3-031-09315-9 978-3-031-09316-6}
\urldef\tempurl%
\url{https://doi.org/10.1007/978-3-031-09316-6_5}
\showDOI{\tempurl}
\newblock
\shownote{Series Title: Communications in Computer and Information Science}.


\bibitem[Le~Merrer et~al\mbox{.}(2023)]%
        {le_merrer_modeling_2023}
\bibfield{author}{\bibinfo{person}{Erwan Le~Merrer}, \bibinfo{person}{Gilles Tredan}, {and} \bibinfo{person}{Ali Yesilkanat}.} \bibinfo{year}{2023}\natexlab{}.
\newblock \showarticletitle{Modeling rabbit-holes on {YouTube}}.
\newblock \bibinfo{journal}{\emph{Social Network Analysis and Mining}} \bibinfo{volume}{13}, \bibinfo{number}{1} (\bibinfo{date}{Aug.} \bibinfo{year}{2023}), \bibinfo{pages}{100}.
\newblock
\showISSN{1869-5469}
\urldef\tempurl%
\url{https://doi.org/10.1007/s13278-023-01105-9}
\showDOI{\tempurl}


\bibitem[Ledwich et~al\mbox{.}(2022)]%
        {ledwich_radical_2022}
\bibfield{author}{\bibinfo{person}{Mark Ledwich}, \bibinfo{person}{Anna Zaitsev}, {and} \bibinfo{person}{Anton Laukemper}.} \bibinfo{year}{2022}\natexlab{}.
\newblock \showarticletitle{Radical bubbles on {YouTube}? {Revisiting} algorithmic extremism with personalised recommendations}.
\newblock \bibinfo{journal}{\emph{First Monday}} (\bibinfo{date}{Dec.} \bibinfo{year}{2022}).
\newblock
\showISSN{1396-0466}
\urldef\tempurl%
\url{https://doi.org/10.5210/fm.v27i12.12552}
\showDOI{\tempurl}


\bibitem[Leidinger and Rogers(2023)]%
        {leidinger_which_2023}
\bibfield{author}{\bibinfo{person}{Alina Leidinger} {and} \bibinfo{person}{Richard Rogers}.} \bibinfo{year}{2023}\natexlab{}.
\newblock \showarticletitle{Which {Stereotypes} {Are} {Moderated} and {Under}-{Moderated} in {Search} {Engine} {Autocompletion}?}. In \bibinfo{booktitle}{\emph{2023 {ACM} {Conference} on {Fairness}, {Accountability}, and {Transparency}}}. \bibinfo{publisher}{ACM}, \bibinfo{address}{Chicago IL USA}, \bibinfo{pages}{1049--1061}.
\newblock
\showISBNx{9798400701924}
\urldef\tempurl%
\url{https://doi.org/10.1145/3593013.3594062}
\showDOI{\tempurl}


\bibitem[Li and Sinnamon(2024)]%
        {li_generative_2024}
\bibfield{author}{\bibinfo{person}{Alice Li} {and} \bibinfo{person}{Luanne Sinnamon}.} \bibinfo{year}{2024}\natexlab{}.
\newblock \showarticletitle{Generative {\textless}span style="font-variant:small-caps;"{\textgreater}{AI}{\textless}/span{\textgreater} {Search} {Engines} as {Arbiters} of {Public} {Knowledge}: {An} {Audit} of {Bias} and {Authority}}.
\newblock \bibinfo{journal}{\emph{Proceedings of the Association for Information Science and Technology}} \bibinfo{volume}{61}, \bibinfo{number}{1} (\bibinfo{date}{Oct.} \bibinfo{year}{2024}), \bibinfo{pages}{205--217}.
\newblock
\showISSN{2373-9231, 2373-9231}
\urldef\tempurl%
\url{https://doi.org/10.1002/pra2.1021}
\showDOI{\tempurl}


\bibitem[Lin et~al\mbox{.}(2023)]%
        {lin_trapped_2023}
\bibfield{author}{\bibinfo{person}{Cong Lin}, \bibinfo{person}{Yuxin Gao}, \bibinfo{person}{Na Ta}, \bibinfo{person}{Kaiyu Li}, {and} \bibinfo{person}{Hongyao Fu}.} \bibinfo{year}{2023}\natexlab{}.
\newblock \showarticletitle{Trapped in the search box: {An} examination of algorithmic bias in search engine autocomplete predictions}.
\newblock \bibinfo{journal}{\emph{Telematics and Informatics}}  \bibinfo{volume}{85} (\bibinfo{date}{Nov.} \bibinfo{year}{2023}), \bibinfo{pages}{102068}.
\newblock
\showISSN{07365853}
\urldef\tempurl%
\url{https://doi.org/10.1016/j.tele.2023.102068}
\showDOI{\tempurl}


\bibitem[Linxen et~al\mbox{.}(2021)]%
        {linxen_how_2021}
\bibfield{author}{\bibinfo{person}{Sebastian Linxen}, \bibinfo{person}{Christian Sturm}, \bibinfo{person}{Florian Brühlmann}, \bibinfo{person}{Vincent Cassau}, \bibinfo{person}{Klaus Opwis}, {and} \bibinfo{person}{Katharina Reinecke}.} \bibinfo{year}{2021}\natexlab{}.
\newblock \showarticletitle{How {WEIRD} is {CHI}?}. In \bibinfo{booktitle}{\emph{Proceedings of the 2021 {CHI} {Conference} on {Human} {Factors} in {Computing} {Systems}}} \emph{(\bibinfo{series}{{CHI} '21})}. \bibinfo{publisher}{Association for Computing Machinery}, \bibinfo{address}{New York, NY, USA}, \bibinfo{pages}{1--14}.
\newblock
\showISBNx{978-1-4503-8096-6}
\urldef\tempurl%
\url{https://doi.org/10.1145/3411764.3445488}
\showDOI{\tempurl}


\bibitem[{Lisa Zieringer} and {Diana Rieger}(2023)]%
        {lisa_zieringer_algorithmic_2023}
\bibfield{author}{\bibinfo{person}{{Lisa Zieringer}} {and} \bibinfo{person}{{Diana Rieger}}.} \bibinfo{year}{2023}\natexlab{}.
\newblock \showarticletitle{Algorithmic {Recommendations}’ {Role} for the {Interrelatedness} of {Counter}-{Messages} and {Polluted} {Content} on {YouTube} – {A} {Network} {Analysis}}.
\newblock \bibinfo{journal}{\emph{Computational Communication Research}} \bibinfo{volume}{5}, \bibinfo{number}{1} (\bibinfo{date}{Jan.} \bibinfo{year}{2023}), \bibinfo{pages}{109}.
\newblock
\showISSN{2665-9085}
\urldef\tempurl%
\url{https://doi.org/10.5117/CCR2023.1.005.ZIER}
\showDOI{\tempurl}


\bibitem[Lurie and Mustafaraj({[n.\,d.]})]%
        {lurie_opening_nodate}
\bibfield{author}{\bibinfo{person}{Emma Lurie} {and} \bibinfo{person}{Eni Mustafaraj}.} \bibinfo{year}{[n.\,d.]}\natexlab{}.
\newblock \showarticletitle{Opening {Up} the {Black} {Box}: {Auditing} {Google}'s {Top} {Stories} {Algorithm}}.
\newblock  (\bibinfo{year}{[n.\,d.]}).
\newblock


\bibitem[Lutz et~al\mbox{.}(2021)]%
        {lutz_examining_2021}
\bibfield{author}{\bibinfo{person}{Michael Lutz}, \bibinfo{person}{Sanjana Gadaginmath}, \bibinfo{person}{Natraj Vairavan}, {and} \bibinfo{person}{Phil Mui}.} \bibinfo{year}{2021}\natexlab{}.
\newblock \showarticletitle{Examining {Political} {Bias} within {YouTube} {Search} and {Recommendation} {Algorithms}}. In \bibinfo{booktitle}{\emph{2021 {IEEE} {Symposium} {Series} on {Computational} {Intelligence} ({SSCI})}}. \bibinfo{publisher}{IEEE}, \bibinfo{address}{Orlando, FL, USA}, \bibinfo{pages}{1--7}.
\newblock
\showISBNx{978-1-72819-048-8}
\urldef\tempurl%
\url{https://doi.org/10.1109/SSCI50451.2021.9660012}
\showDOI{\tempurl}


\bibitem[Ma et~al\mbox{.}(2022)]%
        {ma_enthusiasts_2022}
\bibfield{author}{\bibinfo{person}{Yong Ma}, \bibinfo{person}{Yomna Abdelrahman}, \bibinfo{person}{Barbarella Petz}, \bibinfo{person}{Heiko Drewes}, \bibinfo{person}{Florian Alt}, \bibinfo{person}{Heinrich Hussmann}, {and} \bibinfo{person}{Andreas Butz}.} \bibinfo{year}{2022}\natexlab{}.
\newblock \showarticletitle{Enthusiasts, {Pragmatists}, and {Skeptics}: {Investigating} {Users}’ {Attitudes} {Towards} {Emotion}- and {Personality}-{Aware} {Voice} {Assistants} across {Cultures}}. In \bibinfo{booktitle}{\emph{Proceedings of {Mensch} und {Computer} 2022}} \emph{(\bibinfo{series}{{MuC} '22})}. \bibinfo{publisher}{Association for Computing Machinery}, \bibinfo{address}{New York, NY, USA}, \bibinfo{pages}{308--322}.
\newblock
\showISBNx{978-1-4503-9690-5}
\urldef\tempurl%
\url{https://doi.org/10.1145/3543758.3543776}
\showDOI{\tempurl}


\bibitem[Mahomed et~al\mbox{.}(2024)]%
        {mahomed_auditing_2024}
\bibfield{author}{\bibinfo{person}{Yaaseen Mahomed}, \bibinfo{person}{Charlie~M. Crawford}, \bibinfo{person}{Sanjana Gautam}, \bibinfo{person}{Sorelle~A. Friedler}, {and} \bibinfo{person}{Danaë Metaxa}.} \bibinfo{year}{2024}\natexlab{}.
\newblock \showarticletitle{Auditing {GPT}'s {Content} {Moderation} {Guardrails}: {Can} {ChatGPT} {Write} {Your} {Favorite} {TV} {Show}?}. In \bibinfo{booktitle}{\emph{The 2024 {ACM} {Conference} on {Fairness}, {Accountability}, and {Transparency}}}. \bibinfo{publisher}{ACM}, \bibinfo{address}{Rio de Janeiro Brazil}, \bibinfo{pages}{660--686}.
\newblock
\showISBNx{9798400704505}
\urldef\tempurl%
\url{https://doi.org/10.1145/3630106.3658932}
\showDOI{\tempurl}


\bibitem[Makhortykh et~al\mbox{.}(2022a)]%
        {makhortykh_memory_2022}
\bibfield{author}{\bibinfo{person}{Mykola Makhortykh}, \bibinfo{person}{Aleksandra Urman}, {and} \bibinfo{person}{Roberto Ulloa}.} \bibinfo{year}{2022}\natexlab{a}.
\newblock \showarticletitle{Memory, counter-memory and denialism: {How} search engines circulate information about the {Holodomor}-related memory wars}.
\newblock \bibinfo{journal}{\emph{Memory Studies}} \bibinfo{volume}{15}, \bibinfo{number}{6} (\bibinfo{date}{Dec.} \bibinfo{year}{2022}), \bibinfo{pages}{1330--1345}.
\newblock
\showISSN{1750-6980, 1750-6999}
\urldef\tempurl%
\url{https://doi.org/10.1177/17506980221133732}
\showDOI{\tempurl}


\bibitem[Makhortykh et~al\mbox{.}(2023)]%
        {vakoch_this_2023}
\bibfield{author}{\bibinfo{person}{Mykola Makhortykh}, \bibinfo{person}{Aleksandra Urman}, {and} \bibinfo{person}{Roberto Ulloa}.} \bibinfo{year}{2023}\natexlab{}.
\newblock \showarticletitle{This {Is} {What} {Pandemic} {Looks} {Like}: {Visual} {Framing} of {COVID}-19 on {Search} {Engines}}.
\newblock In \bibinfo{booktitle}{\emph{{COVID} {Communication}}}, \bibfield{editor}{\bibinfo{person}{Douglas~A. Vakoch}, \bibinfo{person}{John~C. Pollock}, {and} \bibinfo{person}{Amanda~M. Caleb}} (Eds.). \bibinfo{publisher}{Springer International Publishing}, \bibinfo{address}{Cham}, \bibinfo{pages}{113--123}.
\newblock
\showISBNx{978-3-031-27664-4 978-3-031-27665-1}
\urldef\tempurl%
\url{https://doi.org/10.1007/978-3-031-27665-1_9}
\showDOI{\tempurl}


\bibitem[Makhortykh et~al\mbox{.}(2022b)]%
        {makhortykh_story_2022}
\bibfield{author}{\bibinfo{person}{Mykola Makhortykh}, \bibinfo{person}{Aleksandra Urman}, {and} \bibinfo{person}{Mariëlle Wijermars}.} \bibinfo{year}{2022}\natexlab{b}.
\newblock \showarticletitle{A story of (non)compliance, bias, and conspiracies: {How} {Google} and {Yandex} represented {Smart} {Voting} during the 2021 parliamentary elections in {Russia}}.
\newblock \bibinfo{journal}{\emph{Harvard Kennedy School Misinformation Review}} (\bibinfo{date}{March} \bibinfo{year}{2022}).
\newblock
\urldef\tempurl%
\url{https://doi.org/10.37016/mr-2020-94}
\showDOI{\tempurl}


\bibitem[Matias(2023)]%
        {matias_influencing_2023}
\bibfield{author}{\bibinfo{person}{J.~Nathan Matias}.} \bibinfo{year}{2023}\natexlab{}.
\newblock \showarticletitle{Influencing recommendation algorithms to reduce the spread of unreliable news by encouraging humans to fact-check articles, in a field experiment}.
\newblock \bibinfo{journal}{\emph{Scientific Reports}} \bibinfo{volume}{13}, \bibinfo{number}{1} (\bibinfo{date}{July} \bibinfo{year}{2023}), \bibinfo{pages}{11715}.
\newblock
\showISSN{2045-2322}
\urldef\tempurl%
\url{https://doi.org/10.1038/s41598-023-38277-5}
\showDOI{\tempurl}


\bibitem[Matias et~al\mbox{.}(2022)]%
        {matias_software-supported_2022}
\bibfield{author}{\bibinfo{person}{J.~Nathan Matias}, \bibinfo{person}{Austin Hounsel}, {and} \bibinfo{person}{Nick Feamster}.} \bibinfo{year}{2022}\natexlab{}.
\newblock \showarticletitle{Software-{Supported} {Audits} of {Decision}-{Making} {Systems}: {Testing} {Google} and {Facebook}'s {Political} {Advertising} {Policies}}.
\newblock \bibinfo{journal}{\emph{Proceedings of the ACM on Human-Computer Interaction}} \bibinfo{volume}{6}, \bibinfo{number}{CSCW1} (\bibinfo{date}{March} \bibinfo{year}{2022}), \bibinfo{pages}{1--19}.
\newblock
\showISSN{2573-0142}
\urldef\tempurl%
\url{https://doi.org/10.1145/3512965}
\showDOI{\tempurl}


\bibitem[Mbalaka(2023)]%
        {mbalaka_epistemically_2023}
\bibfield{author}{\bibinfo{person}{Blessing Mbalaka}.} \bibinfo{year}{2023}\natexlab{}.
\newblock \showarticletitle{Epistemically violent biases in artificial intelligence design: the case of {DALLE}-{E} 2 and {Starry} {AI}}.
\newblock \bibinfo{journal}{\emph{Digital Transformation and Society}} \bibinfo{volume}{2}, \bibinfo{number}{4} (\bibinfo{date}{Oct.} \bibinfo{year}{2023}), \bibinfo{pages}{376--402}.
\newblock
\showISSN{2755-0761, 2755-077X}
\urldef\tempurl%
\url{https://doi.org/10.1108/DTS-01-2023-0003}
\showDOI{\tempurl}


\bibitem[McMahon et~al\mbox{.}(2017)]%
        {mcmahon_substantial_2017}
\bibfield{author}{\bibinfo{person}{Connor McMahon}, \bibinfo{person}{Isaac Johnson}, {and} \bibinfo{person}{Brent Hecht}.} \bibinfo{year}{2017}\natexlab{}.
\newblock \showarticletitle{The {Substantial} {Interdependence} of {Wikipedia} and {Google}: {A} {Case} {Study} on the {Relationship} {Between} {Peer} {Production} {Communities} and {Information} {Technologies}}.
\newblock \bibinfo{journal}{\emph{Proceedings of the International AAAI Conference on Web and Social Media}} \bibinfo{volume}{11}, \bibinfo{number}{1} (\bibinfo{date}{May} \bibinfo{year}{2017}), \bibinfo{pages}{142--151}.
\newblock
\showISSN{2334-0770, 2162-3449}
\urldef\tempurl%
\url{https://doi.org/10.1609/icwsm.v11i1.14883}
\showDOI{\tempurl}


\bibitem[Metaxa et~al\mbox{.}(2021a)]%
        {metaxa_image_2021}
\bibfield{author}{\bibinfo{person}{Danaë Metaxa}, \bibinfo{person}{Michelle~A. Gan}, \bibinfo{person}{Su Goh}, \bibinfo{person}{Jeff Hancock}, {and} \bibinfo{person}{James~A. Landay}.} \bibinfo{year}{2021}\natexlab{a}.
\newblock \showarticletitle{An {Image} of {Society}: {Gender} and {Racial} {Representation} and {Impact} in {Image} {Search} {Results} for {Occupations}}.
\newblock \bibinfo{journal}{\emph{Proceedings of the ACM on Human-Computer Interaction}} \bibinfo{volume}{5}, \bibinfo{number}{CSCW1} (\bibinfo{date}{April} \bibinfo{year}{2021}), \bibinfo{pages}{1--23}.
\newblock
\showISSN{2573-0142}
\urldef\tempurl%
\url{https://doi.org/10.1145/3449100}
\showDOI{\tempurl}


\bibitem[Metaxa et~al\mbox{.}(2019)]%
        {metaxa_search_2019}
\bibfield{author}{\bibinfo{person}{Danaë Metaxa}, \bibinfo{person}{Joon~Sung Park}, \bibinfo{person}{James~A. Landay}, {and} \bibinfo{person}{Jeff Hancock}.} \bibinfo{year}{2019}\natexlab{}.
\newblock \showarticletitle{Search {Media} and {Elections}: {A} {Longitudinal} {Investigation} of {Political} {Search} {Results}}.
\newblock \bibinfo{journal}{\emph{Proceedings of the ACM on Human-Computer Interaction}} \bibinfo{volume}{3}, \bibinfo{number}{CSCW} (\bibinfo{date}{Nov.} \bibinfo{year}{2019}), \bibinfo{pages}{1--17}.
\newblock
\showISSN{2573-0142}
\urldef\tempurl%
\url{https://doi.org/10.1145/3359231}
\showDOI{\tempurl}


\bibitem[Metaxa et~al\mbox{.}(2021b)]%
        {metaxa_auditing_2021}
\bibfield{author}{\bibinfo{person}{Danaë Metaxa}, \bibinfo{person}{Joon~Sung Park}, \bibinfo{person}{Ronald~E. Robertson}, \bibinfo{person}{Karrie Karahalios}, \bibinfo{person}{Christo Wilson}, \bibinfo{person}{Jeff Hancock}, {and} \bibinfo{person}{Christian Sandvig}.} \bibinfo{year}{2021}\natexlab{b}.
\newblock \showarticletitle{Auditing {Algorithms}: {Understanding} {Algorithmic} {Systems} from the {Outside} {In}}.
\newblock \bibinfo{journal}{\emph{Foundations and Trends® in Human–Computer Interaction}} \bibinfo{volume}{14}, \bibinfo{number}{4} (\bibinfo{year}{2021}), \bibinfo{pages}{272--344}.
\newblock
\showISSN{1551-3955, 1551-3963}
\urldef\tempurl%
\url{https://doi.org/10.1561/1100000083}
\showDOI{\tempurl}


\bibitem[Mikians et~al\mbox{.}(2012)]%
        {mikians_detecting_2012}
\bibfield{author}{\bibinfo{person}{Jakub Mikians}, \bibinfo{person}{László Gyarmati}, \bibinfo{person}{Vijay Erramilli}, {and} \bibinfo{person}{Nikolaos Laoutaris}.} \bibinfo{year}{2012}\natexlab{}.
\newblock \showarticletitle{Detecting price and search discrimination on the internet}. In \bibinfo{booktitle}{\emph{Proceedings of the 11th {ACM} {Workshop} on {Hot} {Topics} in {Networks}}}. \bibinfo{publisher}{ACM}, \bibinfo{address}{Redmond Washington}, \bibinfo{pages}{79--84}.
\newblock
\showISBNx{978-1-4503-1776-4}
\urldef\tempurl%
\url{https://doi.org/10.1145/2390231.2390245}
\showDOI{\tempurl}


\bibitem[Mikians et~al\mbox{.}(2013)]%
        {mikians_crowd-assisted_2013}
\bibfield{author}{\bibinfo{person}{Jakub Mikians}, \bibinfo{person}{László Gyarmati}, \bibinfo{person}{Vijay Erramilli}, {and} \bibinfo{person}{Nikolaos Laoutaris}.} \bibinfo{year}{2013}\natexlab{}.
\newblock \showarticletitle{Crowd-assisted search for price discrimination in e-commerce: first results}. In \bibinfo{booktitle}{\emph{Proceedings of the ninth {ACM} conference on {Emerging} networking experiments and technologies}}. \bibinfo{publisher}{ACM}, \bibinfo{address}{Santa Barbara California USA}, \bibinfo{pages}{1--6}.
\newblock
\showISBNx{978-1-4503-2101-3}
\urldef\tempurl%
\url{https://doi.org/10.1145/2535372.2535415}
\showDOI{\tempurl}


\bibitem[Moe(2019)]%
        {moe_comparing_2019}
\bibfield{author}{\bibinfo{person}{Hallvard Moe}.} \bibinfo{year}{2019}\natexlab{}.
\newblock \showarticletitle{Comparing {Platform} “{Ranking} {Cultures}” {Across} {Languages}: {The} {Case} of {Islam} on {YouTube} in {Scandinavia}}.
\newblock \bibinfo{journal}{\emph{Social Media + Society}} \bibinfo{volume}{5}, \bibinfo{number}{1} (\bibinfo{date}{Jan.} \bibinfo{year}{2019}), \bibinfo{pages}{205630511881703}.
\newblock
\showISSN{2056-3051, 2056-3051}
\urldef\tempurl%
\url{https://doi.org/10.1177/2056305118817038}
\showDOI{\tempurl}


\bibitem[Moher et~al\mbox{.}(2009)]%
        {moher_preferred_2009}
\bibfield{author}{\bibinfo{person}{David Moher}, \bibinfo{person}{Alessandro Liberati}, \bibinfo{person}{Jennifer Tetzlaff}, {and} \bibinfo{person}{Douglas~G. Altman}.} \bibinfo{year}{2009}\natexlab{}.
\newblock \showarticletitle{Preferred {Reporting} {Items} for {Systematic} {Reviews} and {Meta}-{Analyses}: {The} {PRISMA} {Statement}}.
\newblock \bibinfo{journal}{\emph{Annals of Internal Medicine}} \bibinfo{volume}{151}, \bibinfo{number}{4} (\bibinfo{date}{Aug.} \bibinfo{year}{2009}), \bibinfo{pages}{264--269}.
\newblock
\showISSN{0003-4819}
\urldef\tempurl%
\url{https://doi.org/10.7326/0003-4819-151-4-200908180-00135}
\showDOI{\tempurl}
\newblock
\shownote{Publisher: American College of Physicians}.


\bibitem[Monk(2023)]%
        {monk_monk_2023}
\bibfield{author}{\bibinfo{person}{Ellis Monk}.} \bibinfo{year}{2023}\natexlab{}.
\newblock \bibinfo{title}{The {Monk} {Skin} {Tone} {Scale}}.
\newblock
\newblock
\urldef\tempurl%
\url{https://doi.org/10.31235/osf.io/pdf4c}
\showDOI{\tempurl}


\bibitem[Murthy(2021)]%
        {murthy_evaluating_2021}
\bibfield{author}{\bibinfo{person}{Dhiraj Murthy}.} \bibinfo{year}{2021}\natexlab{}.
\newblock \showarticletitle{Evaluating {Platform} {Accountability}: {Terrorist} {Content} on {YouTube}}.
\newblock \bibinfo{journal}{\emph{American Behavioral Scientist}} \bibinfo{volume}{65}, \bibinfo{number}{6} (\bibinfo{date}{May} \bibinfo{year}{2021}), \bibinfo{pages}{800--824}.
\newblock
\showISSN{0002-7642, 1552-3381}
\urldef\tempurl%
\url{https://doi.org/10.1177/0002764221989774}
\showDOI{\tempurl}


\bibitem[Mustafaraj et~al\mbox{.}(2020)]%
        {mustafaraj_case_2020}
\bibfield{author}{\bibinfo{person}{Eni Mustafaraj}, \bibinfo{person}{Emma Lurie}, {and} \bibinfo{person}{Claire Devine}.} \bibinfo{year}{2020}\natexlab{}.
\newblock \showarticletitle{The case for voter-centered audits of search engines during political elections}. In \bibinfo{booktitle}{\emph{Proceedings of the 2020 {Conference} on {Fairness}, {Accountability}, and {Transparency}}}. \bibinfo{publisher}{ACM}, \bibinfo{address}{Barcelona Spain}, \bibinfo{pages}{559--569}.
\newblock
\showISBNx{978-1-4503-6936-7}
\urldef\tempurl%
\url{https://doi.org/10.1145/3351095.3372835}
\showDOI{\tempurl}


\bibitem[Mähler and Vonderau(2017)]%
        {mahler_studying_2017}
\bibfield{author}{\bibinfo{person}{Roger Mähler} {and} \bibinfo{person}{Patrick Vonderau}.} \bibinfo{year}{2017}\natexlab{}.
\newblock \showarticletitle{Studying {Ad} {Targeting} with {Digital} {Methods}: {The} {Case} of {Spotify}}.
\newblock \bibinfo{journal}{\emph{Culture Unbound}} \bibinfo{volume}{9}, \bibinfo{number}{2} (\bibinfo{date}{Oct.} \bibinfo{year}{2017}), \bibinfo{pages}{212--221}.
\newblock
\showISSN{2000-1525}
\urldef\tempurl%
\url{https://doi.org/10.3384/cu.2000.1525.1792212}
\showDOI{\tempurl}


\bibitem[Nechushtai and Lewis(2019)]%
        {nechushtai_what_2019}
\bibfield{author}{\bibinfo{person}{Efrat Nechushtai} {and} \bibinfo{person}{Seth~C. Lewis}.} \bibinfo{year}{2019}\natexlab{}.
\newblock \showarticletitle{What kind of news gatekeepers do we want machines to be? {Filter} bubbles, fragmentation, and the normative dimensions of algorithmic recommendations}.
\newblock \bibinfo{journal}{\emph{Computers in Human Behavior}}  \bibinfo{volume}{90} (\bibinfo{date}{Jan.} \bibinfo{year}{2019}), \bibinfo{pages}{298--307}.
\newblock
\showISSN{07475632}
\urldef\tempurl%
\url{https://doi.org/10.1016/j.chb.2018.07.043}
\showDOI{\tempurl}


\bibitem[Nechushtai et~al\mbox{.}(2023)]%
        {nechushtai_more_2023}
\bibfield{author}{\bibinfo{person}{Efrat Nechushtai}, \bibinfo{person}{Rodrigo Zamith}, {and} \bibinfo{person}{Seth~C. Lewis}.} \bibinfo{year}{2023}\natexlab{}.
\newblock \showarticletitle{More of the {Same}? {Homogenization} in {News} {Recommendations} {When} {Users} {Search} on {Google}, {YouTube}, {Facebook}, and {Twitter}}.
\newblock \bibinfo{journal}{\emph{Mass Communication and Society}} (\bibinfo{date}{March} \bibinfo{year}{2023}), \bibinfo{pages}{1--27}.
\newblock
\showISSN{1520-5436, 1532-7825}
\urldef\tempurl%
\url{https://doi.org/10.1080/15205436.2023.2173609}
\showDOI{\tempurl}


\bibitem[Neumann et~al\mbox{.}(2024)]%
        {neumann_data_2024}
\bibfield{author}{\bibinfo{person}{Nico Neumann}, \bibinfo{person}{Catherine~E. Tucker}, \bibinfo{person}{Levi Kaplan}, \bibinfo{person}{Alan Mislove}, {and} \bibinfo{person}{Piotr Sapiezynski}.} \bibinfo{year}{2024}\natexlab{}.
\newblock \showarticletitle{Data {Deserts} and {Black} {Boxes}: {The} {Impact} of {Socio}-{Economic} {Status} on {Consumer} {Profiling}}.
\newblock \bibinfo{journal}{\emph{Management Science}} \bibinfo{volume}{70}, \bibinfo{number}{11} (\bibinfo{date}{Nov.} \bibinfo{year}{2024}), \bibinfo{pages}{8003--8029}.
\newblock
\showISSN{0025-1909, 1526-5501}
\urldef\tempurl%
\url{https://doi.org/10.1287/mnsc.2023.4979}
\showDOI{\tempurl}


\bibitem[Ng et~al\mbox{.}(2023)]%
        {ng_exploring_2023}
\bibfield{author}{\bibinfo{person}{Yee Man~Margaret Ng}, \bibinfo{person}{Katherine Hoffmann~Pham}, {and} \bibinfo{person}{Miguel Luengo-Oroz}.} \bibinfo{year}{2023}\natexlab{}.
\newblock \showarticletitle{Exploring {YouTube}’s {Recommendation} {System} in the {Context} of {COVID}-19 {Vaccines}: {Computational} and {Comparative} {Analysis} of {Video} {Trajectories}}.
\newblock \bibinfo{journal}{\emph{Journal of Medical Internet Research}}  \bibinfo{volume}{25} (\bibinfo{date}{Sept.} \bibinfo{year}{2023}), \bibinfo{pages}{e49061}.
\newblock
\showISSN{1438-8871}
\urldef\tempurl%
\url{https://doi.org/10.2196/49061}
\showDOI{\tempurl}


\bibitem[Nigatu and Raji(2024)]%
        {nigatu_i_2024}
\bibfield{author}{\bibinfo{person}{Hellina~Hailu Nigatu} {and} \bibinfo{person}{Inioluwa~Deborah Raji}.} \bibinfo{year}{2024}\natexlab{}.
\newblock \showarticletitle{“{I} {Searched} for a {Religious} {Song} in {Amharic} and {Got} {Sexual} {Content} {Instead}’’: {Investigating} {Online} {Harm} in {Low}-{Resourced} {Languages} on {YouTube}.}. In \bibinfo{booktitle}{\emph{The 2024 {ACM} {Conference} on {Fairness}, {Accountability}, and {Transparency}}}. \bibinfo{publisher}{ACM}, \bibinfo{address}{Rio de Janeiro Brazil}, \bibinfo{pages}{141--160}.
\newblock
\showISBNx{9798400704505}
\urldef\tempurl%
\url{https://doi.org/10.1145/3630106.3658546}
\showDOI{\tempurl}


\bibitem[Noble(2013)]%
        {noble_google_2013}
\bibfield{author}{\bibinfo{person}{Safiya Noble}.} \bibinfo{year}{2013}\natexlab{}.
\newblock \showarticletitle{Google {Search}: {Hyper}-visibility as a {Means} of {Rendering} {Black} {Women} and {Girls} {Invisible}}.
\newblock  (\bibinfo{date}{Oct.} \bibinfo{year}{2013}).
\newblock
\urldef\tempurl%
\url{https://urresearch.rochester.edu/institutionalPublicationPublicView.action?institutionalItemId=27584}
\showURL{%
\tempurl}


\bibitem[Onyepunuka et~al\mbox{.}({[n.\,d.]})]%
        {onyepunuka_multilingual_nodate}
\bibfield{author}{\bibinfo{person}{Ugochukwu Onyepunuka}, \bibinfo{person}{Mustafa Alassad}, \bibinfo{person}{Lotenna Nwana}, {and} \bibinfo{person}{Nitin Agarwal}.} \bibinfo{year}{[n.\,d.]}\natexlab{}.
\newblock \showarticletitle{Multilingual {Analysis} of {YouTube}’s {Recommendation} {System}: {Examining} {Topic} and {Emotion} {Drift} in the ‘{Cheng} {Ho}’ {Narrative}}.
\newblock  (\bibinfo{year}{[n.\,d.]}).
\newblock


\bibitem[Ovalle et~al\mbox{.}(2023)]%
        {ovalle_im_2023}
\bibfield{author}{\bibinfo{person}{Anaelia Ovalle}, \bibinfo{person}{Palash Goyal}, \bibinfo{person}{Jwala Dhamala}, \bibinfo{person}{Zachary Jaggers}, \bibinfo{person}{Kai-Wei Chang}, \bibinfo{person}{Aram Galstyan}, \bibinfo{person}{Richard Zemel}, {and} \bibinfo{person}{Rahul Gupta}.} \bibinfo{year}{2023}\natexlab{}.
\newblock \showarticletitle{“{I}’m fully who {I} am”: {Towards} {Centering} {Transgender} and {Non}-{Binary} {Voices} to {Measure} {Biases} in {Open} {Language} {Generation}}. In \bibinfo{booktitle}{\emph{2023 {ACM} {Conference} on {Fairness}, {Accountability}, and {Transparency}}}. \bibinfo{publisher}{ACM}, \bibinfo{address}{Chicago IL USA}, \bibinfo{pages}{1246--1266}.
\newblock
\showISBNx{9798400701924}
\urldef\tempurl%
\url{https://doi.org/10.1145/3593013.3594078}
\showDOI{\tempurl}


\bibitem[Papakyriakopoulos and Mboya(2023)]%
        {papakyriakopoulos_beyond_2023}
\bibfield{author}{\bibinfo{person}{Orestis Papakyriakopoulos} {and} \bibinfo{person}{Arwa~M. Mboya}.} \bibinfo{year}{2023}\natexlab{}.
\newblock \showarticletitle{Beyond {Algorithmic} {Bias}: {A} {Socio}-{Computational} {Interrogation} of the {Google} {Search} by {Image} {Algorithm}}.
\newblock \bibinfo{journal}{\emph{Social Science Computer Review}} \bibinfo{volume}{41}, \bibinfo{number}{4} (\bibinfo{date}{Aug.} \bibinfo{year}{2023}), \bibinfo{pages}{1100--1125}.
\newblock
\showISSN{0894-4393, 1552-8286}
\urldef\tempurl%
\url{https://doi.org/10.1177/08944393211073169}
\showDOI{\tempurl}


\bibitem[Patel et~al\mbox{.}(2023)]%
        {patel_assessing_2023}
\bibfield{author}{\bibinfo{person}{Vraj Patel}, \bibinfo{person}{Mason Lovett}, \bibinfo{person}{Ryan Rybarczyk}, {and} \bibinfo{person}{John Hertig}.} \bibinfo{year}{2023}\natexlab{}.
\newblock \showarticletitle{Assessing {Trustworthiness} of {Internet} {Pharmacies} with an {Internet} {Browser} {Extension}}.
\newblock \bibinfo{journal}{\emph{The Journal of Medicine Access}}  \bibinfo{volume}{7} (\bibinfo{date}{Jan.} \bibinfo{year}{2023}), \bibinfo{pages}{27550834231191895}.
\newblock
\showISSN{2755-0834, 2755-0834}
\urldef\tempurl%
\url{https://doi.org/10.1177/27550834231191895}
\showDOI{\tempurl}


\bibitem[Perreault et~al\mbox{.}(2024)]%
        {perreault_algorithmic_2024}
\bibfield{author}{\bibinfo{person}{Brooke Perreault}, \bibinfo{person}{Johanna~Hoonsun Lee}, \bibinfo{person}{Ropafadzo Shava}, {and} \bibinfo{person}{Eni Mustafaraj}.} \bibinfo{year}{2024}\natexlab{}.
\newblock \showarticletitle{Algorithmic {Misjudgement} in {Google} {Search} {Results}: {Evidence} from {Auditing} the {US} {Online} {Electoral} {Information} {Environment}}. In \bibinfo{booktitle}{\emph{The 2024 {ACM} {Conference} on {Fairness}, {Accountability}, and {Transparency}}}. \bibinfo{publisher}{ACM}, \bibinfo{address}{Rio de Janeiro Brazil}, \bibinfo{pages}{433--443}.
\newblock
\showISBNx{9798400704505}
\urldef\tempurl%
\url{https://doi.org/10.1145/3630106.3658916}
\showDOI{\tempurl}


\bibitem[Puschmann(2019)]%
        {puschmann_beyond_2019}
\bibfield{author}{\bibinfo{person}{Cornelius Puschmann}.} \bibinfo{year}{2019}\natexlab{}.
\newblock \showarticletitle{Beyond the {Bubble}: {Assessing} the {Diversity} of {Political} {Search} {Results}}.
\newblock \bibinfo{journal}{\emph{Digital Journalism}} \bibinfo{volume}{7}, \bibinfo{number}{6} (\bibinfo{date}{July} \bibinfo{year}{2019}), \bibinfo{pages}{824--843}.
\newblock
\showISSN{2167-0811, 2167-082X}
\urldef\tempurl%
\url{https://doi.org/10.1080/21670811.2018.1539626}
\showDOI{\tempurl}


\bibitem[Radesky et~al\mbox{.}(2024)]%
        {radesky_algorithmic_2024}
\bibfield{author}{\bibinfo{person}{Jenny Radesky}, \bibinfo{person}{Enrica Bridgewater}, \bibinfo{person}{Shira Black}, \bibinfo{person}{August O’Neil}, \bibinfo{person}{Yilin Sun}, \bibinfo{person}{Alexandria Schaller}, \bibinfo{person}{Heidi~M. Weeks}, {and} \bibinfo{person}{Scott~W. Campbell}.} \bibinfo{year}{2024}\natexlab{}.
\newblock \showarticletitle{Algorithmic {Content} {Recommendations} on a {Video}-{Sharing} {Platform} {Used} by {Children}}.
\newblock \bibinfo{journal}{\emph{JAMA Network Open}} \bibinfo{volume}{7}, \bibinfo{number}{5} (\bibinfo{date}{May} \bibinfo{year}{2024}), \bibinfo{pages}{e2413855}.
\newblock
\showISSN{2574-3805}
\urldef\tempurl%
\url{https://doi.org/10.1001/jamanetworkopen.2024.13855}
\showDOI{\tempurl}


\bibitem[Raji and Buolamwini(2019)]%
        {raji_actionable_2019}
\bibfield{author}{\bibinfo{person}{Inioluwa~Deborah Raji} {and} \bibinfo{person}{Joy Buolamwini}.} \bibinfo{year}{2019}\natexlab{}.
\newblock \showarticletitle{Actionable {Auditing}: {Investigating} the {Impact} of {Publicly} {Naming} {Biased} {Performance} {Results} of {Commercial} {AI} {Products}}. In \bibinfo{booktitle}{\emph{Proceedings of the 2019 {AAAI}/{ACM} {Conference} on {AI}, {Ethics}, and {Society}}} \emph{(\bibinfo{series}{{AIES} '19})}. \bibinfo{publisher}{Association for Computing Machinery}, \bibinfo{address}{New York, NY, USA}, \bibinfo{pages}{429--435}.
\newblock
\showISBNx{978-1-4503-6324-2}
\urldef\tempurl%
\url{https://doi.org/10.1145/3306618.3314244}
\showDOI{\tempurl}


\bibitem[Reber et~al\mbox{.}(2020)]%
        {reber_data_2020}
\bibfield{author}{\bibinfo{person}{Martin Reber}, \bibinfo{person}{Tobias~D. Krafft}, \bibinfo{person}{Roman Krafft}, \bibinfo{person}{Katharina~A. Zweig}, {and} \bibinfo{person}{Anna Couturier}.} \bibinfo{year}{2020}\natexlab{}.
\newblock \showarticletitle{Data {Donations} for {Mapping} {Risk} in {Google} {Search} of {Health} {Queries}: {A} case study of unproven stem cell treatments in {SEM}}. In \bibinfo{booktitle}{\emph{2020 {IEEE} {Symposium} {Series} on {Computational} {Intelligence} ({SSCI})}}. \bibinfo{publisher}{IEEE}, \bibinfo{address}{Canberra, ACT, Australia}, \bibinfo{pages}{2985--2992}.
\newblock
\showISBNx{978-1-72812-547-3}
\urldef\tempurl%
\url{https://doi.org/10.1109/SSCI47803.2020.9308420}
\showDOI{\tempurl}


\bibitem[Ribeiro et~al\mbox{.}(2020)]%
        {ribeiro_auditing_2020}
\bibfield{author}{\bibinfo{person}{Manoel~Horta Ribeiro}, \bibinfo{person}{Raphael Ottoni}, \bibinfo{person}{Robert West}, \bibinfo{person}{Virgílio A.~F. Almeida}, {and} \bibinfo{person}{Wagner Meira}.} \bibinfo{year}{2020}\natexlab{}.
\newblock \showarticletitle{Auditing radicalization pathways on {YouTube}}. In \bibinfo{booktitle}{\emph{Proceedings of the 2020 {Conference} on {Fairness}, {Accountability}, and {Transparency}}}. \bibinfo{publisher}{ACM}, \bibinfo{address}{Barcelona Spain}, \bibinfo{pages}{131--141}.
\newblock
\showISBNx{978-1-4503-6936-7}
\urldef\tempurl%
\url{https://doi.org/10.1145/3351095.3372879}
\showDOI{\tempurl}


\bibitem[Ridgway(2024)]%
        {ridgway_screenshotting_2024}
\bibfield{author}{\bibinfo{person}{Renée Ridgway}.} \bibinfo{year}{2024}\natexlab{}.
\newblock \showarticletitle{Screenshotting partial perspectives: {The} case of {\textless}span style="font-variant:small-caps;"{\textgreater}{Danish}{\textless}/span{\textgreater} mink in {\textless}span style="font-variant:small-caps;"{\textgreater}{Google}{\textless}/span{\textgreater} search results}.
\newblock \bibinfo{journal}{\emph{Journal of the Association for Information Science and Technology}} \bibinfo{volume}{75}, \bibinfo{number}{10} (\bibinfo{date}{Oct.} \bibinfo{year}{2024}), \bibinfo{pages}{1104--1118}.
\newblock
\showISSN{2330-1635, 2330-1643}
\urldef\tempurl%
\url{https://doi.org/10.1002/asi.24892}
\showDOI{\tempurl}


\bibitem[Rieder et~al\mbox{.}(2018)]%
        {rieder_ranking_2018}
\bibfield{author}{\bibinfo{person}{Bernhard Rieder}, \bibinfo{person}{Ariadna Matamoros-Fernández}, {and} \bibinfo{person}{Òscar Coromina}.} \bibinfo{year}{2018}\natexlab{}.
\newblock \showarticletitle{From ranking algorithms to ‘ranking cultures’: {Investigating} the modulation of visibility in {YouTube} search results}.
\newblock \bibinfo{journal}{\emph{Convergence: The International Journal of Research into New Media Technologies}} \bibinfo{volume}{24}, \bibinfo{number}{1} (\bibinfo{date}{Feb.} \bibinfo{year}{2018}), \bibinfo{pages}{50--68}.
\newblock
\showISSN{1354-8565, 1748-7382}
\urldef\tempurl%
\url{https://doi.org/10.1177/1354856517736982}
\showDOI{\tempurl}


\bibitem[Robertson et~al\mbox{.}(2018a)]%
        {robertson_auditing_2018}
\bibfield{author}{\bibinfo{person}{Ronald~E. Robertson}, \bibinfo{person}{Shan Jiang}, \bibinfo{person}{Kenneth Joseph}, \bibinfo{person}{Lisa Friedland}, \bibinfo{person}{David Lazer}, {and} \bibinfo{person}{Christo Wilson}.} \bibinfo{year}{2018}\natexlab{a}.
\newblock \showarticletitle{Auditing {Partisan} {Audience} {Bias} within {Google} {Search}}.
\newblock \bibinfo{journal}{\emph{Proceedings of the ACM on Human-Computer Interaction}} \bibinfo{volume}{2}, \bibinfo{number}{CSCW} (\bibinfo{date}{Nov.} \bibinfo{year}{2018}), \bibinfo{pages}{1--22}.
\newblock
\showISSN{2573-0142}
\urldef\tempurl%
\url{https://doi.org/10.1145/3274417}
\showDOI{\tempurl}


\bibitem[Robertson et~al\mbox{.}(2019)]%
        {robertson_auditing_2019}
\bibfield{author}{\bibinfo{person}{Ronald~E. Robertson}, \bibinfo{person}{Shan Jiang}, \bibinfo{person}{David Lazer}, {and} \bibinfo{person}{Christo Wilson}.} \bibinfo{year}{2019}\natexlab{}.
\newblock \showarticletitle{Auditing {Autocomplete}: {Suggestion} {Networks} and {Recursive} {Algorithm} {Interrogation}}. In \bibinfo{booktitle}{\emph{Proceedings of the 10th {ACM} {Conference} on {Web} {Science}}}. \bibinfo{publisher}{ACM}, \bibinfo{address}{Boston Massachusetts USA}, \bibinfo{pages}{235--244}.
\newblock
\showISBNx{978-1-4503-6202-3}
\urldef\tempurl%
\url{https://doi.org/10.1145/3292522.3326047}
\showDOI{\tempurl}


\bibitem[Robertson et~al\mbox{.}(2018b)]%
        {robertson_auditing_2018-1}
\bibfield{author}{\bibinfo{person}{Ronald~E. Robertson}, \bibinfo{person}{David Lazer}, {and} \bibinfo{person}{Christo Wilson}.} \bibinfo{year}{2018}\natexlab{b}.
\newblock \showarticletitle{Auditing the {Personalization} and {Composition} of {Politically}-{Related} {Search} {Engine} {Results} {Pages}}. In \bibinfo{booktitle}{\emph{Proceedings of the 2018 {World} {Wide} {Web} {Conference} on {World} {Wide} {Web} - {WWW} '18}}. \bibinfo{publisher}{ACM Press}, \bibinfo{address}{Lyon, France}, \bibinfo{pages}{955--965}.
\newblock
\showISBNx{978-1-4503-5639-8}
\urldef\tempurl%
\url{https://doi.org/10.1145/3178876.3186143}
\showDOI{\tempurl}


\bibitem[Rowland et~al\mbox{.}(2023)]%
        {rowland_shaping_2023}
\bibfield{author}{\bibinfo{person}{Jussara Rowland}, \bibinfo{person}{Sergi López‐Asensio}, \bibinfo{person}{Ataberk Bagci}, \bibinfo{person}{Ana Delicado}, {and} \bibinfo{person}{Ana Prades}.} \bibinfo{year}{2023}\natexlab{}.
\newblock \showarticletitle{Shaping information and knowledge on climate change technologies: {A} cross‐country qualitative analysis of carbon capture and storage results on {Google} search}.
\newblock \bibinfo{journal}{\emph{Journal of the Association for Information Science and Technology}} (\bibinfo{date}{Sept.} \bibinfo{year}{2023}), \bibinfo{pages}{asi.24828}.
\newblock
\showISSN{2330-1635, 2330-1643}
\urldef\tempurl%
\url{https://doi.org/10.1002/asi.24828}
\showDOI{\tempurl}


\bibitem[Sambasivan et~al\mbox{.}(2021)]%
        {sambasivan_re-imagining_2021}
\bibfield{author}{\bibinfo{person}{Nithya Sambasivan}, \bibinfo{person}{Erin Arnesen}, \bibinfo{person}{Ben Hutchinson}, \bibinfo{person}{Tulsee Doshi}, {and} \bibinfo{person}{Vinodkumar Prabhakaran}.} \bibinfo{year}{2021}\natexlab{}.
\newblock \showarticletitle{Re-imagining {Algorithmic} {Fairness} in {India} and {Beyond}}. In \bibinfo{booktitle}{\emph{Proceedings of the 2021 {ACM} {Conference} on {Fairness}, {Accountability}, and {Transparency}}}. \bibinfo{publisher}{ACM}, \bibinfo{address}{Virtual Event Canada}, \bibinfo{pages}{315--328}.
\newblock
\showISBNx{978-1-4503-8309-7}
\urldef\tempurl%
\url{https://doi.org/10.1145/3442188.3445896}
\showDOI{\tempurl}


\bibitem[Samuel-Azran et~al\mbox{.}(2024)]%
        {samuel-azran_analyzing_2024}
\bibfield{author}{\bibinfo{person}{Tal Samuel-Azran}, \bibinfo{person}{Ilan Manor}, \bibinfo{person}{Evyatar Yitzhak}, {and} \bibinfo{person}{Yair Galily}.} \bibinfo{year}{2024}\natexlab{}.
\newblock \showarticletitle{Analyzing {AI} {Bias}: {The} {Discourse} of {Terror} and {Sport} {Ahead} of {Paris} 2024 {Olympics}}.
\newblock \bibinfo{journal}{\emph{American Behavioral Scientist}} (\bibinfo{date}{July} \bibinfo{year}{2024}), \bibinfo{pages}{00027642241261265}.
\newblock
\showISSN{0002-7642, 1552-3381}
\urldef\tempurl%
\url{https://doi.org/10.1177/00027642241261265}
\showDOI{\tempurl}


\bibitem[Sandvig et~al\mbox{.}(2014)]%
        {sandvig_auditing_nodate}
\bibfield{author}{\bibinfo{person}{Christian Sandvig}, \bibinfo{person}{Kevin Hamilton}, \bibinfo{person}{Karrie Karahalios}, {and} \bibinfo{person}{Cedric Langbort}.} \bibinfo{year}{2014}\natexlab{}.
\newblock \showarticletitle{Auditing {Algorithms}: {Research} {Methods} for {Detecting} {Discrimination} on {Internet} {Platforms}}.
\newblock  (\bibinfo{year}{2014}).
\newblock


\bibitem[Sanna et~al\mbox{.}(2021)]%
        {lossio-ventura_yttrex_2021}
\bibfield{author}{\bibinfo{person}{Leonardo Sanna}, \bibinfo{person}{Salvatore Romano}, \bibinfo{person}{Giulia Corona}, {and} \bibinfo{person}{Claudio Agosti}.} \bibinfo{year}{2021}\natexlab{}.
\newblock \showarticletitle{{YTTREX}: {Crowdsourced} {Analysis} of {YouTube}’s {Recommender} {System} {During} {COVID}-19 {Pandemic}}.
\newblock In \bibinfo{booktitle}{\emph{Information {Management} and {Big} {Data}}}, \bibfield{editor}{\bibinfo{person}{Juan~Antonio Lossio-Ventura}, \bibinfo{person}{Jorge~Carlos Valverde-Rebaza}, \bibinfo{person}{Eduardo Díaz}, {and} \bibinfo{person}{Hugo Alatrista-Salas}} (Eds.). Vol.~\bibinfo{volume}{1410}. \bibinfo{publisher}{Springer International Publishing}, \bibinfo{address}{Cham}, \bibinfo{pages}{107--121}.
\newblock
\showISBNx{978-3-030-76227-8 978-3-030-76228-5}
\urldef\tempurl%
\url{https://doi.org/10.1007/978-3-030-76228-5_8}
\showDOI{\tempurl}
\newblock
\shownote{Series Title: Communications in Computer and Information Science}.


\bibitem[Santini et~al\mbox{.}(2023)]%
        {santini_recommending_2023}
\bibfield{author}{\bibinfo{person}{Rose~Marie Santini}, \bibinfo{person}{Débora Salles}, {and} \bibinfo{person}{Bruno Mattos}.} \bibinfo{year}{2023}\natexlab{}.
\newblock \showarticletitle{Recommending instead of taking down: {YouTube} hyperpartisan content promotion amid the {Brazilian} general elections}.
\newblock \bibinfo{journal}{\emph{Policy \& Internet}} \bibinfo{volume}{15}, \bibinfo{number}{4} (\bibinfo{date}{Dec.} \bibinfo{year}{2023}), \bibinfo{pages}{512--527}.
\newblock
\showISSN{1944-2866, 1944-2866}
\urldef\tempurl%
\url{https://doi.org/10.1002/poi3.380}
\showDOI{\tempurl}


\bibitem[Sapiezynski et~al\mbox{.}(2024)]%
        {sapiezynski_use_2024}
\bibfield{author}{\bibinfo{person}{Piotr Sapiezynski}, \bibinfo{person}{Levi Kaplan}, \bibinfo{person}{Alan Mislove}, {and} \bibinfo{person}{Aleksandra Korolova}.} \bibinfo{year}{2024}\natexlab{}.
\newblock \showarticletitle{On the {Use} of {Proxies} in {Political} {Ad} {Targeting}}.
\newblock \bibinfo{journal}{\emph{Proceedings of the ACM on Human-Computer Interaction}} \bibinfo{volume}{8}, \bibinfo{number}{CSCW2} (\bibinfo{date}{Nov.} \bibinfo{year}{2024}), \bibinfo{pages}{1--31}.
\newblock
\showISSN{2573-0142}
\urldef\tempurl%
\url{https://doi.org/10.1145/3686917}
\showDOI{\tempurl}


\bibitem[Schellingerhout et~al\mbox{.}(2023)]%
        {schellingerhout_accounting_2023}
\bibfield{author}{\bibinfo{person}{Roan Schellingerhout}, \bibinfo{person}{Davide Beraldo}, {and} \bibinfo{person}{Maarten Marx}.} \bibinfo{year}{2023}\natexlab{}.
\newblock \showarticletitle{Accounting for {Personalization} in {Personalization} {Algorithms}: {YouTube}’s {Treatment} of {Conspiracy} {Content}}.
\newblock \bibinfo{journal}{\emph{Digital Journalism}} (\bibinfo{date}{May} \bibinfo{year}{2023}), \bibinfo{pages}{1--29}.
\newblock
\showISSN{2167-0811, 2167-082X}
\urldef\tempurl%
\url{https://doi.org/10.1080/21670811.2023.2209153}
\showDOI{\tempurl}


\bibitem[Septiandri et~al\mbox{.}(2023)]%
        {septiandri_weird_2023}
\bibfield{author}{\bibinfo{person}{Ali~Akbar Septiandri}, \bibinfo{person}{Marios Constantinides}, \bibinfo{person}{Mohammad Tahaei}, {and} \bibinfo{person}{Daniele Quercia}.} \bibinfo{year}{2023}\natexlab{}.
\newblock \showarticletitle{{WEIRD} {FAccTs}: {How} {Western}, {Educated}, {Industrialized}, {Rich}, and {Democratic} is {FAccT}?}. In \bibinfo{booktitle}{\emph{Proceedings of the 2023 {ACM} {Conference} on {Fairness}, {Accountability}, and {Transparency}}} \emph{(\bibinfo{series}{{FAccT} '23})}. \bibinfo{publisher}{Association for Computing Machinery}, \bibinfo{address}{New York, NY, USA}, \bibinfo{pages}{160--171}.
\newblock
\showISBNx{9798400701924}
\urldef\tempurl%
\url{https://doi.org/10.1145/3593013.3593985}
\showDOI{\tempurl}


\bibitem[Shi and Li(2024)]%
        {shi_new_2024}
\bibfield{author}{\bibinfo{person}{Wen Shi} {and} \bibinfo{person}{Jinhui Li}.} \bibinfo{year}{2024}\natexlab{}.
\newblock \showarticletitle{New {Digital} {Divide} {Shaped} by {Algorithm}? {Evidence} from {Agent}-{Based} {Testing} on {Douyin}’s {Health}-{Related} {Video} {Recommendation}}.
\newblock \bibinfo{journal}{\emph{Communication Research}} (\bibinfo{date}{July} \bibinfo{year}{2024}), \bibinfo{pages}{00936502241262056}.
\newblock
\showISSN{0093-6502, 1552-3810}
\urldef\tempurl%
\url{https://doi.org/10.1177/00936502241262056}
\showDOI{\tempurl}


\bibitem[Shin and Jitkajornwanich(2024)]%
        {shin_how_2024}
\bibfield{author}{\bibinfo{person}{Donghee Shin} {and} \bibinfo{person}{Kulsawasd Jitkajornwanich}.} \bibinfo{year}{2024}\natexlab{}.
\newblock \showarticletitle{How {Algorithms} {Promote} {Self}-{Radicalization}: {Audit} of {TikTok}’s {Algorithm} {Using} a {Reverse} {Engineering} {Method}}.
\newblock \bibinfo{journal}{\emph{Social Science Computer Review}} \bibinfo{volume}{42}, \bibinfo{number}{4} (\bibinfo{date}{Aug.} \bibinfo{year}{2024}), \bibinfo{pages}{1020--1040}.
\newblock
\showISSN{0894-4393, 1552-8286}
\urldef\tempurl%
\url{https://doi.org/10.1177/08944393231225547}
\showDOI{\tempurl}


\bibitem[Shin and Valente(2020)]%
        {shin_algorithms_2020}
\bibfield{author}{\bibinfo{person}{Jieun Shin} {and} \bibinfo{person}{Thomas Valente}.} \bibinfo{year}{2020}\natexlab{}.
\newblock \showarticletitle{Algorithms and {Health} {Misinformation}: {A} {Case} {Study} of {Vaccine} {Books} on {Amazon}}.
\newblock \bibinfo{journal}{\emph{Journal of Health Communication}} \bibinfo{volume}{25}, \bibinfo{number}{5} (\bibinfo{date}{May} \bibinfo{year}{2020}), \bibinfo{pages}{394--401}.
\newblock
\showISSN{1081-0730, 1087-0415}
\urldef\tempurl%
\url{https://doi.org/10.1080/10810730.2020.1776423}
\showDOI{\tempurl}


\bibitem[Silva et~al\mbox{.}(2020)]%
        {silva_facebook_2020}
\bibfield{author}{\bibinfo{person}{Márcio Silva}, \bibinfo{person}{Lucas Santos De~Oliveira}, \bibinfo{person}{Athanasios Andreou}, \bibinfo{person}{Pedro~Olmo Vaz De~Melo}, \bibinfo{person}{Oana Goga}, {and} \bibinfo{person}{Fabricio Benevenuto}.} \bibinfo{year}{2020}\natexlab{}.
\newblock \showarticletitle{Facebook {Ads} {Monitor}: {An} {Independent} {Auditing} {System} for {Political} {Ads} on {Facebook}}. In \bibinfo{booktitle}{\emph{Proceedings of {The} {Web} {Conference} 2020}}. \bibinfo{publisher}{ACM}, \bibinfo{address}{Taipei Taiwan}, \bibinfo{pages}{224--234}.
\newblock
\showISBNx{978-1-4503-7023-3}
\urldef\tempurl%
\url{https://doi.org/10.1145/3366423.3380109}
\showDOI{\tempurl}


\bibitem[Singh et~al\mbox{.}(2024)]%
        {singh_language_2024}
\bibfield{author}{\bibinfo{person}{Vivek~K Singh}, \bibinfo{person}{Pamela Valera}, \bibinfo{person}{Ishaan Singh}, \bibinfo{person}{Ritesh Sawant}, {and} \bibinfo{person}{Yisel Breton}.} \bibinfo{year}{2024}\natexlab{}.
\newblock \showarticletitle{Language disparities in pandemic information: {Autocomplete} analysis of {COVID}-19 searches in {New} {York}}.
\newblock \bibinfo{journal}{\emph{Health Informatics Journal}} \bibinfo{volume}{30}, \bibinfo{number}{4} (\bibinfo{date}{Oct.} \bibinfo{year}{2024}), \bibinfo{pages}{14604582241307836}.
\newblock
\showISSN{1460-4582, 1741-2811}
\urldef\tempurl%
\url{https://doi.org/10.1177/14604582241307836}
\showDOI{\tempurl}


\bibitem[Smets et~al\mbox{.}(2019)]%
        {smets_does_2019}
\bibfield{author}{\bibinfo{person}{Annelien Smets}, \bibinfo{person}{Eladio Montero}, {and} \bibinfo{person}{Pieter Ballon}.} \bibinfo{year}{2019}\natexlab{}.
\newblock \showarticletitle{Does the {Bubble} {Go} {Beyond}?}
\newblock  (\bibinfo{year}{2019}).
\newblock


\bibitem[Snickars(2017)]%
        {snickars_more_2017}
\bibfield{author}{\bibinfo{person}{Pelle Snickars}.} \bibinfo{year}{2017}\natexlab{}.
\newblock \showarticletitle{More of the {Same} – {On} {Spotify} {Radio}}.
\newblock \bibinfo{journal}{\emph{Culture Unbound}} \bibinfo{volume}{9}, \bibinfo{number}{2} (\bibinfo{date}{Oct.} \bibinfo{year}{2017}), \bibinfo{pages}{184--211}.
\newblock
\showISSN{2000-1525}
\urldef\tempurl%
\url{https://doi.org/10.3384/cu.2000.1525.1792184}
\showDOI{\tempurl}


\bibitem[Soeller et~al\mbox{.}(2016)]%
        {soeller_mapwatch_2016}
\bibfield{author}{\bibinfo{person}{Gary Soeller}, \bibinfo{person}{Karrie Karahalios}, \bibinfo{person}{Christian Sandvig}, {and} \bibinfo{person}{Christo Wilson}.} \bibinfo{year}{2016}\natexlab{}.
\newblock \showarticletitle{{MapWatch}: {Detecting} and {Monitoring} {International} {Border} {Personalization} on {Online} {Maps}}. In \bibinfo{booktitle}{\emph{Proceedings of the 25th {International} {Conference} on {World} {Wide} {Web}}}. \bibinfo{publisher}{International World Wide Web Conferences Steering Committee}, \bibinfo{address}{Montréal Québec Canada}, \bibinfo{pages}{867--878}.
\newblock
\showISBNx{978-1-4503-4143-1}
\urldef\tempurl%
\url{https://doi.org/10.1145/2872427.2883016}
\showDOI{\tempurl}


\bibitem[Spyridou et~al\mbox{.}(2022)]%
        {spyridou_modeling_2022}
\bibfield{author}{\bibinfo{person}{Paschalia~(Lia) Spyridou}, \bibinfo{person}{Constantinos Djouvas}, {and} \bibinfo{person}{Dimitra Milioni}.} \bibinfo{year}{2022}\natexlab{}.
\newblock \showarticletitle{Modeling and {Validating} a {News} {Recommender} {Algorithm} in a {Mainstream} {Medium}-{Sized} {News} {Organization}: {An} {Experimental} {Approach}}.
\newblock \bibinfo{journal}{\emph{Future Internet}} \bibinfo{volume}{14}, \bibinfo{number}{10} (\bibinfo{date}{Sept.} \bibinfo{year}{2022}), \bibinfo{pages}{284}.
\newblock
\showISSN{1999-5903}
\urldef\tempurl%
\url{https://doi.org/10.3390/fi14100284}
\showDOI{\tempurl}


\bibitem[Stanusch(2023)]%
        {stanusch_relationship_2023}
\bibfield{author}{\bibinfo{person}{Natalia Stanusch}.} \bibinfo{year}{2023}\natexlab{}.
\newblock \showarticletitle{The {Relationship} {Between} {Knowledge} {Production} and {Google} in {Framing} and {Reframing} {AI} {Imaginary}. {A} {Comparative} {Algorithmic} {Audit} between the {US} and {Italy}}.
\newblock  (\bibinfo{year}{2023}).
\newblock


\bibitem[Sun et~al\mbox{.}(2023)]%
        {sun_smiling_2023}
\bibfield{author}{\bibinfo{person}{Luhang Sun}, \bibinfo{person}{Mian Wei}, \bibinfo{person}{Yibing Sun}, \bibinfo{person}{Yoo~Ji Suh}, \bibinfo{person}{Liwei Shen}, {and} \bibinfo{person}{Sijia Yang}.} \bibinfo{year}{2023}\natexlab{}.
\newblock \showarticletitle{Smiling women pitching down: auditing representational and presentational gender biases in image-generative {AI}}.
\newblock \bibinfo{journal}{\emph{Journal of Computer-Mediated Communication}} \bibinfo{volume}{29}, \bibinfo{number}{1} (\bibinfo{date}{Nov.} \bibinfo{year}{2023}), \bibinfo{pages}{zmad045}.
\newblock
\showISSN{1083-6101}
\urldef\tempurl%
\url{https://doi.org/10.1093/jcmc/zmad045}
\showDOI{\tempurl}


\bibitem[Sun et~al\mbox{.}(2024)]%
        {sun_value_2024}
\bibfield{author}{\bibinfo{person}{Tianshu Sun}, \bibinfo{person}{Zhe Yuan}, \bibinfo{person}{Chunxiao Li}, \bibinfo{person}{Kaifu Zhang}, {and} \bibinfo{person}{Jun Xu}.} \bibinfo{year}{2024}\natexlab{}.
\newblock \showarticletitle{The {Value} of {Personal} {Data} in {Internet} {Commerce}: {A} {High}-{Stakes} {Field} {Experiment} on {Data} {Regulation} {Policy}}.
\newblock \bibinfo{journal}{\emph{Management Science}} \bibinfo{volume}{70}, \bibinfo{number}{4} (\bibinfo{date}{April} \bibinfo{year}{2024}), \bibinfo{pages}{2645--2660}.
\newblock
\showISSN{0025-1909, 1526-5501}
\urldef\tempurl%
\url{https://doi.org/10.1287/mnsc.2023.4828}
\showDOI{\tempurl}


\bibitem[Sweeney(2013)]%
        {sweeney_discrimination_2013}
\bibfield{author}{\bibinfo{person}{Latanya Sweeney}.} \bibinfo{year}{2013}\natexlab{}.
\newblock \showarticletitle{Discrimination in {Online} {Ad} {Delivery}: {Google} ads, black names and white names, racial discrimination, and click advertising}.
\newblock \bibinfo{journal}{\emph{Queue}} \bibinfo{volume}{11}, \bibinfo{number}{3} (\bibinfo{date}{March} \bibinfo{year}{2013}), \bibinfo{pages}{10--29}.
\newblock
\showISSN{1542-7730}
\urldef\tempurl%
\url{https://doi.org/10.1145/2460276.2460278}
\showDOI{\tempurl}


\bibitem[Tennant(2020)]%
        {tennant_web_2020}
\bibfield{author}{\bibinfo{person}{Jonathan~P. Tennant}.} \bibinfo{year}{2020}\natexlab{}.
\newblock \showarticletitle{Web of {Science} and {Scopus} are not global databases of knowledge}.
\newblock \bibinfo{journal}{\emph{European Science Editing}}  \bibinfo{volume}{46} (\bibinfo{date}{Oct.} \bibinfo{year}{2020}), \bibinfo{pages}{e51987}.
\newblock
\showISSN{2518-3354}
\urldef\tempurl%
\url{https://doi.org/10.3897/ese.2020.e51987}
\showDOI{\tempurl}
\newblock
\shownote{Publisher: European Association of Science Editors (EASE)}.


\bibitem[Tharakan et~al\mbox{.}(2024)]%
        {tharakan_chatgpt_2024}
\bibfield{author}{\bibinfo{person}{Shebin Tharakan}, \bibinfo{person}{Brandon Klein}, \bibinfo{person}{Lucas Bartlett}, \bibinfo{person}{Aaron Atlas}, \bibinfo{person}{Stephen~A. Parada}, {and} \bibinfo{person}{Randy~M. Cohn}.} \bibinfo{year}{2024}\natexlab{}.
\newblock \showarticletitle{Do {ChatGPT} and {Google} differ in answers to commonly asked patient questions regarding total shoulder and total elbow arthroplasty?}
\newblock \bibinfo{journal}{\emph{Journal of Shoulder and Elbow Surgery}} \bibinfo{volume}{33}, \bibinfo{number}{8} (\bibinfo{date}{Aug.} \bibinfo{year}{2024}), \bibinfo{pages}{e429--e437}.
\newblock
\showISSN{10582746}
\urldef\tempurl%
\url{https://doi.org/10.1016/j.jse.2023.11.014}
\showDOI{\tempurl}


\bibitem[Thorson et~al\mbox{.}(2021)]%
        {thorson_algorithmic_2021}
\bibfield{author}{\bibinfo{person}{Kjerstin Thorson}, \bibinfo{person}{Kelley Cotter}, \bibinfo{person}{Mel Medeiros}, {and} \bibinfo{person}{Chankyung Pak}.} \bibinfo{year}{2021}\natexlab{}.
\newblock \showarticletitle{Algorithmic inference, political interest, and exposure to news and politics on {Facebook}}.
\newblock \bibinfo{journal}{\emph{Information, Communication \& Society}} \bibinfo{volume}{24}, \bibinfo{number}{2} (\bibinfo{date}{Jan.} \bibinfo{year}{2021}), \bibinfo{pages}{183--200}.
\newblock
\showISSN{1369-118X, 1468-4462}
\urldef\tempurl%
\url{https://doi.org/10.1080/1369118X.2019.1642934}
\showDOI{\tempurl}


\bibitem[Toepfl et~al\mbox{.}(2023a)]%
        {toepfl_who_2023}
\bibfield{author}{\bibinfo{person}{Florian Toepfl}, \bibinfo{person}{Daria Kravets}, \bibinfo{person}{Anna Ryzhova}, {and} \bibinfo{person}{Arista Beseler}.} \bibinfo{year}{2023}\natexlab{a}.
\newblock \showarticletitle{Who are the plotters behind the pandemic? {Comparing} {Covid}-19 conspiracy theories in {Google} search results across five key target countries of {Russia}’s foreign communication}.
\newblock \bibinfo{journal}{\emph{Information, Communication \& Society}} \bibinfo{volume}{26}, \bibinfo{number}{10} (\bibinfo{date}{July} \bibinfo{year}{2023}), \bibinfo{pages}{2033--2051}.
\newblock
\showISSN{1369-118X, 1468-4462}
\urldef\tempurl%
\url{https://doi.org/10.1080/1369118X.2022.2065213}
\showDOI{\tempurl}


\bibitem[Toepfl et~al\mbox{.}(2023b)]%
        {toepfl_googling_2023}
\bibfield{author}{\bibinfo{person}{Florian Toepfl}, \bibinfo{person}{Anna Ryzhova}, \bibinfo{person}{Daria Kravets}, {and} \bibinfo{person}{Arista Beseler}.} \bibinfo{year}{2023}\natexlab{b}.
\newblock \showarticletitle{Googling in {Russian} {Abroad}: {How} {Kremlin}-{Affiliated} {Websites} {Contribute} to the {Visibility} of {COVID}-19 {Conspiracy} {Theories} in {Search} {Results}}.
\newblock \bibinfo{journal}{\emph{International Journal of Communication}} \bibinfo{volume}{17}, \bibinfo{number}{0} (\bibinfo{date}{Jan.} \bibinfo{year}{2023}), \bibinfo{pages}{21}.
\newblock
\showISSN{1932-8036}
\urldef\tempurl%
\url{https://ijoc.org/index.php/ijoc/article/view/19423}
\showURL{%
\tempurl}
\newblock
\shownote{Number: 0}.


\bibitem[Tomlein et~al\mbox{.}(2021)]%
        {tomlein_audit_2021}
\bibfield{author}{\bibinfo{person}{Matus Tomlein}, \bibinfo{person}{Branislav Pecher}, \bibinfo{person}{Jakub Simko}, \bibinfo{person}{Ivan Srba}, \bibinfo{person}{Robert Moro}, \bibinfo{person}{Elena Stefancova}, \bibinfo{person}{Michal Kompan}, \bibinfo{person}{Andrea Hrckova}, \bibinfo{person}{Juraj Podrouzek}, {and} \bibinfo{person}{Maria Bielikova}.} \bibinfo{year}{2021}\natexlab{}.
\newblock \showarticletitle{An {Audit} of {Misinformation} {Filter} {Bubbles} on {YouTube}: {Bubble} {Bursting} and {Recent} {Behavior} {Changes}}. In \bibinfo{booktitle}{\emph{Fifteenth {ACM} {Conference} on {Recommender} {Systems}}}. \bibinfo{publisher}{ACM}, \bibinfo{address}{Amsterdam Netherlands}, \bibinfo{pages}{1--11}.
\newblock
\showISBNx{978-1-4503-8458-2}
\urldef\tempurl%
\url{https://doi.org/10.1145/3460231.3474241}
\showDOI{\tempurl}


\bibitem[Trielli and Diakopoulos(2019)]%
        {trielli_search_2019}
\bibfield{author}{\bibinfo{person}{Daniel Trielli} {and} \bibinfo{person}{Nicholas Diakopoulos}.} \bibinfo{year}{2019}\natexlab{}.
\newblock \showarticletitle{Search as {News} {Curator}: {The} {Role} of {Google} in {Shaping} {Attention} to {News} {Information}}. In \bibinfo{booktitle}{\emph{Proceedings of the 2019 {CHI} {Conference} on {Human} {Factors} in {Computing} {Systems}}}. \bibinfo{publisher}{ACM}, \bibinfo{address}{Glasgow Scotland Uk}, \bibinfo{pages}{1--15}.
\newblock
\showISBNx{978-1-4503-5970-2}
\urldef\tempurl%
\url{https://doi.org/10.1145/3290605.3300683}
\showDOI{\tempurl}


\bibitem[Tschantz et~al\mbox{.}(2018)]%
        {tschantz_accuracy_2018}
\bibfield{author}{\bibinfo{person}{Michael~Carl Tschantz}, \bibinfo{person}{Serge Egelman}, \bibinfo{person}{Jaeyoung Choi}, \bibinfo{person}{Nicholas Weaver}, {and} \bibinfo{person}{Gerald Friedland}.} \bibinfo{year}{2018}\natexlab{}.
\newblock \showarticletitle{The {Accuracy} of the {Demographic} {Inferences} {Shown} on {Google}'s {Ad} {Settings}}. In \bibinfo{booktitle}{\emph{Proceedings of the 2018 {Workshop} on {Privacy} in the {Electronic} {Society}}}. \bibinfo{publisher}{ACM}, \bibinfo{address}{Toronto Canada}, \bibinfo{pages}{33--41}.
\newblock
\showISBNx{978-1-4503-5989-4}
\urldef\tempurl%
\url{https://doi.org/10.1145/3267323.3268962}
\showDOI{\tempurl}


\bibitem[Ulloa et~al\mbox{.}(2023)]%
        {ulloa_novelty_2023}
\bibfield{author}{\bibinfo{person}{Roberto Ulloa}, \bibinfo{person}{Mykola Makhortykh}, \bibinfo{person}{Aleksandra Urman}, {and} \bibinfo{person}{Juhi Kulshrestha}.} \bibinfo{year}{2023}\natexlab{}.
\newblock \showarticletitle{Novelty in {News} {Search}: {A} {Longitudinal} {Study} of the 2020 {US} {Elections}}.
\newblock \bibinfo{journal}{\emph{Social Science Computer Review}} (\bibinfo{date}{Aug.} \bibinfo{year}{2023}), \bibinfo{pages}{08944393231195471}.
\newblock
\showISSN{0894-4393, 1552-8286}
\urldef\tempurl%
\url{https://doi.org/10.1177/08944393231195471}
\showDOI{\tempurl}


\bibitem[Ungless et~al\mbox{.}(2023)]%
        {ungless_stereotypes_2023}
\bibfield{author}{\bibinfo{person}{Eddie Ungless}, \bibinfo{person}{Bjorn Ross}, {and} \bibinfo{person}{Anne Lauscher}.} \bibinfo{year}{2023}\natexlab{}.
\newblock \showarticletitle{Stereotypes and {Smut}: {The} ({Mis})representation of {Non}-cisgender {Identities} by {Text}-to-{Image} {Models}}. In \bibinfo{booktitle}{\emph{Findings of the {Association} for {Computational} {Linguistics}: {ACL} 2023}}, \bibfield{editor}{\bibinfo{person}{Anna Rogers}, \bibinfo{person}{Jordan Boyd-Graber}, {and} \bibinfo{person}{Naoaki Okazaki}} (Eds.). \bibinfo{publisher}{Association for Computational Linguistics}, \bibinfo{address}{Toronto, Canada}, \bibinfo{pages}{7919--7942}.
\newblock
\urldef\tempurl%
\url{https://doi.org/10.18653/v1/2023.findings-acl.502}
\showDOI{\tempurl}


\bibitem[Unkel and Haim(2021)]%
        {unkel_googling_2021}
\bibfield{author}{\bibinfo{person}{Julian Unkel} {and} \bibinfo{person}{Mario Haim}.} \bibinfo{year}{2021}\natexlab{}.
\newblock \showarticletitle{Googling {Politics}: {Parties}, {Sources}, and {Issue} {Ownerships} on {Google} in the 2017 {German} {Federal} {Election} {Campaign}}.
\newblock \bibinfo{journal}{\emph{Social Science Computer Review}} \bibinfo{volume}{39}, \bibinfo{number}{5} (\bibinfo{date}{Oct.} \bibinfo{year}{2021}), \bibinfo{pages}{844--861}.
\newblock
\showISSN{0894-4393, 1552-8286}
\urldef\tempurl%
\url{https://doi.org/10.1177/0894439319881634}
\showDOI{\tempurl}


\bibitem[Urman and Makhortykh(2022)]%
        {urman_foreign_2022}
\bibfield{author}{\bibinfo{person}{Aleksandra Urman} {and} \bibinfo{person}{Mykola Makhortykh}.} \bibinfo{year}{2022}\natexlab{}.
\newblock \showarticletitle{“{Foreign} beauties want to meet you”: {The} sexualization of women in {Google}’s organic and sponsored text search results}.
\newblock \bibinfo{journal}{\emph{New Media \& Society}} (\bibinfo{date}{June} \bibinfo{year}{2022}), \bibinfo{pages}{146144482210995}.
\newblock
\showISSN{1461-4448, 1461-7315}
\urldef\tempurl%
\url{https://doi.org/10.1177/14614448221099536}
\showDOI{\tempurl}


\bibitem[Urman et~al\mbox{.}(2021)]%
        {urman_auditing_2021}
\bibfield{author}{\bibinfo{person}{Aleksandra Urman}, \bibinfo{person}{Mykola Makhortykh}, {and} \bibinfo{person}{Roberto Ulloa}.} \bibinfo{year}{2021}\natexlab{}.
\newblock \showarticletitle{Auditing {Source} {Diversity} {Bias} in {Video} {Search} {Results} {Using} {Virtual} {Agents}}. In \bibinfo{booktitle}{\emph{Companion {Proceedings} of the {Web} {Conference} 2021}}. \bibinfo{publisher}{ACM}, \bibinfo{address}{Ljubljana Slovenia}, \bibinfo{pages}{232--236}.
\newblock
\showISBNx{978-1-4503-8313-4}
\urldef\tempurl%
\url{https://doi.org/10.1145/3442442.3452306}
\showDOI{\tempurl}


\bibitem[Urman et~al\mbox{.}(2022a)]%
        {urman_auditing_2022}
\bibfield{author}{\bibinfo{person}{Aleksandra Urman}, \bibinfo{person}{Mykola Makhortykh}, {and} \bibinfo{person}{Roberto Ulloa}.} \bibinfo{year}{2022}\natexlab{a}.
\newblock \showarticletitle{Auditing the representation of migrants in image web search results}.
\newblock \bibinfo{journal}{\emph{Humanities and Social Sciences Communications}} \bibinfo{volume}{9}, \bibinfo{number}{1} (\bibinfo{date}{April} \bibinfo{year}{2022}), \bibinfo{pages}{130}.
\newblock
\showISSN{2662-9992}
\urldef\tempurl%
\url{https://doi.org/10.1057/s41599-022-01144-1}
\showDOI{\tempurl}


\bibitem[Urman et~al\mbox{.}(2022b)]%
        {urman_matter_2022}
\bibfield{author}{\bibinfo{person}{Aleksandra Urman}, \bibinfo{person}{Mykola Makhortykh}, {and} \bibinfo{person}{Roberto Ulloa}.} \bibinfo{year}{2022}\natexlab{b}.
\newblock \showarticletitle{The {Matter} of {Chance}: {Auditing} {Web} {Search} {Results} {Related} to the 2020 {U}.{S}. {Presidential} {Primary} {Elections} {Across} {Six} {Search} {Engines}}.
\newblock \bibinfo{journal}{\emph{Social Science Computer Review}} \bibinfo{volume}{40}, \bibinfo{number}{5} (\bibinfo{date}{Oct.} \bibinfo{year}{2022}), \bibinfo{pages}{1323--1339}.
\newblock
\showISSN{0894-4393, 1552-8286}
\urldef\tempurl%
\url{https://doi.org/10.1177/08944393211006863}
\showDOI{\tempurl}


\bibitem[Urman et~al\mbox{.}(2022c)]%
        {urman_where_2022}
\bibfield{author}{\bibinfo{person}{Aleksandra Urman}, \bibinfo{person}{Mykola Makhortykh}, \bibinfo{person}{Roberto Ulloa}, {and} \bibinfo{person}{Juhi Kulshrestha}.} \bibinfo{year}{2022}\natexlab{c}.
\newblock \showarticletitle{Where the earth is flat and 9/11 is an inside job: {A} comparative algorithm audit of conspiratorial information in web search results}.
\newblock \bibinfo{journal}{\emph{Telematics and Informatics}}  \bibinfo{volume}{72} (\bibinfo{date}{Aug.} \bibinfo{year}{2022}), \bibinfo{pages}{101860}.
\newblock
\showISSN{07365853}
\urldef\tempurl%
\url{https://doi.org/10.1016/j.tele.2022.101860}
\showDOI{\tempurl}


\bibitem[Urman et~al\mbox{.}(2023)]%
        {urman_right_2023}
\bibfield{author}{\bibinfo{person}{Aleksandra Urman}, \bibinfo{person}{Ivan Smirnov}, {and} \bibinfo{person}{Jana Lasser}.} \bibinfo{year}{2023}\natexlab{}.
\newblock \bibinfo{title}{The right to audit and power asymmetries in algorithm auditing}.
\newblock
\newblock
\urldef\tempurl%
\url{https://doi.org/10.48550/arXiv.2302.08301}
\showDOI{\tempurl}
\newblock
\shownote{arXiv:2302.08301 [cs]}.


\bibitem[Van~Es et~al\mbox{.}(2021)]%
        {van_es_gendered_2021}
\bibfield{author}{\bibinfo{person}{Karin Van~Es}, \bibinfo{person}{Daniel Everts}, {and} \bibinfo{person}{Iris Muis}.} \bibinfo{year}{2021}\natexlab{}.
\newblock \showarticletitle{Gendered language and employment {Web} sites: {How} search algorithms can cause allocative harm}.
\newblock \bibinfo{journal}{\emph{First Monday}} (\bibinfo{date}{July} \bibinfo{year}{2021}).
\newblock
\showISSN{1396-0466}
\urldef\tempurl%
\url{https://doi.org/10.5210/fm.v26i8.11717}
\showDOI{\tempurl}


\bibitem[Van~Hoof et~al\mbox{.}(2024)]%
        {van_hoof_it_2024}
\bibfield{author}{\bibinfo{person}{Marieke Van~Hoof}, \bibinfo{person}{Damian Trilling}, \bibinfo{person}{Judith Moeller}, {and} \bibinfo{person}{Corine~S Meppelink}.} \bibinfo{year}{2024}\natexlab{}.
\newblock \showarticletitle{It matters how you google it? {Using} agent-based testing to assess the impact of user choices in search queries and algorithmic personalization on political {Google} {Search} results}.
\newblock \bibinfo{journal}{\emph{Journal of Computer-Mediated Communication}} \bibinfo{volume}{29}, \bibinfo{number}{6} (\bibinfo{date}{Sept.} \bibinfo{year}{2024}), \bibinfo{pages}{zmae020}.
\newblock
\showISSN{1083-6101}
\urldef\tempurl%
\url{https://doi.org/10.1093/jcmc/zmae020}
\showDOI{\tempurl}


\bibitem[Venkatadri et~al\mbox{.}(2019)]%
        {venkatadri_auditing_2019}
\bibfield{author}{\bibinfo{person}{Giridhari Venkatadri}, \bibinfo{person}{Piotr Sapiezynski}, \bibinfo{person}{Elissa~M. Redmiles}, \bibinfo{person}{Alan Mislove}, \bibinfo{person}{Oana Goga}, \bibinfo{person}{Michelle Mazurek}, {and} \bibinfo{person}{Krishna~P. Gummadi}.} \bibinfo{year}{2019}\natexlab{}.
\newblock \showarticletitle{Auditing {Offline} {Data} {Brokers} via {Facebook}'s {Advertising} {Platform}}. In \bibinfo{booktitle}{\emph{The {World} {Wide} {Web} {Conference}}}. \bibinfo{publisher}{ACM}, \bibinfo{address}{San Francisco CA USA}, \bibinfo{pages}{1920--1930}.
\newblock
\showISBNx{978-1-4503-6674-8}
\urldef\tempurl%
\url{https://doi.org/10.1145/3308558.3313666}
\showDOI{\tempurl}


\bibitem[Vincent et~al\mbox{.}(2019)]%
        {vincent_measuring_2019}
\bibfield{author}{\bibinfo{person}{Nicholas Vincent}, \bibinfo{person}{Isaac Johnson}, \bibinfo{person}{Patrick Sheehan}, {and} \bibinfo{person}{Brent Hecht}.} \bibinfo{year}{2019}\natexlab{}.
\newblock \showarticletitle{Measuring the {Importance} of {User}-{Generated} {Content} to {Search} {Engines}}.
\newblock \bibinfo{journal}{\emph{Proceedings of the International AAAI Conference on Web and Social Media}}  \bibinfo{volume}{13} (\bibinfo{date}{July} \bibinfo{year}{2019}), \bibinfo{pages}{505--516}.
\newblock
\showISSN{2334-0770}
\urldef\tempurl%
\url{https://doi.org/10.1609/icwsm.v13i01.3248}
\showDOI{\tempurl}


\bibitem[Vombatkere et~al\mbox{.}(2024)]%
        {vombatkere_tiktok_2024}
\bibfield{author}{\bibinfo{person}{Karan Vombatkere}, \bibinfo{person}{Sepehr Mousavi}, \bibinfo{person}{Savvas Zannettou}, \bibinfo{person}{Franziska Roesner}, {and} \bibinfo{person}{Krishna~P. Gummadi}.} \bibinfo{year}{2024}\natexlab{}.
\newblock \showarticletitle{{TikTok} and the {Art} of {Personalization}: {Investigating} {Exploration} and {Exploitation} on {Social} {Media} {Feeds}}. In \bibinfo{booktitle}{\emph{Proceedings of the {ACM} on {Web} {Conference} 2024}}. \bibinfo{publisher}{ACM}, \bibinfo{address}{Singapore Singapore}, \bibinfo{pages}{3789--3797}.
\newblock
\showISBNx{9798400701719}
\urldef\tempurl%
\url{https://doi.org/10.1145/3589334.3645600}
\showDOI{\tempurl}


\bibitem[W3Techs(2023)]%
        {w3techs_usage_2023}
\bibfield{author}{\bibinfo{person}{W3Techs}.} \bibinfo{year}{2023}\natexlab{}.
\newblock \bibinfo{title}{Usage {Statistics} and {Market} {Share} of {Content} {Languages} for {Websites}, {December} 2023}.
\newblock
\newblock
\urldef\tempurl%
\url{https://w3techs.com/technologies/overview/content_language}
\showURL{%
\tempurl}


\bibitem[Wang et~al\mbox{.}(2024)]%
        {wang_lower_2024}
\bibfield{author}{\bibinfo{person}{Stephanie Wang}, \bibinfo{person}{Shengchun Huang}, \bibinfo{person}{Alvin Zhou}, {and} \bibinfo{person}{Danaë Metaxa}.} \bibinfo{year}{2024}\natexlab{}.
\newblock \showarticletitle{Lower {Quantity}, {Higher} {Quality}: {Auditing} {News} {Content} and {User} {Perceptions} on {Twitter}/{X} {Algorithmic} versus {Chronological} {Timelines}}.
\newblock \bibinfo{journal}{\emph{Proceedings of the ACM on Human-Computer Interaction}} \bibinfo{volume}{8}, \bibinfo{number}{CSCW2} (\bibinfo{date}{Nov.} \bibinfo{year}{2024}), \bibinfo{pages}{1--25}.
\newblock
\showISSN{2573-0142}
\urldef\tempurl%
\url{https://doi.org/10.1145/3687046}
\showDOI{\tempurl}


\bibitem[Weber and Kosterich(2018)]%
        {weber_coding_2018}
\bibfield{author}{\bibinfo{person}{Matthew~S. Weber} {and} \bibinfo{person}{Allie Kosterich}.} \bibinfo{year}{2018}\natexlab{}.
\newblock \showarticletitle{Coding the {News}: {The} role of computer code in filtering and distributing news}.
\newblock \bibinfo{journal}{\emph{Digital Journalism}} \bibinfo{volume}{6}, \bibinfo{number}{3} (\bibinfo{date}{March} \bibinfo{year}{2018}), \bibinfo{pages}{310--329}.
\newblock
\showISSN{2167-0811, 2167-082X}
\urldef\tempurl%
\url{https://doi.org/10.1080/21670811.2017.1366865}
\showDOI{\tempurl}


\bibitem[Xivuri and Twinomurinzi(2021)]%
        {xivuri_systematic_2021}
\bibfield{author}{\bibinfo{person}{Khensani Xivuri} {and} \bibinfo{person}{Hossana Twinomurinzi}.} \bibinfo{year}{2021}\natexlab{}.
\newblock \showarticletitle{A {Systematic} {Review} of {Fairness} in {Artificial} {Intelligence} {Algorithms}}. In \bibinfo{booktitle}{\emph{Responsible {AI} and {Analytics} for an {Ethical} and {Inclusive} {Digitized} {Society}}} \emph{(\bibinfo{series}{Lecture {Notes} in {Computer} {Science}})}, \bibfield{editor}{\bibinfo{person}{Denis Dennehy}, \bibinfo{person}{Anastasia Griva}, \bibinfo{person}{Nancy Pouloudi}, \bibinfo{person}{Yogesh~K. Dwivedi}, \bibinfo{person}{Ilias Pappas}, {and} \bibinfo{person}{Matti Mäntymäki}} (Eds.). \bibinfo{publisher}{Springer International Publishing}, \bibinfo{address}{Cham}, \bibinfo{pages}{271--284}.
\newblock
\showISBNx{978-3-030-85447-8}
\urldef\tempurl%
\url{https://doi.org/10.1007/978-3-030-85447-8_24}
\showDOI{\tempurl}


\bibitem[Yang et~al\mbox{.}(2023)]%
        {yang_bubbles_2023}
\bibfield{author}{\bibinfo{person}{Can Yang}, \bibinfo{person}{Xinyuan Xu}, \bibinfo{person}{Bernardo~Pereira Nunes}, {and} \bibinfo{person}{Sean Wolfgand~Matsui Siqueira}.} \bibinfo{year}{2023}\natexlab{}.
\newblock \showarticletitle{Bubbles bursting: {Investigating} and measuring the personalisation of social media searches}.
\newblock \bibinfo{journal}{\emph{Telematics and Informatics}}  \bibinfo{volume}{82} (\bibinfo{date}{Aug.} \bibinfo{year}{2023}), \bibinfo{pages}{101999}.
\newblock
\showISSN{07365853}
\urldef\tempurl%
\url{https://doi.org/10.1016/j.tele.2023.101999}
\showDOI{\tempurl}


\end{thebibliography}

\appendix

\section{Further methodological details}
\label{appendix:methodology}
\subsection{Exclusion after the eligibility check}
The reasons for exclusion were the following: \textbf{Not public-facing} - 50 studies were excluded due to not fitting our narrower definition of a public(-facing) system. \textbf{Theory or methods} - 20 studies dealt with theoretical or methodological contributions and not specific systems. \textbf{User study} - 12 studies focused on users and their experiences with a system but not the system itself. \textbf{Non-algorithm} - 11 papers 
    dealt with some aspects of a system but not its algorithm. \textbf{Overview} - 2 studies were overviews of algorithm auditing-related research but not empirical studies. \textbf{Development process} - 1 study dealt with the system development. One paper was a work-in-progress as it outlined the research the author was planning to do within their dissertation. Finally, in 7 cases, we could not retrieve the full text of the articles. Altogether, \textbf{176 papers were included in the final analysis}. We provide the full list of papers along with the coding of key categories in the Appendix \ref{appendix:papers}.

\subsection{Specific Problem Categories: Definitions}
\begin{itemize}
    \item Personalization - 
    factors that affect content personalization on a platform and/or distortions and disparities in content delivery arising from personalization. 
    \item Filter bubble - the presence/absence of so-called filter bubbles. While filter bubbles might be seen as a specific case of personalization, we decided to include this as a separate option due to the amount of scientific attention this specific phenomenon has attracted.
    \item News distribution - issues regarding news distribution by algorithmic systems.
    \item Harmful content - algorithmic distribution (e.g., amplification) of harmful content (see below for the Harmful content types specification).
    \item Group misrepresentation - misrepresentation (e.g., stereotypical representations) of specific groups of people by algorithmic systems/in their outputs.
    \item Price discrimination - potential price discrimination by algorithmic systems.
    \item Discrimination (other) - other types of discrimination that are not related to price discrimination and/or group misrepresentation (e.g., gender-based discrimination in job ad delivery).
    \item Information quality - the quality of information provided by algorithmic systems without focusing on harmful content or news specifically.
    \item User categorization - automatically classifying the user into categories based on their behavior/characteristics.
\end{itemize}

\section{Figure descriptions (for accessibility for the submitted draft)}
\label{appendix:figuredesc}
\subsection{Figure \ref{fig:problems}}
Bubble chart showing the distribution of four categories—Discrimination, Distortion, Exploitation, and Misjudgment—across the years 2012 to 2024. Each bubble represents the count of issues, with larger bubbles indicating higher counts. Discrimination starts with low numbers in earlier years but gradually increases to peak around 2022–2023. Distortion shows a consistent increase over time, with a notable peak of 28 issues in both 2023 and 2024. Exploitation has sporadic occurrences, peaking in 2017 with 2 issues. Misjudgment is infrequent, with only isolated instances in a few years. The chart uses pink shading for the bubbles, with increasing intensity and size corresponding to higher values.

\subsection{Figure \ref{fig:specificproblems}}
Stacked bar chart depicting the number of studies addressing various specific problems from 2012 to 2024. Each bar represents a year, with sections of the bar indicating the number of studies focused on specific problems. The problems, identified in the legend, include discrimination, discrimination (other), filter bubble, group misrepresentation, harmful content, information quality, news distribution, personalization, price discrimination, and user categorization. Colors represent the different problems, and the chart shows an increasing trend in the total number of studies over time, peaking in 2023. A note at the top clarifies that one study could address multiple problems, so the total number of studies exceeds the total number of reviewed studies. The y-axis indicates the number of studies, and the x-axis lists the years from 2012 to 2024.

\subsection{Figure \ref{fig:domains}}
Bubble chart showing the distribution of studies across various domains from 2012 to 2024. The y-axis lists domains such as E-commerce, Ad delivery, Search, Recommendation, User categorization, Monetization, Spam, Translation, and Generative AI, while the x-axis represents years. Each bubble indicates the number of studies for a given domain and year, with larger bubbles representing higher counts. Search and Recommendation see steady growth, peaking around 2023 with counts of 16. E-commerce and Ad delivery show sporadic activity. Generative AI emerges in 2023 and 2024 with increasing focus (4 studies in 2023 and 8 in 2024). Overall, the chart illustrates growing interest in certain domains over time, especially in later years.

\subsection{Figure \ref{fig:platforms}}
Bubble chart showing the distribution of studies across different platforms from 2012 to 2024. The y-axis lists platforms, including Bing, Google, Twitter, Facebook, Google News, YouTube, DuckDuckGo, Yahoo, Yandex, and ChatGPT. The x-axis represents years. Each bubble represents the count of studies focused on a platform in a specific year, with larger bubbles indicating higher counts. Google has the most consistent and prominent presence, peaking in 2022 with 13 studies. Platforms like YouTube, DuckDuckGo, and Yandex show increased focus in later years, while ChatGPT emerges in 2023 and 2024 with 2 and 5 studies, respectively. Facebook and Twitter have smaller, sporadic bubbles, indicating less focus overall. The chart highlights trends in platform-specific studies over time.

\subsection{Figure \ref{fig:countries}}
Choropleth map showing the total number of audits conducted per country globally. The countries are shaded in varying intensities of green, with darker shades representing a higher number of audits. The United States stands out with the darkest shade, indicating the highest number of audits (100). Other countries with notable audit activity include several in Europe, such as Germany and the United Kingdom, as well as Russia. Most countries are shaded in light green, indicating between 1 and 5 audits, or are unshaded, representing zero audits. The legend on the right provides a scale: 0 (white), 1–5 (light green), 6–10 (medium green), 24 (darker green), and 100 (darkest green).

\subsection{Figure \ref{fig:langs}}
Bar and line chart showing the count and percentage of studies or data points by language. The x-axis lists languages (e.g., English, German, Russian, French, etc.), while the left y-axis shows the count, and the right y-axis shows the corresponding percentage. English dominates the dataset with 132 instances, accounting for 75.4\%, shown as the tallest bar and highest point on the red line. Other languages, such as German (15, 8.6\%), Russian (9, 5.1\%), and French (7, 4\%), have much smaller counts, with the remaining languages having counts of 5 or fewer and percentages below 3\%. The red line illustrates the cumulative percentage across languages, decreasing sharply after English and tapering off for the less-represented languages.

\section{List of reviewed auditing studies}
\label{appendix:papers}

\begin{table}[]
\resizebox{\textwidth}{!}{%
\begin{tabular}{@{}llllllllll@{}}
\toprule
Paper &
  Year &
  Platform &
  Problem &
  Specific problem &
  Method &
  Domain &
  Language &
  Country-context &
  Author affiliation \\ \midrule
  \cite{ridgway_screenshotting_2024} &
  2024 &
  Google &
  Distortion &
  Personalization &
  Crowdsourcing &
  Search &
  English &
  Denmark &
  Denmark \\
  \cite{wang_lower_2024} &
  2024 &
  Twitter &
  Distortion &
  News distribution &
  Crowdsourcing &
  Recommendation &
  English &
  US &
  US \\
  \cite{van_hoof_it_2024} &
  2024 &
  Google &
  Distortion &
  Personalization &
  Persona scrape &
  Search &
  Dutch &
  Netherlands &
  Netherlands, Germany \\
  \cite{li_generative_2024} &
  2024 &
  ChatGPT, BingChat, Perplexity &
  Distortion &
  Information quality &
  Non-persona scrape &
  Search, Generative AI &
  English &
  Canada &
  Canada \\
  \cite{kacperski_examining_2024} &
  2024 &
  Google Scholar, Semantic Scholar &
  Distortion &
  Information quality &
  Non-persona scrape &
  Search &
  English &
  US, Germany &
  Germany, Switzerland, Austria \\
  \cite{neumann_data_2024} &
  2024 &
  Data brokers &
  Discrimination &
  User categorization &
  Repurposing &
  Ad delivery, User categorization &
  English &
  US &
  US, Australia \\
  \cite{dash_investigating_2024} &
  2024 &
  Amazon &
  Discrimination &
  Discrimination (other) &
  Non-persona scrape &
  E-commerce &
  Mixed &
  India, US, Germany, France &
  India, Germany \\
  \cite{sapiezynski_use_2024} &
  2024 &
  Facebook &
  Discrimination &
  Discrimination (other) &
  Repurposing &
  Ad delivery &
  English &
  US &
  US \\
  \cite{godinez_youtube_2024} &
  2024 &
  YouTube &
  Distortion &
  Information quality, Harmful content, News distribution &
  Non-persona scrape &
  Search &
  English &
  US &
  US \\
  \cite{shin_how_2024} &
  2024 &
  TikTok &
  Distortion &
  Harmful content, Filter bubble, Personalization &
  Crowdsourcing &
  Recommendation &
  English &
  US &
  US \\
  \cite{singh_language_2024} &
  2024 &
  Google &
  Distortion &
  Information quality &
  Non-persona scrape &
  Search &
  English, Spanish &
  US &
  US \\
  \cite{cakmak_bias_2024} &
  2024 &
  YouTube &
  Distortion &
  Information quality &
  Non-persona scrape &
  Search, Recommendation &
  English, Indonesian &
  US &
  US \\
\cite{tharakan_chatgpt_2024} &
  2024 &
  Google, ChatGPT &
  Distortion &
  Information quality &
  Non-persona scrape &
  Search, Generative AI &
  English &
  US &
  US \\
\cite{apiola_first_2024} &
  2024 &
  Midjourney, DALL-E, Bing, Stable Diffusion &
  Distortion &
  Group misrepresentation &
  Crowdsourcing &
  Generative AI &
  English &
  Mixed &
  Finland \\
\cite{nigatu_i_2024} &
  2024 &
  YouTube &
  Distortion &
  Harmful content &
  Non-persona scrape &
  Search &
  Amharic &
  Ethiopia, US, UK, UAE, Saudi Arabia &
  US \\
\cite{mahomed_auditing_2024} &
  2024 &
  ChatGPT &
  Distortion &
  Harmful content &
  Non-persona scrape &
  Generative AI &
  English &
  Mixed &
  US \\
\cite{kaplan_comprehensively_2024} &
  2024 &
  TikTok &
  Distortion &
  Personalization &
  Persona scrape &
  Recommendation &
  Mixed &
  US &
  US \\
\cite{vombatkere_tiktok_2024} &
  2024 &
  TikTok &
  Distortion &
  Personalization &
  Persona scrape &
  Recommendation &
  Mixed &
  Mixed &
  US, Germany, Netherlands \\
\cite{duskin_echo_2024} &
  2024 &
  Twitter &
  Distortion &
  Filter bubble &
  Persona scrape &
  Recommendation &
  English &
  US &
  US \\
\cite{perreault_algorithmic_2024} &
  2024 &
  Google &
  Misjudgement &
  Personalization &
  Persona scrape &
  Search &
  English &
  US &
  US \\
\cite{gleason_perceptions_2024} &
  2024 &
  Google, Bing &
  Distortion &
  Group misrepresentation &
  Non-persona scrape &
  Search &
  English &
  US &
  US \\
\cite{imana_auditing_2024} &
  2024 &
  Meta &
  Discrimination &
  Discrimination &
  Repurposing &
  Ad delivery &
  English &
  US &
  US \\
\cite{koronska_fact_2024} &
  2024 &
  Google &
  Distortion &
  Harmful content &
  Non-persona scrape &
  Search &
  English &
  US &
  Netherlands \\
\cite{radesky_algorithmic_2024} &
  2024 &
  YouTube &
  Distortion &
  Harmful content &
  Non-persona scrape &
  Search &
  English &
  US &
  US \\
\cite{amirova_framework-based_2024} &
  2024 &
  ChatGPT &
  Distortion &
  Information quality &
  Non-persona scrape &
  Generative AI &
  English &
  Mixed &
  UK, Cyprus \\
\cite{hosseinmardi_causally_2024} &
  2024 &
  YouTube &
  Distortion &
  Personalization, Information quality, filter bubble &
  Persona scrape &
  Recommendation &
  Mixed &
  Mixed &
  US, Switzerland \\
\cite{gong_does_2024} &
  2024 &
  Google &
  Distortion &
  Personalization, filter bubble &
  Persona scrape &
  Search &
  English &
  US &
  US \\
\cite{shi_new_2024} &
  2024 &
  Douyin &
  Discrimination &
  Discrimination &
  Persona scrape &
  Recommendation &
  Chinese &
  China &
  China \\
\cite{kim_exploring_2024} &
  2024 &
  ChatGPT &
  Distortion &
  Information quality &
  Non-persona scrape &
  Generative AI &
  English &
  US &
  US \\
\cite{samuel-azran_analyzing_2024} &
  2024 &
  Character ai &
  Distortion &
  Group misrepresentation &
  Non-persona scrape &
  Generative AI &
  English &
  Mixed &
  Israel \\
\cite{lambrecht_apparent_2024} &
  2024 &
  Google &
  Discrimination &
  Discrimination &
  Repurposing &
  Ad delivery &
  English &
  US &
  UK, US \\
\cite{george_use_2024} &
  2024 &
  YouTube &
  Distortion &
  Information quality &
  Non-persona scrape &
  Search &
  English, Italian &
  US, Italy &
  UK, Italy \\
\cite{sun_value_2024} &
  2024 &
  Alibaba &
  Exploitation &
  Personalization &
  Platform-led experiment &
  Recommendation &
  Chinese &
  China &
  US, China \\
\cite{cherifi_algorithmic_2024} &
  2024 &
  Twitter &
  Distortion &
  Personalization &
  Persona scrape &
  Recommendation &
  English &
  Mixed &
  US \\
\cite{huang_auditing_2024} &
  2024 &
  YouTube &
  Distortion &
  Information quality, News distribution &
  Non-persona scrape &
  Recommendation &
  English, French, German, Italian &
  Mixed &
  US, Hong Kong \\
\cite{santini_recommending_2023} &
  2023 &
  YouTube &
  Distortion &
  Harmful content &
  Non-persona scrape &
  Recommendation &
  Portuguese &
  Brazil &
  Brazil \\
\cite{sun_smiling_2023} &
  2024 &
  DALL-E, Google &
  Distortion &
  Group misrepresentation &
  Non-persona scrape &
  Generative AI, Search &
  English &
  US &
  US \\
\cite{jokubauskaite_winner-take-all_2023} &
  2023 &
  Webcam platforms &
  Distortion &
  Information quality &
  Non-persona scrape &
  Recommendation &
  Mixed &
  Mixed &
  Netherlands \\
\cite{mbalaka_epistemically_2023} &
  2023 &
  DALL-E, Starry AI &
  Distortion &
  Group misrepresentation &
  Non-persona scrape &
  Generative AI &
  English &
  Mixed &
  South Africa \\
\cite{lin_trapped_2023} &
  2023 &
  Google, Bing, Baidu &
  Distortion &
  Group misrepresentation &
  Non-persona scrape &
  Search &
  English, Chinese &
  Mixed &
  China, Canada \\
\cite{ibrahim_youtubes_2023} &
  2023 &
  YouTube &
  Distortion &
  Personalization, Information quality, Harmful content &
  Persona scrape &
  Recommendation &
  English &
  US &
  UAE \\
\cite{jaidka_silenced_2023} &
  2023 &
  Twitter &
  Discrimination &
  Discrimination, User categorization &
  Non-persona scrape &
  Recommendation &
  English &
  US &
  US, Singapore \\
\cite{stanusch_relationship_2023} &
  2023 &
  Google &
  Distortion &
  Information quality &
  Non-persona scrape &
  Search &
  English, Italian &
  US, Italy &
  Netherlands \\
\cite{kuznetsova_blame_2023} &
  2023 &
  Facebook &
  Distortion &
  Information quality, Personalization, News distribution &
  Persona scrape &
  Search, Recommendation &
  English &
  US &
  Germany, Switzerland \\
\cite{fujimoto_revisiting_2023} &
  2023 &
  ChatGPT &
  Distortion &
  Information quality &
  Non-persona scrape &
  Generative AI &
  English &
  Mixed &
  Japan \\
\cite{ungless_stereotypes_2023} &
  2023 &
  DALL-E, Stable Diffusion &
  Distortion &
  Group misrepresentation &
  Non-persona scrape &
  Generative AI &
  English &
  Mixed &
  UK, Germany \\
\cite{vakoch_this_2023} &
  2023 &
  Bing, DuckDuckGo, Google, Yandex, Yahoo &
  Distortion &
  Information quality &
  Non-persona scrape &
  Search &
  English, Russian, Chinese &
  Mixed &
  Switzerland, Germany \\
\cite{ng_exploring_2023} &
  2023 &
  YouTube &
  Distortion &
  Information quality &
  Crowdsourcing, Non-persona scrape &
  Recommendation &
  English &
  Mixed &
  US, Spain \\
\cite{toepfl_googling_2023} &
  2023 &
  Google &
  Distortion &
  Harmful content, News distribution &
  Non-persona scrape &
  Search &
  English, Russian &
  Russia, US, Germany, Ukraine, and Belarus &
  Germany \\
\cite{ieracitano_im_2023} &
  2023 &
  Google &
  Discrimination &
  Group misrepresentation &
  Non-persona scrape &
  Search &
  French, German, Italian, English &
  France, Germany, Italy, UK &
  Italy \\
\cite{papakyriakopoulos_beyond_2023} &
  2023 &
  Google &
  Discrimination &
  Group misrepresentation &
  Non-persona scrape &
  Search &
  Mixed &
  Mixed &
  US \\
\cite{ovalle_im_2023} &
  2023 &
  ChatGPT, other LLM &
  Discrimination &
  Group misrepresentation &
  Non-persona scrape &
  Generative AI &
  English &
  Mixed &
  US \\
\cite{kravets_different_2023} &
  2023 &
  Google, Yandex &
  Distortion &
  Harmful content &
  Non-persona scrape &
  Search &
  Russian &
  Belarus &
  Germany \\
\cite{rowland_shaping_2023} &
  2023 &
  Google &
  Distortion &
  Information quality &
  Non-persona scrape &
  Search &
  French, Spanish, Portuguese &
  France, Spain, Portugal &
  France, Spain, Portugal \\
\cite{lisa_zieringer_algorithmic_2023} &
  2023 &
  YouTube &
  Distortion &
  Harmful content &
  Non-persona scrape &
  Recommendation &
  German &
  Germany &
  Germany \\
\cite{leidinger_which_2023} &
  2023 &
  Google, DuckDuckGo, Yahoo &
  Discrimination &
  Group misrepresentation &
  Non-persona scrape &
  Search &
  English &
  Mixed &
  Netherlands \\
\cite{onyepunuka_multilingual_nodate} &
  2023 &
  YouTube &
  Distortion &
  Filter bubble &
  Non-persona scrape &
  Recommendation &
  Indonesian, English &
  Indonesia &
  US \\
\cite{le_merrer_modeling_2023} &
  2023 &
  Youtube &
  Distortion &
  Harmful content &
  Persona scrape &
  Recommendation &
  English &
  Mixed &
  France \\
\cite{matias_influencing_2023} &
  2023 &
  Reddit &
  Distortion &
  News distribution &
  Non-persona scrape &
  Recommendation &
  English &
  Mixed &
  US \\
\cite{yang_bubbles_2023} &
  2023 &
  Twitter &
  Distortion &
  Filter bubble &
  Persona scrape &
  Search &
  English &
  Mixed &
  Australia, Brazil \\
\cite{karan_your_2023}&
  2023 &
  Online market &
  Discrimination &
  Price discrimination &
  Persona scrape &
  E-commerce &
  English &
  US &
  US \\
\cite{schellingerhout_accounting_2023} &
  2023 &
  YouTube &
  Distortion &
  Harmful content &
  Persona scrape &
  Recommendation &
  English &
  Mixed &
  Netherlands \\
\cite{bandy_facebooks_2023} &
  2023 &
  Facebook &
  Distortion &
  News distribution &
  Crowdsourcing &
  Recommendation &
  English &
  US &
  US \\
\cite{juneja_assessing_2023} &
  2023 &
  YouTube &
  Distortion &
  Harmful content &
  Crowdsourcing &
  Search, Recommendation &
  English &
  US &
  US \\
\cite{patel_assessing_2023} &
  2023 &
  DuckDuckGo &
  Distortion &
  Harmful content &
  Non-persona scrape &
  Search &
  English &
  US &
  US \\
\cite{hagar_algorithmic_2023} &
  2023 &
  TikTok &
  Distortion &
  News distribution &
  Persona scrape &
  Recommendation &
  English &
  US &
  US \\
\cite{becerril-arreola_method_2023} &
  2023 &
  Online market &
  Discrimination &
  Price discrimination &
  Non-persona scrape &
  E-commerce &
  English &
  US &
  US \\
\cite{nechushtai_more_2023} &
  2023 &
  Google, Google News, Facebook, YouTube, Twitter &
  Distortion &
  News distribution &
  Crowdsourcing &
  Search &
  English &
  US &
  US \\
\cite{ulloa_novelty_2023} &
  2023 &
  Baidu, Bing, DuckDuckGo, Google, Yahoo &
  Distortion &
  News distribution &
  Non-persona scrape &
  Search &
  English &
  US, Germany &
  Germany, Switzerland, Finland \\
\cite{toepfl_who_2023} &
  2023 &
  Google &
  Distortion &
  Harmful content &
  Non-persona scrape &
  Search &
  English, German, Estonian, Belarusian, Russian, Ukrainian &
  US, Germany, Belarus, Russia, Estonia, Ukraine &
  Germany \\
\cite{dabran-zivan_is_2023} &
  2023 &
  Google &
  Distortion &
  Harmful content &
  Non-persona scrape &
  Recommendation &
  English, Arabic, Russian, Hebrew &
  US, UK, Nigeria, Philippines, Russia, Belarus, Kazakhstan, Egypt, Iraq, Israel &
  Israel, Switzerland \\
\cite{urman_auditing_2022} &
  2022 &
  Google, Bing, Yahoo, Baidu, Yandex, DuckDuckGo &
  Discrimination &
  Group misrepresentation &
  Non-persona scrape &
  Search &
  English, German &
  Germany &
  Switzerland, Germany \\
\cite{dunna_paying_2022} &
  2022 &
  YouTube &
  Discrimination &
  Discrimination &
  Non-persona scrape &
  Monetization &
  English &
  Mixed &
  US, Spain \\
\cite{buda_foundations_2022} &
  2022 &
  Foundations app &
  Discrimination &
  Discrimination &
  Crowdsourcing &
  Recommendation &
  English &
  Mixed &
  Spain \\
\cite{spyridou_modeling_2022} &
  2022 &
  News website &
  Distortion &
  News distribution &
  Crowdsourcing &
  Recommendation &
  Greek &
  Cyprus &
  Cyprus \\
\cite{makhortykh_memory_2022} &
  2022 &
  Bing, DuckDuckGo, Google, Yandex &
  Distortion &
  Information quality &
  Non-persona scrape &
  Search &
  English, Russian, Ukrainian, German &
  Germany &
  Switzerland, Germany \\
\cite{haak_auditing_2022} &
  2022 &
  Google, Bing &
  Distortion &
  Information quality &
  Non-persona scrape &
  Search &
  German &
  Germany &
  Germany \\
\cite{boratto_crowdsourcing_2022} &
  2022 &
  Google &
  Distortion &
  Personalization &
  Crowdsourcing &
  Search &
  English &
  Germany, Brazil, US, India, Spain &
  Australia \\
\cite{farkas_how_2022} &
  2022 &
  Google Translate &
  Discrimination &
  Group misrepresentation &
  Non-persona scrape &
  Translation &
  English, Hungarian &
  Mixed &
  Hungary \\
\cite{albadi_deradicalizing_2022} &
  2022 &
  YouTube &
  Distortion &
  Harmful content &
  Persona scrape &
  Recommendation &
  Arabic &
  Mixed &
  Saudi Arabia, US \\
\cite{iqbal_left_2022} &
  2022 &
  Gmail, Outlook, Yahoo &
  Discrimination &
  Discrimination &
  Persona scrape &
  Spam &
  English &
  US &
  US \\
\cite{kravets_gauging_2022} &
  2022 &
  Google, Yandex, Google News, Yandex News &
  Distortion &
  Information quality, News distribution &
  Non-persona scrape &
  Search &
  Russian &
  Mixed &
  Germany \\
\cite{makhortykh_story_2022} &
  2022 &
  Google, Yandex &
  Distortion &
  Information quality &
  Non-persona scrape &
  Search &
  Russian &
  Russia &
  Switzerland, Netherlands \\
\cite{evans_google_2023} &
  2022 &
  Google News &
  Distortion &
  Personalization &
  Crowdsourcing &
  Search &
  English &
  UK &
  UK \\
\cite{urman_matter_2022} &
  2022 &
  Google, Baidu, Bing, DuckDuckGo, Yahoo, Yandex &
  Distortion &
  Information quality &
  Non-persona scrape &
  Search &
  English &
  US &
  Switzerland, Germany \\
\cite{matias_software-supported_2022} &
  2022 &
  Facebook, Google &
  Discrimination &
  Personalization &
  Repurposing &
  Ad delivery &
  English &
  US &
  US \\
\cite{ledwich_radical_2022} &
  2022 &
  YouTube &
  Distortion &
  Filter bubble &
  Persona scrape &
  Recommendation &
  English &
  US &
  US \\
\cite{urman_foreign_2022} &
  2022 &
  Google &
  Discrimination &
  Group misrepresentation &
  Non-persona scrape &
  Search &
  English &
  US, Ireland &
  Switzerland \\
\cite{boeker_empirical_2022} &
  2022 &
  TikTok &
  Distortion &
  Personalization &
  Persona scrape &
  Recommendation &
  English, German, French &
  US, Canada, Germany &
  Switzerland, Germany \\
\cite{urman_where_2022} &
  2022 &
  Google, DuckDuckGo, Bing, Yahoo, Yandex &
  Distortion &
  Harmful content &
  Non-persona scrape &
  Search &
  English &
  US, UK &
  Switzerland, Germany \\
\cite{huszar_algorithmic_2022} &
  2022 &
  Twitter &
  Distortion &
  News distribution, Information quality &
  Platform-led experiment &
  Recommendation &
  English, Japanese, French, Spanish, German &
  US, UK, Japan, France, Spain, Canada, Germany &
  US, UK \\
\cite{fabris_algorithmic_2021} &
  2021 &
  Comparison website &
  Discrimination &
  Price discrimination &
  Persona scrape &
  E-commerce &
  Italian &
  Italy &
  Italy, US \\
\cite{unkel_googling_2021} &
  2021 &
  Google &
  Distortion &
  Information quality &
  Persona scrape &
  Search &
  German &
  Germany &
  Germany \\
\cite{heuer_auditing_2021} &
  2021 &
  YouTube &
  Distortion &
  Information quality &
  Non-persona scrape &
  Recommendation &
  German &
  Germany &
  Germany \\
\cite{urman_auditing_2021} &
  2021 &
  Google, Bing, Yahoo, Yandex, DuckDuckGo &
  Distortion &
  Information quality &
  Non-persona scrape &
  Search &
  English, Russian &
  Germany &
  Switzerland, Germany \\
\cite{bartley_auditing_2021} &
  2021 &
  Twitter &
  Distortion &
  Information quality &
  Persona scrape &
  Recommendation &
  English &
  Mixed &
  US, Chile \\
\cite{van_es_gendered_2021} &
  2021 &
  Job websites &
  Discrimination &
  Discrimination &
  Non-persona scrape &
  Search &
  Dutch &
  Netherlands &
  Netherlands \\
\cite{imana_auditing_2021} &
  2021 &
  Facebook, Linkedin &
  Discrimination &
  Discrimination &
  Repurposing &
  Ad delivery &
  English &
  US &
  US \\
\cite{tomlein_audit_2021} &
  2021 &
  YouTube &
  Distortion &
  Filter bubble, Harmful content &
  Persona scrape &
  Recommendation &
  English &
  Mixed &
  Slovakia \\
\cite{murthy_evaluating_2021} &
  2021 &
  YouTube &
  Distortion &
  Harmful content &
  Non-persona scrape &
  Recommendation &
  Mixed &
  Mixed &
  US \\
\cite{dash_when_2021} &
  2021 &
  Amazon &
  Distortion &
  Information quality &
  Non-persona scrape &
  Recommendation &
  English &
  Mixed &
  India, Germany \\
\cite{kollyri_-coding_2021} &
  2021 &
  Instagram &
  Distortion &
  Filter bubble &
  Persona scrape &
  Recommendation &
  English &
  Mixed &
  Netherlands \\
\cite{lossio-ventura_yttrex_2021} &
  2021 &
  YouTube &
  Distortion &
  Filter bubble &
  Crowdsourcing &
  Recommendation &
  English, Spanish, Chinese, Portuguese, Arabic &
  Mixed &
  Italy \\
\cite{metaxa_image_2021} &
  2021 &
  Google &
  Discrimination &
  Group misrepresentation &
  Non-persona scrape &
  Search &
  English &
  US &
  US \\
\cite{glaesener_exploring_2022} &
  2021 &
  Siri &
  Distortion &
  Information quality &
  Crowdsourcing &
  Search &
  English &
  US &
  Sweden \\
\cite{juneja_auditing_2021} &
  2021 &
  Amazon &
  Distortion &
  Harmful content &
  Persona scrape, non-persona scrape &
  Search, Recommendation &
  English &
  US &
  US \\
\cite{bandy_more_2021} &
  2021 &
  Twitter &
  Distortion &
  Information quality, News distribution &
  Persona scrape &
  Recommendation &
  English &
  US &
  US \\
\cite{gezici_evaluation_2021} &
  2021 &
  Google, Bing &
  Distortion &
  Information quality &
  Non-persona scrape &
  Search &
  English &
  US &
  Turkey, UK \\
\cite{dambanemuya_auditing_2021} &
  2021 &
  Alexa &
  Distortion &
  Information quality, News distribution &
  Non-persona scrape &
  Search &
  English &
  US &
  US \\
\cite{lutz_examining_2021} &
  2021 &
  YouTube &
  Distortion &
  Information quality &
  Non-persona scrape &
  Search, Recommendation &
  English &
  US &
  US \\
\cite{bandy_curating_2021} &
  2021 &
  Twitter &
  Distortion &
  News distribution, Harmful content &
  Persona scrape &
  Recommendation &
  English &
  US &
  US \\
\cite{thorson_algorithmic_2021} &
  2021 &
  Facebook &
  Distortion &
  Personalization &
  Crowdsourcing &
  Ad delivery &
  English &
  US &
  US, Netherlands \\
\cite{ali_ad_2021} &
  2021 &
  Facebook &
  Distortion &
  Personalization &
  Repurposing &
  Ad delivery &
  English &
  US &
  US \\
\cite{fosch-villaronga_little_2021} &
  2021 &
  Twitter &
  Misjudgement &
  Group misrepresentation &
  Crowdsourcing &
  User categorization &
  English &
  Mixed &
  Netherlands, Australia, Norway \\
\cite{bandy_errors_2021} &
  2021 &
  Google &
  Misjudgement &
  Personalization &
  Repurposing &
  Ad delivery &
  English &
  US &
  US \\
\cite{bonart_investigation_2019} &
  2020 &
  Google, Bing, DuckDuckGo &
  Distortion &
  Information quality &
  Non-persona scrape &
  Search &
  German &
  Germany &
  Germany \\
\cite{reber_data_2020} &
  2020 &
  Google &
  Distortion &
  Harmful content &
  Crowdsourcing &
  Ad delivery &
  English &
  Mixed &
  Germany \\
\cite{abul-fottouh_examining_2020} &
  2020 &
  YouTube &
  Distortion &
  Harmful content &
  Non-persona scrape &
  Recommendation &
  English &
  Mixed &
  Canada \\
\cite{faddoul_longitudinal_2020} &
  2020 &
  YouTube &
  Distortion &
  Harmful content &
  Non-persona scrape &
  Recommendation &
  English &
  Mixed &
  US \\
\cite{shin_algorithms_2020} &
  2020 &
  Amazon &
  Distortion &
  Harmful content &
  Non-persona scrape &
  Search &
  English &
  US &
  US \\
\cite{hussein_measuring_2020} &
  2020 &
  YouTube &
  Distortion &
  Harmful content, Personalization &
  Persona scrape &
  Recommendation &
  English &
  US &
  US \\
\cite{mustafaraj_case_2020} &
  2020 &
  Google &
  Distortion &
  Information quality &
  Non-persona scrape, Persona scrape &
  Search &
  English &
  US &
  US \\
\cite{ribeiro_auditing_2020} &
  2020 &
  YouTube &
  Distortion &
  Harmful content &
  Non-persona scrape &
  Recommendation &
  English &
  US &
  Switzerland, Brazil \\
\cite{asplund_auditing_2020} &
  2020 &
  Google &
  Discrimination &
  Discrimination &
  Persona scrape &
  Ad delivery &
  English &
  US &
  US \\
\cite{bandy_auditing_2020} &
  2020 &
  Apple News &
  Distortion &
  News distribution, Information quality &
  Persona scrape, Crowdsourcing &
  Recommendation &
  English &
  US &
  US \\
\cite{fischer_auditing_2020} &
  2020 &
  Google News &
  Distortion &
  News distribution, Personalization, Information quality &
  Persona scrape &
  Search &
  English &
  US &
  US \\
\cite{kaiser_birds_2020} &
  2020 &
  YouTube &
  Distortion &
  Harmful content, Filter bubble &
  Persona scrape &
  Recommendation &
  English, German &
  US, Germany &
  US, Germany, Taiwan \\
\cite{silva_facebook_2020}. &
  2020 &
  Facebook &
  Misjudgement &
  Information quality &
  Crowdsourcing &
  Ad delivery &
  Portuguese &
  Brazil &
  Brazil, France \\
\cite{smets_does_2019} &
  2019 &
  Google Maps &
  Distortion &
  Filter bubble &
  Persona scrape &
  Search &
  Dutch, French &
  Belgium &
  Belgium \\
\cite{moe_comparing_2019} &
  2019 &
  YouTube &
  Distortion &
  Information quality &
  Non-persona scrape &
  Search &
  Danish, Norwegian, Swedish &
  Denmark, Norway, Sweden &
  Norway \\
\cite{puschmann_beyond_2019} &
  2019 &
  Google, Google News &
  Distortion &
  News distribution, Personalization, Information quality &
  Crowdsourcing &
  Search &
  German &
  Germany &
  Germany \\
\cite{weber_you_2019} &
  2019 &
  Google &
  Distortion &
  Personalization &
  Crowdsourcing &
  Search &
  English &
  New Zealand &
  New Zealand \\
\cite{hu_auditing_2019} &
  2019 &
  Google &
  Distortion &
  Information quality &
  Non-persona scrape &
  Search &
  English &
  US &
  US \\
\cite{lurie_opening_nodate} &
  2019 &
  Google &
  Distortion &
  Information quality, News distribution &
  Non-persona scrape &
  Search &
  English &
  US &
  US \\
\cite{nechushtai_what_2019} &
  2019 &
  Google News &
  Distortion &
  Information quality, News distribution &
  Crowdsourcing &
  Search &
  English &
  US &
  US \\
\cite{kulshrestha_search_2019} &
  2019 &
  Twitter, Google &
  Distortion &
  Information quality &
  Non-persona scrape &
  Search &
  English &
  US &
  US, Germany \\
\cite{trielli_search_2019} &
  2019 &
  Google &
  Distortion &
  News distribution, Information quality &
  Non-persona scrape &
  Search &
  English &
  US &
  US \\
\cite{lambrecht_algorithmic_2019} &
  2019 &
  Facebook, Google, Instagram, Twitter &
  Discrimination &
  Discrimination &
  Repurposing &
  Ad delivery &
  English &
  Comprehensive &
  US \\
\cite{geyik_fairness-aware_2019} &
  2019 &
  LinkedIn &
  Discrimination &
  Discrimination &
  Platform-led experiment &
  Search &
  English &
  US &
  US \\
\cite{metaxa_search_2019} &
  2019 &
  Google &
  Distortion &
  Information quality &
  Non-persona scrape &
  Search &
  English &
  US &
  US \\
\cite{robertson_auditing_2019} &
  2019 &
  Google, Bing &
  Distortion &
  Information quality &
  Non-persona scrape &
  Search &
  English &
  US &
  US \\
\cite{cano-oron_dr_2019} &
  2019 &
  Google &
  Distortion &
  Information quality &
  Non-persona scrape &
  Search &
  English, Spanish, French &
  Spain, France, Mexico, US, UK &
  Spain \\
\cite{venkatadri_auditing_2019} &
  2019 &
  Facebook, Acxiom, Epsiolon, Experian, Oracle (Datalogix) &
  Misjudgement &
  Group misrepresentation &
  Crowdsourcing &
  Ad delivery &
  Mixed &
  US, Australia, UK, Germany, France, Brazil, Japan &
  US, Germany \\
\cite{bashir_quantity_2019} &
  2019 &
  Google, Facebook, Oracle BlueKai, and Neilsen eXelate &
  Misjudgement &
  Group misrepresentation &
  Crowdsourcing &
  Ad delivery &
  Mixed &
  US, Pakistan &
  US, Pakistan \\
\cite{vincent_measuring_2019} &
  2019 &
  Google &
  Exploitation &
  Information quality &
  Persona scrape &
  Search &
  English &
  US &
  US \\
\cite{tschantz_accuracy_2018} &
  2018 &
  Google &
  Misjudgement &
  Group misrepresentation &
  Crowdsourcing &
  Ad delivery &
  English &
  US &
  US \\
\cite{courtois_challenging_2018} &
  2018 &
  Google &
  Distortion &
  Filter bubble &
  Crowdsourcing &
  Search &
  Dutch &
  Belgium &
  Belgium \\
\cite{bechmann_are_2018} &
  2018 &
  Facebook &
  Distortion &
  Filter bubble, News distribution &
  Crowdsourcing &
  Recommendation &
  Danish &
  Denmark &
  Denmark \\
\cite{haim_burst_2018} &
  2018 &
  Google News &
  Distortion &
  News distribution, Filter bubble &
  Persona scrape &
  Recommendation &
  German &
  Germany &
  Germany \\
\cite{rieder_ranking_2018} &
  2018 &
  YouTube &
  Distortion &
  Information quality &
  Non-persona scrape &
  Search &
  English &
  Mixed &
  Netherlands, Australia, Spain \\
\cite{robertson_auditing_2018} &
  2018 &
  Google &
  Distortion &
  Filter bubble, Information quality &
  Crowdsourcing &
  Search &
  English &
  US &
  US \\
\cite{robertson_auditing_2018-1} &
  2018 &
  Google &
  Distortion &
  Personalization, Information quality &
  Crowdsourcing &
  Search &
  English &
  US &
  US \\
\cite{chakraborty_analyzing_2018} &
  2018 &
  New York Times &
  Distortion &
  News distribution, personalization &
  Persona scrape &
  Recommendation &
  English &
  US &
  India \\
\cite{hupperich_empirical_2018} &
  2018 &
  Booking, Hotels, Avis, Hrs, Orbitz &
  Discrimination &
  Price discrimination &
  Persona scrape &
  E-commerce &
  Mixed &
  France, Georgia, Germany, Pakistan, Russia, US &
  Netherlands, Germany \\
\cite{chen_investigating_2018} &
  2018 &
  Indeed, Monster, CareerBuilder &
  Discrimination &
  Discrimination &
  Persona scrape &
  Search &
  English &
  US &
  US \\
\cite{cabanas_unveiling_2018} &
  2018 &
  Facebook &
  Exploitation &
  Group misrepresentation &
  Repurposing &
  Ad delivery &
  Mixed &
  EU &
  Spain \\
\cite{eriksson_tracking_2017} &
  2017 &
  Spotify &
  Distortion &
  Personalization &
  Persona scrape &
  Recommendation &
  Mixed &
  Mixed &
  Sweden \\
\cite{snickars_more_2017} &
  2017 &
  Spotify &
  Distortion &
  Personalization &
  Persona scrape &
  Recommendation &
  Mixed &
  Mixed &
  Sweden \\
\cite{eslami_be_2017} &
  2017 &
  Booking &
  Distortion &
  Information quality &
  Non-persona scrape &
  E-commerce &
  English &
  US &
  US \\
\cite{kulshrestha_quantifying_2017} &
  2017 &
  Twitter, Google &
  Distortion &
  Information quality &
  Non-persona scrape &
  Search &
  English &
  US &
  US, Germany \\
\cite{weber_coding_2018} &
  2017 &
  News apps (wide range) &
  Distortion &
  News distribution &
  Code audit &
  Recommendation &
  Mixed &
  US &
  US \\
\cite{hannak_bias_2017} &
  2017 &
  TaskRabbit, Fiverr &
  Discrimination &
  Discrimination &
  Non-persona scrape &
  Search &
  English &
  US &
  US \\
\cite{mahler_studying_2017} &
  2017 &
  Spotify &
  Exploitation &
  Group misrepresentation &
  Non-persona scrape &
  Ad delivery &
  Mixed &
  Mixed &
  Sweden \\
\cite{mcmahon_substantial_2017} &
  2017 &
  Google &
  Exploitation &
  Information quality &
  Crowdsourcing &
  Search &
  English &
  US &
  US \\
\cite{soeller_mapwatch_2016} &
  2016 &
  Google Maps, Bing Maps &
  Distortion &
  Personalization &
  Persona scrape &
  Search &
  NA &
  Morocco, Argentina, China, India, Russia, Ukraine &
  US \\
\cite{spiro_identifying_2016} &
  2016 &
  Google, Bing &
  Discrimination &
  Group misrepresentation &
  Non-persona scrape &
  Search &
  Mixed &
  Diverse (shortened here for readability) &
  Brazil \\
\cite{chen_empirical_2016} &
  2016 &
  Amazon &
  Discrimination &
  Price discrimination &
  Non-persona scrape &
  E-commerce &
  English &
  US &
  US \\
\cite{kliman-silver_location_2015} &
  2015 &
  Google &
  Distortion &
  Personalization &
  Persona scrape &
  Search &
  English &
  US &
  US \\
\cite{datta_automated_nodate} &
  2015 &
  Google &
  Exploitation &
  Personalization &
  Persona scrape &
  Ad delivery &
  English &
  US &
  US \\
\cite{kay_unequal_2015} &
  2015 &
  Google &
  Discrimination &
  Group misrepresentation &
  Non-persona scrape &
  Search &
  English &
  US &
  US \\
\cite{hannak_measuring_2014} &
  2014 &
  E-commerce (wide range) &
  Discrimination &
  Price discrimination &
  Persona scrape, Crowdsourcing &
  E-commerce &
  English &
  US &
  US \\
\cite{hannak_measuring_2013} &
  2013 &
  Google &
  Distortion &
  Personalization &
  Persona scrape, Crowdsourcing &
  Search &
  English &
  US &
  US \\
\cite{mikians_crowd-assisted_2013} &
  2013 &
  E-commerce (wide range) &
  Discrimination &
  Price discrimination &
  Crowdsourcing &
  E-commerce &
  Mixed &
  Mixed &
  Spain \\
\cite{sweeney_discrimination_2013} &
  2013 &
  Google &
  Discrimination &
  Group misrepresentation &
  Non-persona scrape &
  Ad delivery &
  English &
  US &
  US \\
\cite{noble_google_2013} &
  2013 &
  Google &
  Discrimination &
  Group misrepresentation &
  Non-persona scrape &
  Search &
  English &
  US &
  US \\
\cite{mikians_detecting_2012} &
  2012 &
  Google, Bing, Amazon, E-commerce (wide range) &
  Discrimination &
  Price discrimination &
  Persona scrape &
  E-commerce &
  Mixed &
  Greece, Hungary, Italy, Germany, US, Austria, UK, Poland, Spain &
  US, Spain \\ \bottomrule
\end{tabular}%
}
\end{table}

\end{document}